\documentclass[useAMS,usenatbib]{mn2e}
\usepackage{aas_macros}
\usepackage{graphicx}
\usepackage{lscape}
\usepackage{mathptmx}
\usepackage{times}
\usepackage{amssymb}
\usepackage{hyperref}

\def\p0{\phantom{0}}

\begin{document}  
\title[IFRSs: A New Population of High-$z$ Radio Galaxies]{Infrared-Faint Radio Sources: A New Population of High-redshift Radio Galaxies}

\author[J.~D.~Collier et al. 2013]
{J.~D.~Collier$^{1,2}$\thanks{E-mail: j.collier@uws.edu.au},
J.~K.~Banfield$^{2,3}$,
R.~P.~Norris$^{2}$,
D.~H.~F.~M.~Schnitzeler$^{4}$,
A.~E.~Kimball$^{2,5}$
\newauthor M.~D.~Filipovi\'c$^{1}$,
T.~H.~Jarrett$^{6}$,
C.~J.~Lonsdale$^{5}$,
and N.~F.~H. Tothill$^{1}$.
\\
$^{1}$University of Western Sydney, Locked Bag 1797, Penrith South, DC, NSW
1797, Australia\\
$^{2}$CSIRO Astronomy and Space Science, Marsfield, NSW 2122, Australia\\
$^{3}$Research School of Astronomy and Astrophysics, Australian National University, Weston Creek, Australian Capital Territory 2611, Australia\\
$^{4}$Max Planck Institut f\"{u}r Radioastronomie, Auf dem H\"{u}gel 69, 53121 Bonn, Germany\\
$^{5}$National Radio Astronomy Observatory, 520 Edgemont Rd., Charlottesville, VA, 22903, USA\\
$^{6}$Astronomy Department, University of Cape Town, Rondebosch 7700, South Africa\\
}

\date{Accepted 2013 December 20. Received 2013 December 17; in original form 2013 October 11}

\pagerange{\pageref{firstpage}--\pageref{lastpage}} \pubyear{2014}

\maketitle

\label{firstpage}

\begin{abstract}
We present a sample of 1317 Infrared-Faint Radio Sources (IFRSs) that, for the first time, are reliably detected in the infrared, generated by cross-correlating the {\it Wide-Field Infrared Survey Explorer} ({\it WISE}) all-sky survey with major radio surveys. Our IFRSs are brighter in both radio and infrared than the {\it first generation} IFRSs that were undetected in the infrared by the {\it Spitzer Space Telescope}. We present the first spectroscopic redshifts of IFRSs, and find that all but one of the IFRSs with spectroscopy has $z > 2$. We also report the first X-ray counterparts of IFRSs, and present an analysis of radio spectra and polarization, and show that they include Gigahertz-Peaked Spectrum, Compact Steep Spectrum, and Ultra-Steep Spectrum sources. These results, together with their {\it WISE} infrared colours and radio morphologies, imply that our sample of IFRSs represents a population of radio-loud Active Galactic Nuclei at $z > 2$. We conclude that our sample consists of lower-redshift counterparts of the extreme first generation IFRSs, suggesting that the fainter IFRSs are at even higher redshift.
\end{abstract}

%**********************************************************************************************************************************************************

%**********************************************************************************************************************************************************
%**********************************************************************************************************************************************************

%**********************************************************************************************************************************************************
\section{Introduction}

Infrared-Faint Radio Sources (IFRSs) are rare objects that were first identified by \citet{Norris2006} as radio sources which were detected at $\lambda = 20\,$cm ($\nu = 1.4$ GHz) in the deep radio observations of the Australia Telescope Large Area Survey (ATLAS) Chandra Deep Field South \citep[CDFS;][]{2002ApJ...566..667R}, but which were not detected in the {\it Spitzer} Wide-area InfraRed Extragalactic Survey \citep[SWIRE;][]{2003PASP..115..897L} at 3.6, 4.5, 5.8, 8 and $24\,\mu$m. A total of 22 IFRSs were identified by \citet{Norris2006}, which were undetected at 3.6$\,\mu$m down to a 3$\sigma$ level of 3$\,\mu$Jy. At the time, this was an unexpected discovery since it was believed that SWIRE would detect all Active Galactic Nuclei (AGN) or Star-Forming Galaxies (SFGs) that were observable at radio frequencies. Similarly, \citet{Middelberg2008} identified a sample of 31 IFRSs when cross-matching the ATLAS radio observations of the European Large Area ISO Survey South 1 \citep[ELAIS-S1;][]{2000MNRAS.316..749O} field with the co-spatial SWIRE observations, which had similar infrared (IR) sensitivities to the SWIRE CDFS observations. 

Most of these sources were found to have flux densities at $20\,$cm of a few hundred $\mu$Jy, but some were as bright as $20\,$mJy. All 53 IFRSs discovered in ATLAS also lacked optical counterparts. Therefore, IFRSs may be extreme counterparts of the Optically Invisible Radio Sources (OIRS) identified by \citet{2005ApJ...626...58H}, which are compact radio sources undetected in the optical up to an {\it R}-band magnitude of $\sim$25.7. \cite{2008ApJ...688..885H} show that 34 per cent of their OIRS are not detected at 3.6 $\mu$m, and conclude that these undetected sources appear to embody a sample of powerful radio galaxies at $z > 2$. We refer to the IFRSs discovered in ATLAS which lacked optical and IR counterparts as {\it first generation} IFRSs. Since IFRSs were originally discovered, eight IR and nine optical counterparts have been potentially detected \citep{2008MNRAS.391.1000G,2010ApJ...710..698H}, all of which are extremely faint. 

The nature of IFRSs remains unconfirmed, given their non-detections at optical and infrared wavelengths. Putative explanations of their nature have included: (1) high-redshift radio-loud galaxies; (2) extremely obscured radio galaxies at moderate redshifts ($1 < z < 2$); (3) lobes or hotspots of nearby unidentified radio galaxies; (4) very obscured, luminous starburst galaxies; (5) AGN or starburst galaxies in a transitory phase; (6) high-latitude pulsars; (7) misidentifications; (8) an unknown type of object; (9) a combination of these \citep[][and references therein]{Norris2011}. The study of IFRSs has been almost entirely limited to their properties at radio frequencies.

\subsection{Infrared-Faint Radio Sources}
\label{Overview}

The first steps taken toward probing the nature of IFRSs were the Very Long Baseline Interferometry (VLBI) observations undertaken by \cite{NorrisVLBI} and \cite{MiddelbergVLBI}, who respectively observed two and four IFRSs originally identified in ATLAS. \cite{NorrisVLBI} identified an AGN within one of the IFRSs at a flux density of 5.0 mJy, implying a core size of $<$ 0.03 arcsec, corresponding to a linear size of $\le$ 260 pc at any redshift. \cite{NorrisVLBI} suggested that if such an AGN were at a redshift of $z = 1$, it would be detected by SWIRE. However, they proposed that at a redshift of $z = 7$, an AGN like this would most likely elude deep infrared detection, but could still be detected in deep radio observations. 

\cite{MiddelbergVLBI} also detected only one IFRS with VLBI. The detected source had a flux density on the longest baselines of $7\,$ mJy, which they inferred corresponded to a brightness temperature of $T_{\rm{B,min}} = 3.6 \times 10^6$ K, indicating non-thermal emission from an AGN. They showed that the detected IFRS had properties consistent with that of a high redshift ($z > 1$) Compact Steep Spectrum (CSS) source.

A further study of IFRSs was conducted by \cite{2008MNRAS.391.1000G}, who analysed 14 IFRSs in the {\it Spitzer} First Look Survey (FLS) field, using {\it Spitzer} Infrared Array Camera \citep[IRAC;][]{2004ApJS..154...10F} and Multiband Imaging Photometer \citep[MIPS;][]{2004ApJS..154...25R} data, as well as 20 cm Very Large Array (VLA) data from \citet{2003AJ....125.2411C}. Their sample is complementary to that of the previous samples of ATLAS IFRSs, since their $3.6\,\mu$m 3$\sigma$ sensitivity is $\sim 9\,\mu$Jy. Eight optical detections with a median AB magnitude of $R_{\rm AB} = 24.4$ suggest that the sample is a much brighter population of IFRSs, and may be at lower redshift. \cite{2008MNRAS.391.1000G} suggested that since they did not see an increase in the population of IFRSs at low flux densities ($\sim$ 1 mJy), at which point the contribution of SFGs becomes significant, their sources were unlikely to be obscured SFGs. Additionally, upper limits of the flux density ratio $q_{\rm IR} = \log_{10}(S_{\rm IR} / S_{1.4})$ \citep{2004ApJS..154..147A} were calculated to be $q_{24} < -0.7$ and $q_{70} < 1$, well below the typical values derived for SFGs. Through source stacking in the four IRAC bands,  \cite{2008MNRAS.391.1000G} showed that six sources had possible infrared counterparts below the detection threshold. Using a stacked image of the remaining eight sources that were not identified as having a potential counterpart, they found an upper limit of the median 3.6 $\mu$m flux density of $3\sigma/\sqrt{8} = 3.1~\mu$Jy. \cite{2008MNRAS.391.1000G} showed that IFRSs are made up of a population of flat, steep and Ultra-Steep Spectrum (USS) sources, from which they suggested that IFRSs ought not to be treated as a single source population. By modelling the Spectral Energy Distributions (SEDs) of the IFRSs from each of these three classes of radio spectra separately, and by placing upper limits on their linear size, they showed that all their IFRSs could be modelled as well-known Fanaroff and Riley Type II \citep[FR II;][]{1974MNRAS.167P..31F} radio galaxies which are less luminous and placed at high redshift.

Possible infrared detections of two ATLAS IFRSs were made by \citet{2010ApJ...710..698H}, who found 3.6 $\mu$m flux densities of 5.5 $\pm$ 0.3 and 6.6 $\pm$ 0.3 $\mu$Jy for the two sources, using ultra-deep {\it Spitzer} imaging of the extended Chandra Deep Field South (eCDFS). The fainter of these sources also contained an optical Advanced Camera for Surveys \citep[ACS;][]{2004ApJ...600L..93G} counterpart at ${\rm {\it V}_{AB}} = 26.27$ and ${\rm {\it z}_{AB}} = 25.62$ magnitudes. \citet{2010ApJ...710..698H} conducted detailed modelling of the SEDs of these two IFRSs, and two others that were undetected. \citet{2010ApJ...710..698H} found that the data could be reproduced by a 3C 273-like object which, when detected in the infrared, was redshifted to $z = 2$, and when not detected in the infrared, was redshifted to $z > 4$. Furthermore, no non-detected IFRSs could be explained by any SED template at redshifts smaller than $z = 4$. \citet{2010ApJ...710..698H} concluded that their four IFRSs lie well beyond the radio-infrared correlation, since none of them was detected at 24 $\mu$m, down to a 5$\sigma$ level of 50 $\mu$Jy. Hence, their radio emission cannot be accounted for by star formation and must be due to the presence of an AGN.

The deepest {\it Spitzer} imaging to date of the larger regions of the CDFS and ELAIS-S1 fields comes from the {\it Spitzer} Extragalactic Representative Volume Survey \citep[SERVS;][]{2012PASP..124..714M}, which has a 3$\sigma$ noise level of $\sim$ 1.5 $\mu$Jy, and which mostly overlaps with the ATLAS regions where the first IFRSs were identified. Using the 3.6 $\mu$m SERVS data, \citet{Norris2011} found three candidate detections of IFRSs at levels of $\sim$ 2 $\mu$Jy. However, they concluded that 2$-$3 of these detections could be spurious detections due to confusion. After producing a stacked infrared image, they found a median flux density of $\sim$0.2 $\mu$Jy or less, attributing very extreme radio to infrared flux density ratios to these objects. \citet{Norris2011} found no evidence of a cross-identification for the two candidate detections from \citet{2010ApJ...710..698H}, which would have appeared at $\sim$11$\sigma$ and $\sim$13$\sigma$ in SERVS. However, new radio data with greater positional accuracy from \citet{2013ApJS..205...13M} has revealed that while one of the \citet{2010ApJ...710..698H} IR counterparts is probably due to confusion, the other coincides well with the updated radio position and consequently, we consider it to be a reliable match.

\citet{Maini2013} identified a further 21 IFRSs using more recent SERVS data, which went down to an even deeper 3$\sigma$ level of $\sim$1$\,\mu$Jy at 3.6$\,\mu$m, and which extended to the Lockman Hole. They also found a number of new candidate IR detections, and showed that these IFRSs are well modelled as Quasi Stellar Objects (QSOs) at $3 < z < 5$, based on their 3.6 and 4.5$\,\mu$m tracks as compared to the SED tracks of other classes of object. Using median stacking, \citet{Maini2013} found that the undetected IFRSs had flux density upper limits of 0.3 and $0.4\,\mu$Jy, respectively at 3.6$\,\mu$m and 4.5$\,\mu$m.

\citet{Herzog2013} presented a separate sample of IFRSs from the CDFS, and found them to have very similar properties to HzRGs.

\begin{figure*}
\centering
\includegraphics[width=0.73\textwidth]{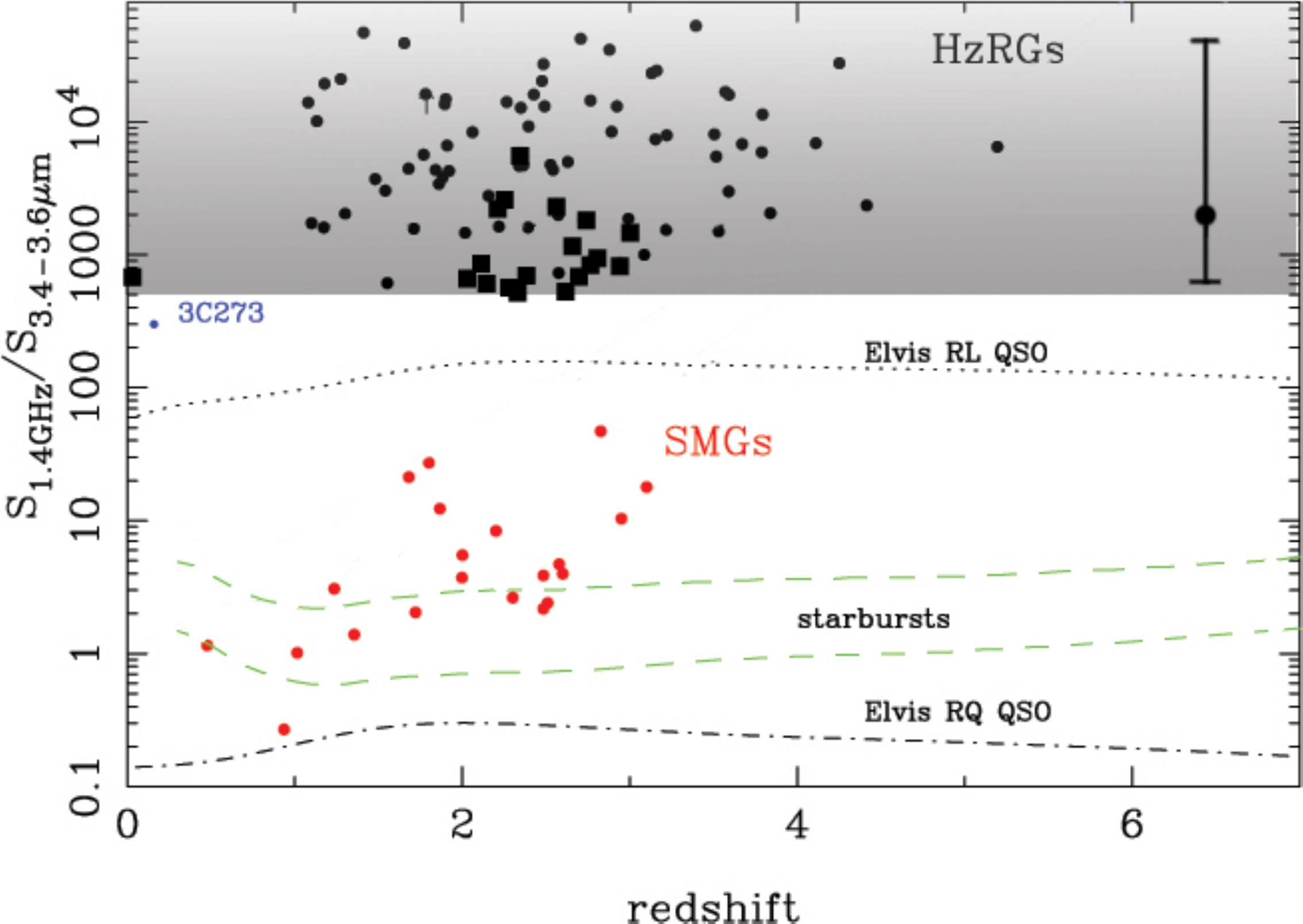}
\caption{The flux density ratio between 20 cm and 3.4$-$3.6 $\mu$m as a function of redshift, for a number of different models, adapted from \citet{Norris2011}. The area in grey represents the ratio range above 500 which all IFRSs occupy. The black squares represent the IFRSs with redshifts from this paper (see \S~\ref{SampleSelection} and \ref{Redshifts}). The filled circles within the grey area are the HzRGs from \citet{2007ASPC..380..393S}. The large black dot and error bar on the right marks the likely range of the first generation IFRSs. The area defined by the dashed green lines either side of the label ``starbursts" represents the expected loci of Luminous Infrared Galaxies (LIRG) and Ultra-Luminous Infrared Galaxies (ULIRG) (using the SED template from \citealt{2009ApJ...692..556R}) and the dotted and dot-dashed lines respectively indicate the loci of a classical radio-loud and radio-quiet QSOs (from \citealt{1994IAUS..159...25E}). The red dots show the locations of classical submillimetre galaxies.}
\label{Olay1}
\end{figure*}

\begin{figure*}
\centering
\includegraphics[width=0.73\textwidth]{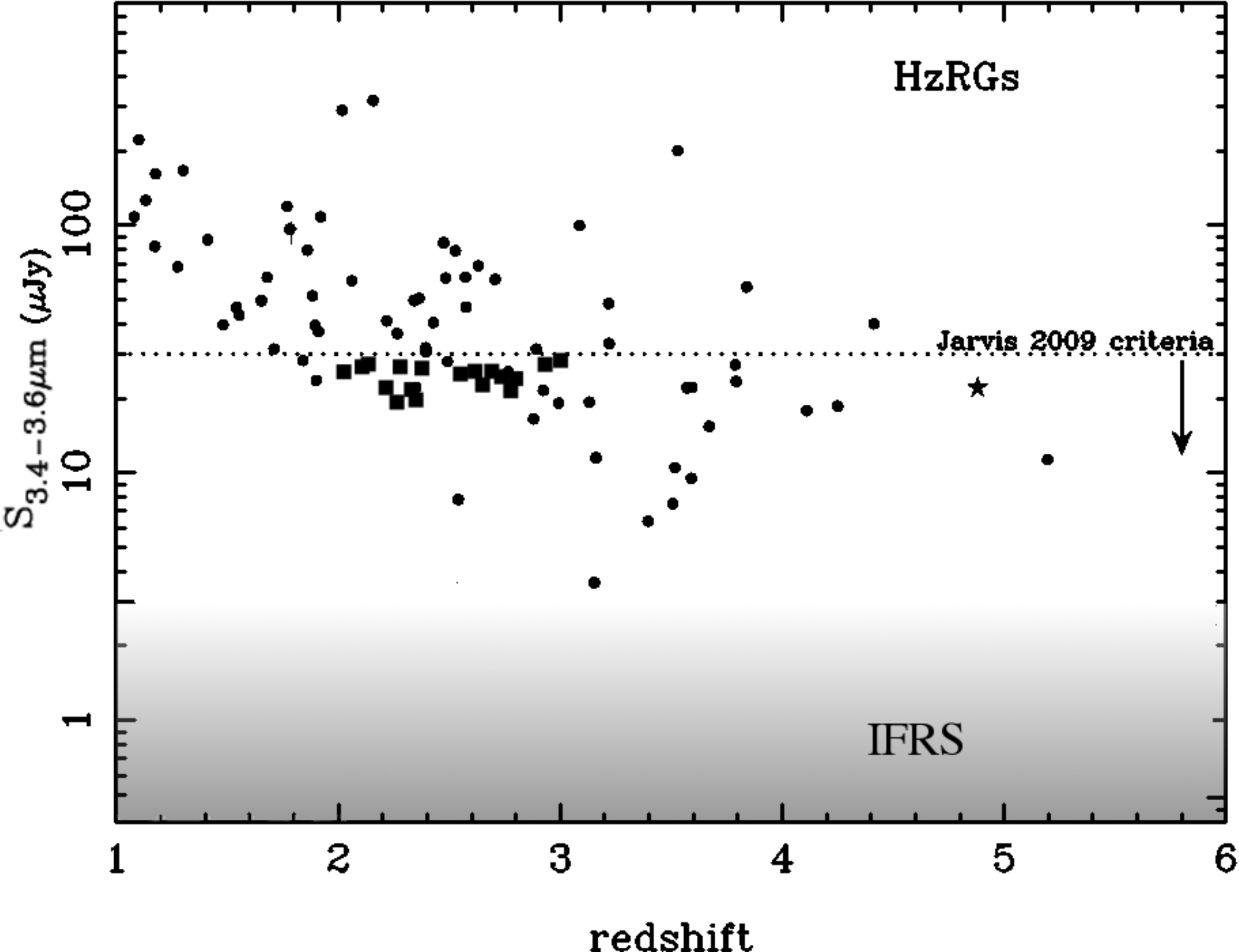}
\caption{The 3.4$-$3.6 $\mu$m flux density as a function of redshift for our IFRSs and the HzRGs from  \citet{2007ASPC..380..393S}, adapted from \citet{Norris2011}. The area in grey represents the range of upper limits on the flux density for the IFRSs discussed in \citet{Norris2011}, as indicated by the IFRS label. The black squares represent the IFRSs with spectroscopic redshifts discussed in this paper (see \S~\ref{SampleSelection} and \ref{Redshifts}). The black dots are the HzRGs from \citet{2007ASPC..380..393S}. The star represents the $z$ = 4.88 radio galaxy discovered by \citet{2009MNRAS.398L..83J}, and the line represents the $S_{\rm 3.6\mu m} < 30~\mu$Jy criterion they used to select their candidate HzRGs, which is also adopted by \citet{Zinn2011} to select IFRSs.}
\label{Olay2}
\end{figure*}

\citet{Norris2011} showed that IFRSs span a range in flux density ratio $S_{\rm20 cm} / S_{\rm3.6 \mu m}$ that is unoccupied except for High-$z$ Radio Galaxies (HzRGs), as illustrated in Fig.~\ref{Olay1}. They showed that if IFRSs follow the observed relation for HzRGs between ${\rm S_{3.6\mu m}}$ and redshift, similar to the well-known $K-z$ relation that holds for other radio galaxies \citep{2003MNRAS.339..173W}, then all first generation IFRSs occupy a place in the relation at $z \sim 5$, as illustrated in Fig.~\ref{Olay2}. \citet{Norris2011} concluded that while there is a possibility that more than one class of object may represent IFRSs, the evidence suggests that a significant fraction, if not all of them, are radio-loud AGN at $z \gtrsim 3$, but could also be made up of a new class of radio-loud AGN at lower redshift ($1 < z < 3$), in which the IR luminosity of the entire host galaxy must be reduced by several magnitudes ($A_V \ge 10$ mag) of dust extinction.

\cite{Middelberg2011} studied the high-resolution characteristics, spectral indices and polarization properties of a sample of 17 IFRSs identified in the ATLAS ELAIS-S1 field. They concluded from the high resolution 4.8 and 8.6 GHz radio-continuum data that the observed sources are smaller than 4.5 kpc $\times$ 2.1 kpc, much smaller than the projected linear sizes of classic high redshift radio galaxies, which range from a few to many hundreds of kiloparsecs. \cite{Middelberg2011} suggested that these IFRSs could therefore be intrinsically much smaller, or have their extended emission resolved out, even at lower resolution. None the less, the hypothesis that IFRSs are simply radio lobes of nearby galaxies was ruled out for these particular sources, since they were too compact to be lobes. \citet{Middelberg2011} found a median radio spectral index of $\alpha = -1.4$\footnote{The spectral index is defined as $S \sim \nu^{\alpha}$.} for their sample of IFRSs, with no indices larger than $-0.7$, as compared to the respective medians of $-0.86$ and $-0.82$ for the general source population and the AGN source population in the ATLAS ELAIS-S1 field. Additionally, they found a curvature in the radio spectra as seen in Gigahertz Peaked Spectrum (GPS) and CSS sources, rather than the power-law spectra of classical AGN. GPS and CSS sources are galaxies that contain small yet powerful AGN at their centres. GPS sources, which show a turnover at a few GHz, are generally $<$~1~kpc in size, and are believed to represent the earliest evolutionary stage of large-scale radio sources \citep{2003PASA...20...69P,2006A&A...445..889T,2009AN....330..120F,2011MNRAS.416.1135R}. CSS sources tend to peak around 200~MHz and reveal steep ($\alpha \lesssim -0.8$) spectral indices across the GHz range, are one to several tens of kpc in size, and are believed to bridge the evolutionary phase between the early GPS sources and the older and larger FR I and FR II galaxies \citep{1998PASP..110..493O}. This implies that some IFRSs are very young and evolving AGN with very small jets, which is in good agreement with their observed sizes.

\cite{Middelberg2011} also showed that the properties of their IFRSs are strikingly similar to those of a sample of HzRGs from \cite{2007ASPC..380..393S}, which had a median radio spectral index of $-1.02$. Furthermore, the flux density ratio ${\rm S_{20cm} / S_{3.6\mu m}}$ of the IFRSs from \cite{Middelberg2011} significantly overlapped with this sample of HzRGs, both of which had values of several hundred up to several tens of thousands, at the very tail end of the general source population, which peaked at ${\rm S_{20cm} / S_{3.6\mu m}}$ $\approx 5$. \citet{Middelberg2011} also found three of the sources to be significantly polarized at 20$\,$cm, with fractional polarizations between 7 -- 12 per cent. Given the evidence, \citet{Middelberg2011} classified 10 of their sources as AGN and the other 7 as most likely AGN based on their 24~$\mu$m non-detections alone.

\citet{Banfield2011} found similar spectral indices and fractional polarizations for a sample of 18 IFRSs in the ELAIS-N1 field. They found a median spectral index of $\alpha = -1.1$ for five sources which had detectable polarizations, ranging from about 6 -- 16 per cent fractional polarization. \citet{Banfield2011} found a steeper median spectral index of $\alpha = -1.5$ for the 13 unpolarized sources. Additionally, four of the polarized sources showed structure on arcsecond scales, while only two unpolarized sources showed resolved structure.

\cite{2011MNRAS.415..845C} observed 16 IFRSs to test the hypothesis that IFRSs are pulsars. After searching for short-term radio pulsations coming from the IFRSs, it was found that pulsed emission could not account for their observed flux densities. \cite{2011MNRAS.415..845C} concluded that it is unlikely that any IFRSs are simply pulsars.

\citet{Zinn2011} showed that the X-ray radiation from IFRSs may contribute significantly to the Cosmic X-ray Background. They estimated the X-ray emission of IFRSs and showed that it is consistent with the missing unresolved components of the Cosmic X-ray Background.

Despite the significant work undertaken in uncovering the nature of IFRSs, it still remains unconfirmed exactly what they are. However, good progress has been made toward ruling out some explanations of what makes up their majority, including pulsars, radio-lobes, and obscured SFGs. Additionally, it has been shown that it is not necessary for some new type of object to explain the existence of IFRSs. The evidence is mounting up that suggests the majority, if not almost all IFRSs are high-redshift ($z > 3$) radio-loud AGN. While it is possible that they could be suffering from significant dust extinction, such extinction is not necessary to explain the observed data. However, a minority of what we are calling IFRSs could be made up of several types of objects, given their non-uniform characteristics such as their radio spectral index. The most likely such objects include: (1) very obscured radio galaxies at moderate redshifts ($1 < z < 2$); (2) hotspots of nearby unidentified radio galaxies; (3) misidentifications. In this paper we address these possibilities and whether they can explain IFRSs.

\subsection{WISE IFRSs}

\citet{Norris2006} defined an IFRS as `a radio source with no detectable IR counterpart'. \citet{Zinn2011} proposed a new set of selection criteria to enhance searching for IFRSs, since the previous selection criterion was survey-specific. They defined IFRSs as sources that have: 

\begin{enumerate}
\item a flux density ratio $S_{\rm20\,cm} / S_{\rm3.6\,\mu m} > 500$
\item a $3.6\,\mu$m flux density $< 30\,\mu$Jy
\end{enumerate}

Although these limits are somewhat arbitrary, they encompass all known IFRSs. The first criterion ensures that the selected sources are outliers in the radio-IR correlation, minimising contamination from SFGs and foreground stars. The second criterion reduces the chance of selecting low-redshift AGN, although it does not rule out the possibility of selecting low-redshift AGN obscured by heavy dust extinction. Using these criteria, \citet{Zinn2011} compiled a catalogue of 55 known IFRSs from four deep radio surveys (CDFS, ELAIS-S1, FLS, and COSMOS), which remains until now the largest catalogue of IFRSs. 

The \citet{Zinn2011} criteria enable the selection of a larger, brighter population of IFRSs with detectable infrared and optical emission. Therefore, while previous studies focused on very sensitive observations of a few small regions on the sky, we followed the strategy of combining radio data with IR data from the {\it Wide-field Infrared Survey Explorer} \citep[{\it WISE};][]{2009AAS...21421701W} for a large region of the sky, albeit at poorer sensitivity, and selecting detectable IFRSs from these data using the \citet{Zinn2011} criteria. With this data set we can not only study a statistically significant number of sources, we can learn how the brighter {\it WISE} IFRSs connect to the first generation of IFRSs.

We present the first spectroscopic redshifts of IFRSs, as well as the first X-ray counterparts of IFRSs. Additionally, we examine the properties of all IFRSs with detectable polarization, significantly increasing the number of known polarized IFRSs. We outline the data and sample selection in \S~2, present our results and analysis in \S~3, discuss the implications in \S~4, and present our conclusions in \S~5. The cosmological parameters used throughout this paper are $\Omega_{\rm \Lambda}=0.7$, $\Omega_{\rm M}=0.3$, and $H_{\rm 0}=70\; {\rm km s}^{-1}\; {\rm Mpc}^{-1}$. 

%**********************************************************************************************************************************************************

\section{Data}

\subsection{20 cm radio data}

The 20 cm radio data come from the Unified Radio Catalog (URC) compiled by \citet{Kimball2008}. This radio catalogue combines data from the NRAO VLA Sky Survey \citep[NVSS;][]{Condon1998}, Faint Images of the Radio Sky at Twenty Centimeters \citep[FIRST;][]{Becker1995}, Green Bank $6\,$cm survey \citep[GB6;][]{Gregory1996}, the Westerbork Northern Sky Survey \citep[WENSS;][]{Rengelink1997,2000yCat.8062....0D}, and the Sloan Digital Sky Survey Data Release 6 \citep[SDSS DR6;][]{SDSSdr6}. We use updated NVSS and FIRST data from the URC version 2.0 (Kimball \& Ivezi\'c, in preparation), which includes a number of new sources as well as updated positions and flux densities.

\subsection{Infrared data}

The IR data come from {\it WISE} \citep{2009AAS...21421701W}, which is an all-sky survey centred at 3.4, 4.6, 12, and $22\,\mu$m (referred to as bands W1, W2, W3 and W4), with respective angular resolutions of 6.1$\arcsec$, 6.4$\arcsec$, 6.5$\arcsec$, and 12.0$\arcsec$ (FWHM), and typical $5\sigma$ sensitivity levels of 0.08, 0.11, 1, and $6\,$ mJy, with sensitivity increasing towards the ecliptic poles. The majority of our sources fall in these strips of greater sensitivity, because we require their 3.4$\mu$m flux density to be $< 30\,\mu$Jy. To convert {\it WISE} magnitudes to flux density in Janskys (Jy), $S_{\rm band}$, we used

\begin{equation}
S_{\rm band} = F_{\rm band}({\rm iso}) \times 10^{(-M_{\rm band}/2.5)} \, {\rm Jy} \, ,
\end{equation}
where $F_{\rm band}({\rm iso})$ is the flux correction factor at the given {\it WISE} band from \citet{Jarrett2011} and $M_{\rm band}$ is the observed magnitude at the corresponding {\it WISE} band.

Additionally, we need to apply a colour correction, which depends on the assumed SED, which we don't know in detail. Hence, we initially assume the correction factor of 0.991 in band W1 given by \cite{Jarrett2011} for a source that scales as $F_{\nu} \propto \nu^{0}$, which is generally suitable for galaxies, and which results in the conversion

\begin{equation}
S_{ \rm{3.4\mu m}} = 309.540 \times 10^{(-M_{\rm{3.4\mu m}}/2.5)} \, {\rm Jy}
\end{equation}

We initially selected the IFRSs using this conversion factor. We then measured the {\it WISE} colours of this pre-selection in order to determine if our colour correction was suitable. Band W4 was not considered in this calculation, and hence neither the colour ${\rm [W3-W4]}$, since only seven sources were detected at  $\ge 5\sigma$ in band W4. The respective median found for each colour was ${\rm [W1 - W2] = 1.376}$ and ${\rm [W2 - W3] = 4.277}$. According to \citet{2010AJ....140.1868W}, these colours are closest to that of a source that scales as $F_{\nu} \propto \nu^{-2}$, which gives an SED with {\it WISE} colours ${\rm [W1 - W2] = 1.3246}$ and ${\rm [W2 - W3] = 3.9225}$, and for which there is no colour correction. Hence, when converting 3.4~$\mu$m from mags to Jy, we use no colour correction factor, which results in the conversion

\begin{equation}
S_{ \rm{3.4\mu m}} =  306.682 \times 10^{(-M_{\rm{3.4\mu m}}/2.5)} \, {\rm Jy} \,
\end{equation}

The 30 $\mu$Jy cutoff therefore corresponds to a lower limit in band W1 of $\sim$17.5 magnitudes.
Using this more appropriate conversion, we then reselected the IFRSs, adding less than 5 per cent to the original.

\subsection{Positional uncertainties and confusion}

\label{MisIDs}

\begin{figure}
\includegraphics[scale=0.45]{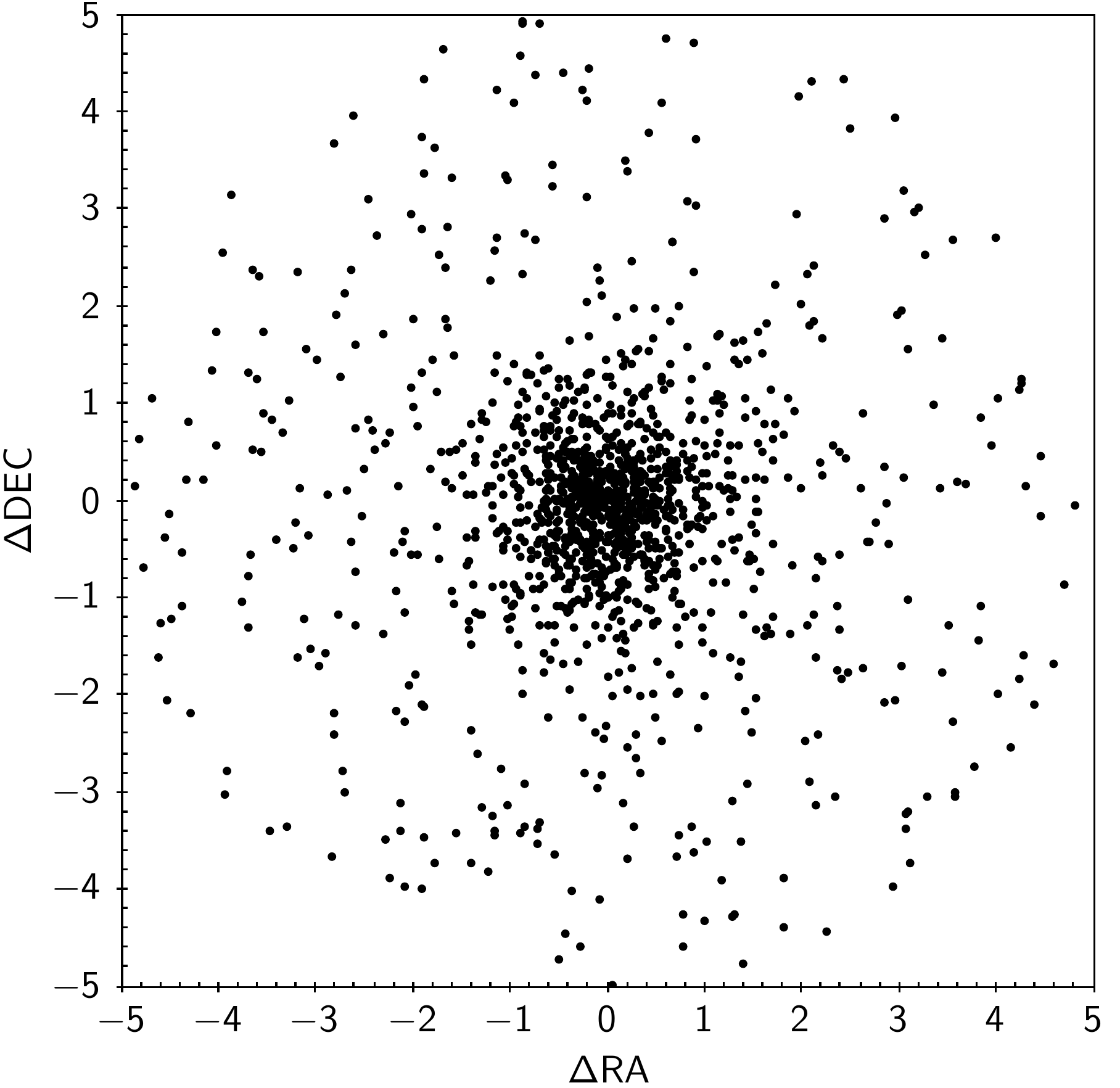}
\caption{The sky separation between the FIRST and {\it WISE} positions for our 1317 IFRSs, where $\Delta$RA and $\Delta$DEC are in arcseconds. The standard deviation of both $\Delta$RA and $\Delta$DEC is 1.5 arcsec.}
\label{SkySep}
\end{figure}

The FIRST survey has an astrometric precision of $\lesssim 1 \arcsec$ \citep{Becker1995}. {\it WISE} has an astrometric precision of $< 0.15 \arcsec$ with a further error of $\sim$FWHM/(2 $\times$ SNR) added in quadrature, where SNR is the signal-to-noise ratio \citep{2010AJ....140.1868W}. Since we require a 5$\sigma$ detection, all {\it WISE} positions in our sample are accurate to $< 1\arcsec$. The mean sky separation between the {\it WISE} positions and the FIRST positions is 1.5 arcsec, with the mean $\Delta$RA and $\Delta$DEC both $< 1$ arcsec, and a standard deviation of $\sigma = 1.4$ arcsec for both the RA and DEC. We find 65 per cent of our sources have a FIRST/{\it WISE} separation $<$ 1.5 arcsec. The sky separation for all of our IFRSs is shown in Fig.~\ref{SkySep}. 

Since the {\it WISE} 3.4 $\mu$m data approach the confusion limit, it is necessary to estimate the misidentifications that are due to confusion. We do so by taking an unbiased subsample of 312 (0.1 per cent) of the NVSS sources with FIRST positions from the URC, which are located all across the sky. We then shift their radio positions by an amount (typically $\sim 15\arcsec$) that is greater than the beam size of both the radio and IR data, but smaller than the scale of the variations in the image sampling. We then apply our matching and selection procedure (see \S~\ref{SampleSelection}), to estimate the confusion. This process was repeated eight times, each time shifting the radio positions by a different amount. It was found that a mean of 25.5 $\pm$ 4.8 ($\sim$8 per cent) of the shifted radio sources contained {\it WISE} counterparts. Of these, a mean of 0.75 $\pm$ 0.83 matched the criteria of being an IFRS, giving a misidentification rate of our final sample of 0.24 $\pm$ 0.27 per cent. Hence, the vast majority of our sample are genuine cross-matches, with an expectation of about 3 false-positives in our total of 1317 IFRSs. 

\subsection{Sample selection}

\label{SampleSelection}

Our sample of 1317 IFRSs has been selected with the following criteria:

\begin{itemize}
\item Single NVSS source with $S_{\rm 20 cm} > 7.5$ mJy
\item At least one FIRST counterpart within 30$\arcsec$ of NVSS source
\item {\it WISE} counterpart within 5$\arcsec$ from FIRST position
\item {\it WISE} $S_{\rm 3.4\,\mu m} < 30\,\mu {\rm Jy}$
\item The NVSS to {\it WISE} flux ratio, $S_{\rm 20\,cm}/S_{\rm 3.4\,\mu m} > $ 500
\item Signal-to-noise ratio (SNR) at $3.4\,\mu$m is $\ge 5$
\item Visually do not appear as radio lobe matched to IR source
\end{itemize}

\noindent We now discuss the selection criteria in detail.

\subsubsection{$S_{\rm 20 cm} > 7.5$ mJy}

\begin{figure}
 \includegraphics[width=84mm]{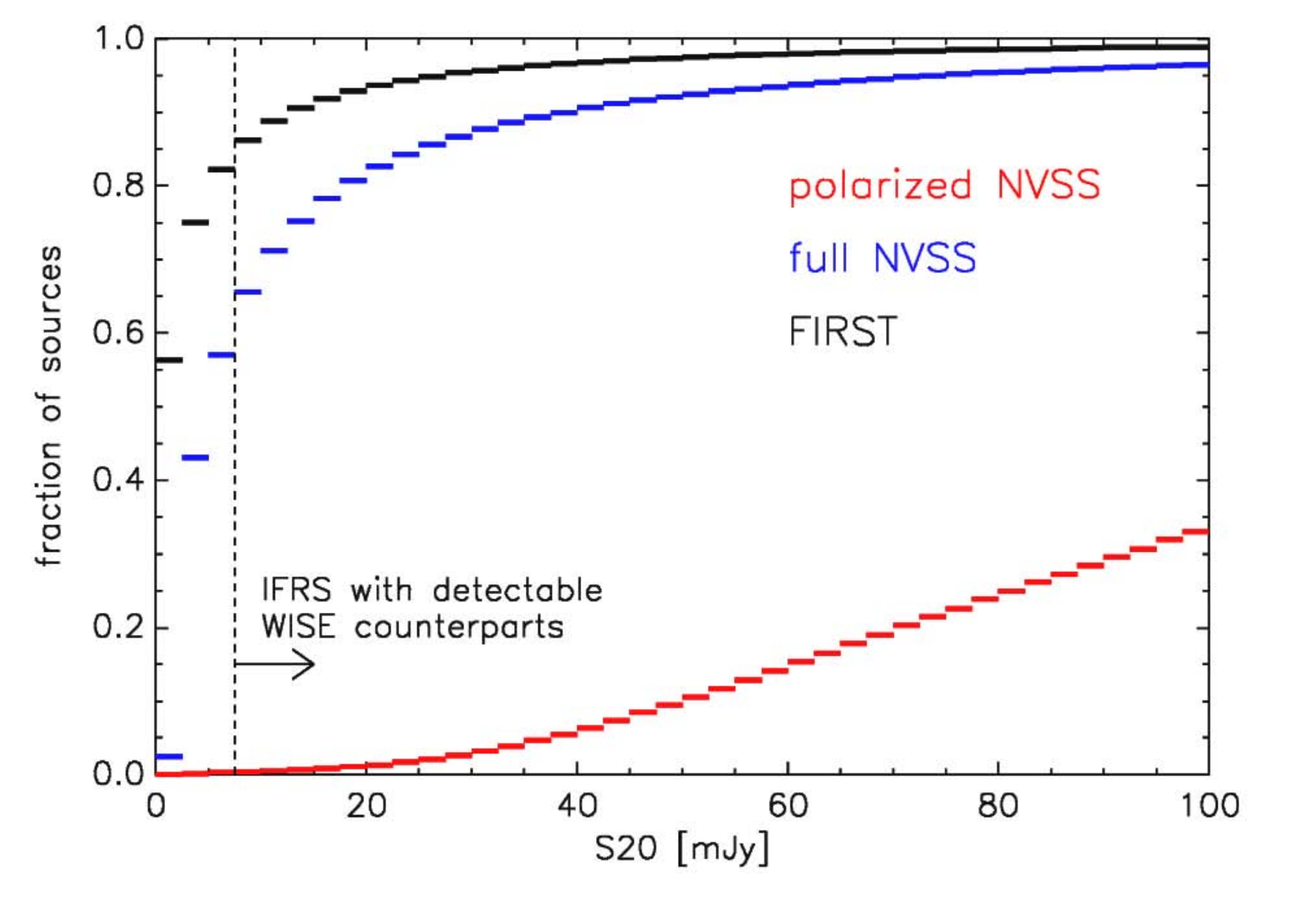}
 \caption{Cumulative distribution of flux densities for three different radio surveys at $20\,$cm. An IFRS could have a radio counterpart as faint as $7.5\,$mJy, if {\it WISE} detected it at $5\sigma$ at $3.4\,\mu$m and the flux density ratio $S_{\rm 20cm}/S_{\rm 3.4\mu m} = 500$ (assuming the best {\it WISE} noise level of $3\,\mu$Jy). At these flux densities the catalogue of polarized NVSS sources (red) is far from complete, while the other two catalogues contain many faint radio sources.}
 \label{fig1}
\end{figure}

\begin{figure}
 \includegraphics[width=80mm]{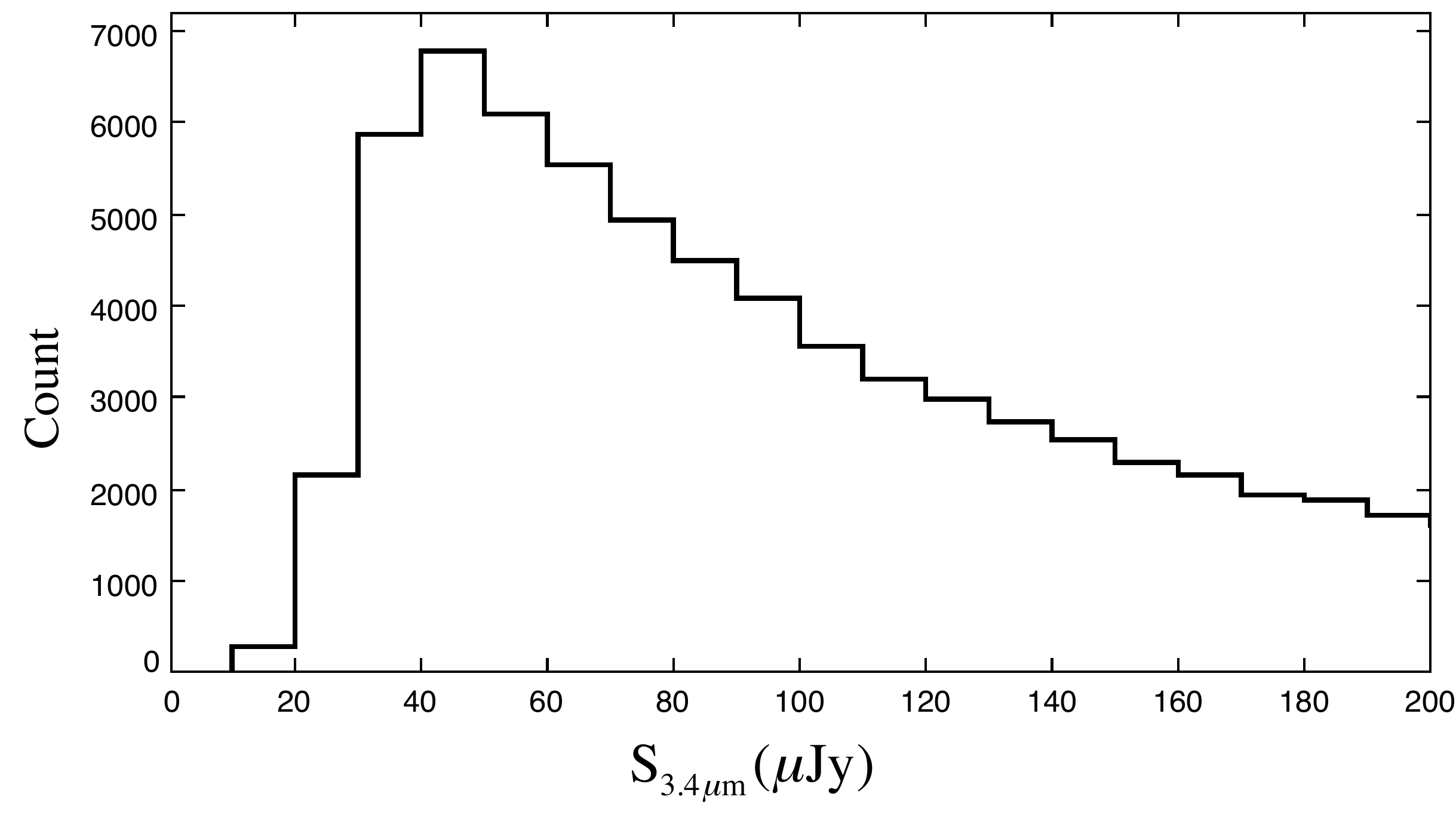}
 \caption{Histogram of the {\it WISE} 3.4 $\mu$m flux density for all {\it WISE} sources detected at $\ge 5\sigma$ with FIRST/NVSS radio counterparts (from the URC) above 7.5 mJy. The sharp drop in the number of sources below 40 $\mu$Jy signifies that the catalogue is far from complete below this level.}
 \label{WISEcompleteness}
\end{figure}

To maximise completeness and minimize polarization bias, we applied a 7.5 mJy cutoff to our NVSS sources from the URC. This flux density corresponds to a 5-sigma detection in the {\it WISE} $3.4\,\mu$m band and a flux ratio $S_{\rm 20\,cm}/S_{\rm 3.4\,\mu m} = 500$, assuming the best {\it WISE} noise level is 3 $\mu$Jy (see Fig.~\ref{fig1}). 
At this level, {\it WISE} is far from complete (see Fig.~\ref{WISEcompleteness}), due to its non-uniform depth across the sky.

\subsubsection{FIRST counterparts}

We extracted from the URC all NVSS radio sources with at least one FIRST counterpart. 
The angular resolution of NVSS is $45\arcsec$, while the angular resolution of FIRST is $5.4\arcsec$ at 20 cm, making the FIRST position more accurate than the NVSS position when matching with counterparts in other surveys. We use the higher angular resolution data of FIRST for the accurate positional information, whilst using the NVSS flux density as a measure of the total integrated flux density of all of the FIRST components, which generally number between 1 and 3. Using NVSS data also allows us to match sources from the NVSS rotation measure (RM) catalogue \citep{Taylor2009}, which consists of RMs for $37~543$ polarized sources from the NVSS catalogue, from which we extracted 41 matches.

After applying an NVSS radio flux density cutoff at $7.5\,$mJy and selecting only sources with at least one FIRST counterpart in the URC, we had $312~514$ radio sources. 

\begin{figure}
\includegraphics[scale=0.3,trim=0mm 0mm 0mm 0mm,angle=-90]{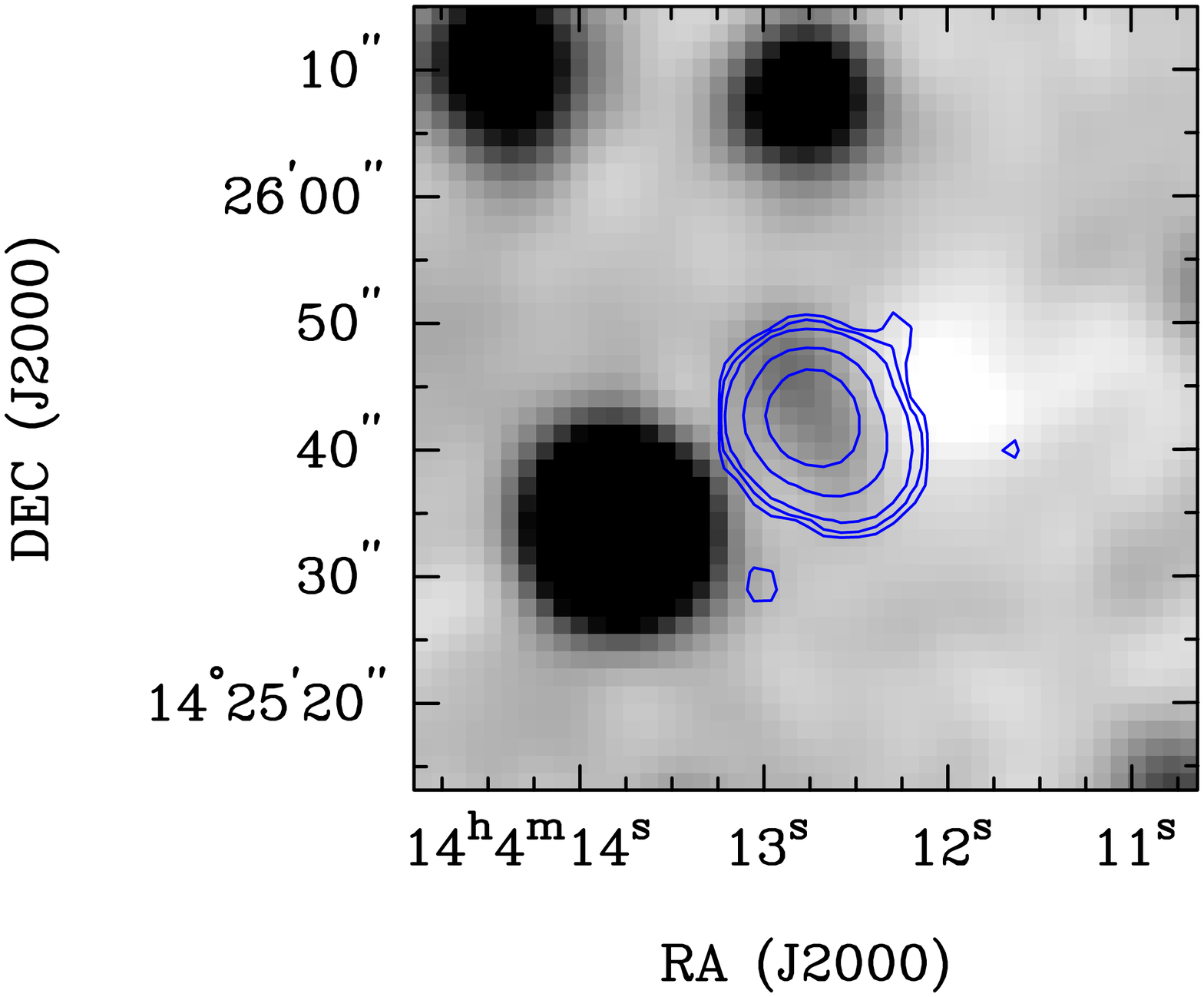}
\includegraphics[scale=0.3,trim=0mm 0mm 0mm 0mm,angle=-90]{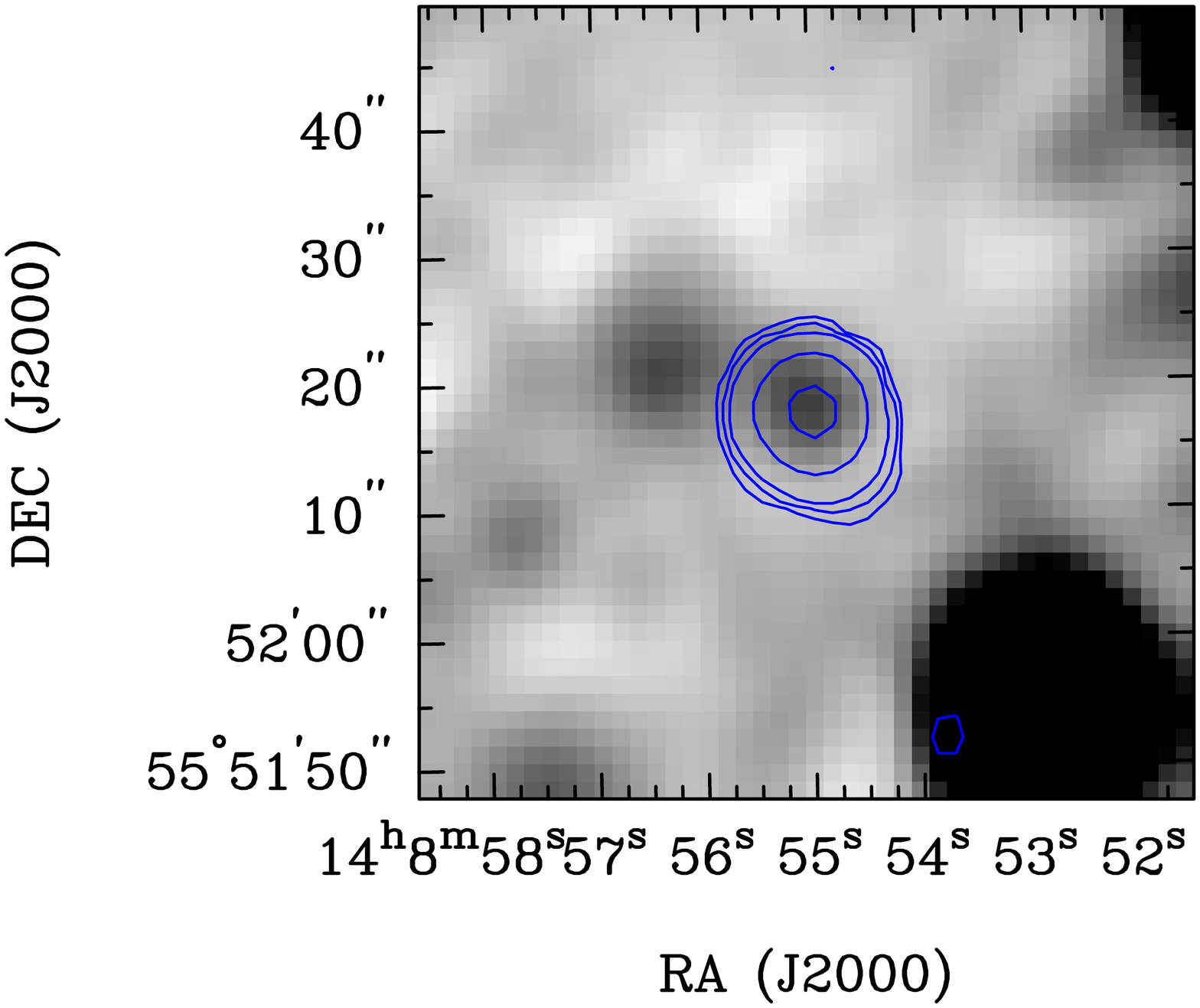}
\caption{Polarized NVSS sources detected by {\it WISE} and classified as IFRSs.  These two sources were found to have cataloged redshifts of 0.99 (top image) and 2.56 (bottom image). The background image shows the {\it WISE} $3.4\,\mu$m detection, and the contours mark the FIRST source at $20\,$cm. Contour levels are: 3, 6, 12, 64, 256, 1024 times the local noise level. The {\it WISE} sources are detected at the 5.5--6$\sigma_{\rm WISE}$ level, while their total intensity counterparts in NVSS are detected at the 33$\sigma_{\rm NVSS}$ level.}
\label{postage}
\end{figure}

\subsubsection{{\it WISE} counterpart within 5$\arcsec$ of FIRST position}

When matching the radio sources to {\it WISE}, we want to ensure that at z~$>$~0.5, the corresponding linear separation is $\lesssim$~30 kpc, smaller than the size of large spiral galaxies. This limit reduces confusion and is consistent with the observed limits on the projected sizes of previous IFRSs. Therefore, {\it WISE} counterparts were searched for in a $5\arcsec$ radius from the FIRST positions. Of the 312~514 preselected radio sources, 137~154 {\it WISE} matches detected at $\ge5\sigma$ in band W1 were found.

\subsubsection{IFRS selection}

We then applied our IFRS selection, using the \citet{Zinn2011} criteria (i.e. $S_{\rm 3.4\,\mu m} < 30\,\mu {\rm Jy}$ and $S_{\rm 20\,cm}/S_{\rm 3.4\,\mu m} > $ 500),  as well as the criterion that the SNR at $3.4\,\mu$m is $\ge 5$.
This resulted in a total of 1471 candidate IFRSs.

\subsubsection{Visual inspection}

\begin{figure}
\includegraphics{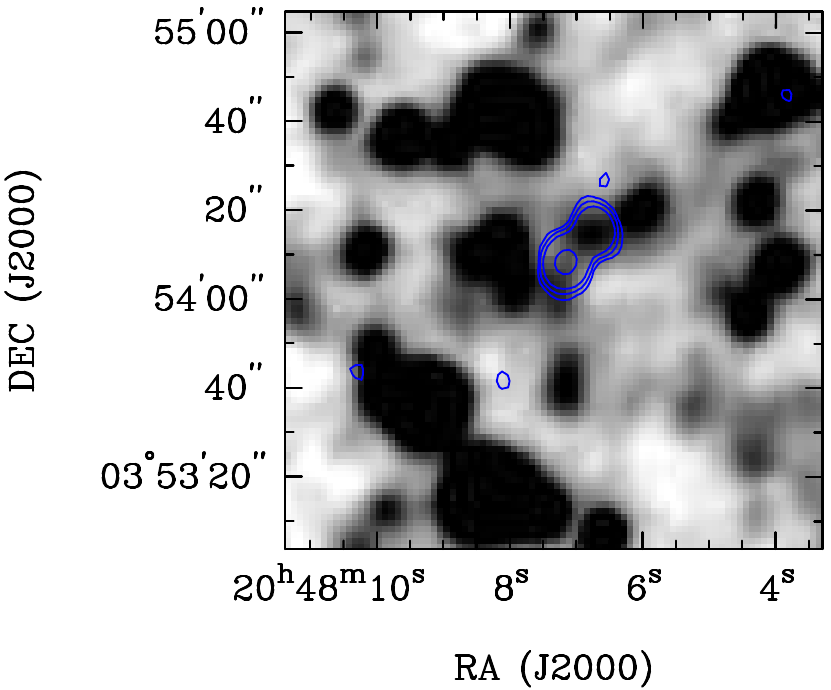}
\caption{An example of a source that was discarded during visual inspection. The background image shows the {\it WISE} $3.4\,\mu$m intensity, and the contours mark the FIRST source at $20\,$cm. Contour levels are: 3, 6, 12, 64, 256, 1024 times the local noise level.}
\label{discard}
\end{figure}

As the radio emission can originate from the the lobes or the central area of the host galaxy, the candidate IFRSs were visually inspected, to ensure that no radio lobes from a nearby source were spuriously matched to a {\it WISE} source. In many cases, the IFRSs were resolved into several FIRST components, which could be identified as nuclei or lobes when overlaid as contours on a {\it WISE} greyscale image. If the radio emission came from the lobe of the radio source and overlapped with a {\it WISE} galaxy, this was not considered an IFRS, and so was discarded from the sample, to reduce the number of misidentifications. 154 sources were discarded during the visual inspection. Fig.~\ref{postage} shows two detected IFRSs and Fig.~\ref{discard} shows a discarded IFRS. 

\subsubsection{Final IFRS catalogue}

The final IFRS catalogue consists of 1317 sources, 41 of which have matches in the \citet{Taylor2009} RM catalogue. Table \ref{ifrstable} lists the FIRST RA and DEC, the radio and infrared flux densities and their ratio (${\rm S_{20cm} / S_{3.4\mu m}}$), the bias corrected fractional polarization and RM, and the radio spectral index for the 41 polarized IFRSs. The full table of 1317 IFRSs is available in the electronic version, which includes a number of additional columns.

\begin{table*}
 \centering
  \caption{The 41 IFRSs which have matches in the \citet{Taylor2009} RM catalogue. Listed is the FIRST RA and DEC, the NVSS 20$\,$cm flux density, the {\it WISE} 3.4$\,\mu$m flux density, the radio-IR flux density ratio, the bias corrected fractional polarization and RM from the \citet{Taylor2009} RM catalogue, and the radio spectral index, calculated using flux densities at 92, 20 and 6 cm, as discussed in \S~\ref{specIndx}. The full table of 1317 IFRSs is available in the electronic version, which includes a number of additional columns. }  
\label{ifrstable}
  \begin{tabular}{ccllllll}
  \hline
FIRST RA & FIRST DEC & $S_{\rm 20\,cm}$ & $S_{\rm 3.4\,\mu m}$ & $S_{\rm 20\,cm}/S_{\rm 3.4\,\mu m}$ & Fractional & Rotation Measure & $\alpha$\\
\multicolumn{2}{c}{(J2000)} & (mJy) & ($\mu$Jy) & & Polarization & (rad m$^{-2}$) & \\
 \hline
02:16:36.78 & $-$00:24:51.36 & 44.47 $\pm$ 1.40 & 28.10 $\pm$ 5.90 & 1583 $\pm$ 393 & 11.57\% & 19.3 $\pm$ 11.9 & \\
02:17:06.19 & $+$05:31:09.58 & 131.66 $\pm$ 4.65 & 29.05 $\pm$ 5.30 & 4533 $\pm$ 1016 & 5.35\% & $-$20.6 $\pm$ 9.7 & \\
02:25:45.07 & $-$06:37:53.99 & 23.81 $\pm$ 0.83 & 25.60 $\pm$ 5.38 & 930 $\pm$ 235 & 12.19\% & $-$11.5 $\pm$ 16.7 & \\
02:29:29.97 & $+$04:53:19.90 & 72.70 $\pm$ 2.22 & 28.67 $\pm$ 5.23 & 2536 $\pm$ 554 & 8.63\% & $-$17.9 $\pm$ 10.1 & \\
02:31:52.06 & $-$02:26:47.48 & 60.19 $\pm$ 2.24 & 28.70 $\pm$ 5.61 & 2097 $\pm$ 504 & 6.92\% & 13.7 $\pm$ 15.8 & \\
11:13:54.32 & $+$35:12:33.97 & 45.00 $\pm$ 1.73 & 26.78 $\pm$ 5.57 & 1680 $\pm$ 427 & 9.78\% & 50.7 $\pm$ 15.1 & \\
11:44:06.17 & $+$56:54:39.63 & 171.63 $\pm$ 5.16 & 30.00 $\pm$ 5.44 & 5721 $\pm$ 1241 & 3.92\% & 9.4 $\pm$ 9.3 & $-$0.69\\
13:25:13.02 & $-$05:48:10.59 & 94.28 $\pm$ 2.86 & 28.44 $\pm$ 5.63 & 3315 $\pm$ 776 & 4.25\% & 4.3 $\pm$ 15.8 & \\
13:54:03.81 & $+$05:05:44.21 & 91.31 $\pm$ 3.30 & 27.97 $\pm$ 5.35 & 3264 $\pm$ 765 & 5.28\% & $-$17.5 $\pm$ 13.2 & \\
13:55:02.25 & $+$16:10:23.64 & 197.37 $\pm$ 5.93 & 25.94 $\pm$ 5.13 & 7610 $\pm$ 1780 & 3.02\% & 7.3 $\pm$ 13.3 & \\
14:04:12.73 & $+$14:25:42.47 & 137.28 $\pm$ 4.14 & 26.86 $\pm$ 4.96 & 5111 $\pm$ 1126 & 6.37\% & $-$8.6 $\pm$ 5.5 & \\
14:08:54.99 & $+$55:52:17.62 & 62.86 $\pm$ 1.93 & 26.42 $\pm$ 4.99 & 2380 $\pm$ 537 & 5.51\% & $-$9.5 $\pm$ 15.4 & $-$0.69\\
14:12:22.77 & $+$28:54:02.20 & 127.16 $\pm$ 4.36 & 27.11 $\pm$ 5.48 & 4691 $\pm$ 1142 & 5.64\% & 13.3 $\pm$ 9.2 & $-$0.98\\
14:24:21.25 & $+$20:14:55.25 & 186.60 $\pm$ 5.61 & 22.90 $\pm$ 4.94 & 8148 $\pm$ 2055 & 3.21\% & 8.8 $\pm$ 10.9 & \\
14:29:33.30 & $+$12:26:01.06 & 66.54 $\pm$ 2.04 & 22.90 $\pm$ 4.61 & 2905 $\pm$ 691 & 6.94\% & 12.8 $\pm$ 12.3 & \\
14:30:18.96 & $+$08:26:12.22 & 96.19 $\pm$ 2.91 & 26.91 $\pm$ 5.29 & 3575 $\pm$ 833 & 3.85\% & 16.0 $\pm$ 15.0 & \\
14:36:53.44 & $+$42:44:16.99 & 276.94 $\pm$ 9.81 & 25.94 $\pm$ 4.93 & 10678 $\pm$ 2480 & 3.59\% & $-$0.1 $\pm$ 6.9 & \\
14:38:32.97 & $+$29:33:54.70 & 290.19 $\pm$ 8.71 & 28.00 $\pm$ 4.59 & 10365 $\pm$ 2064 & 1.77\% & 35.4 $\pm$ 12.0 & $-$0.77\\
14:41:57.04 & $+$41:51:01.80 & 72.62 $\pm$ 2.22 & 25.04 $\pm$ 5.29 & 2900 $\pm$ 719 & 6.08\% & $-$10.9 $\pm$ 12.9 & $-$1.01\\
14:48:55.62 & $+$53:52:10.43 & 109.01 $\pm$ 3.29 & 22.86 $\pm$ 4.80 & 4768 $\pm$ 1176 & 3.49\% & 53.1 $\pm$ 14.9 & $-$0.87\\
14:54:13.84 & $+$53:38:04.11 & 135.50 $\pm$ 4.08 & 28.54 $\pm$ 4.68 & 4747 $\pm$ 946 & 2.81\% & 29.5 $\pm$ 13.8 & $-$0.90\\
15:01:36.29 & $+$05:27:25.64 & 248.49 $\pm$ 7.47 & 27.61 $\pm$ 5.16 & 9000 $\pm$ 2003 & 2.73\% & $-$17.2 $\pm$ 9.4 & \\
15:02:51.14 & $+$60:09:41.68 & 128.96 $\pm$ 3.89 & 28.46 $\pm$ 3.98 & 4531 $\pm$ 789 & 5.59\% & $-$4.8 $\pm$ 8.1 & $-$0.91\\
15:04:22.70 & $-$05:58:16.47 & 127.94 $\pm$ 3.86 & 29.80 $\pm$ 5.93 & 4292 $\pm$ 1009 & 5.10\% & $-$8.4 $\pm$ 10.8 & \\
15:08:46.27 & $+$41:27:50.23 & 407.60 $\pm$ 14.38 & 27.36 $\pm$ 3.82 & 14898 $\pm$ 2681 & 6.25\% & 6.9 $\pm$ 2.8 & $-$0.82\\
15:09:04.89 & $-$03:09:17.66 & 49.32 $\pm$ 2.13 & 25.44 $\pm$ 5.46 & 1939 $\pm$ 517 & 13.69\% & 16.1 $\pm$ 11.2 & \\
15:10:04.47 & $+$60:29:24.64 & 77.46 $\pm$ 2.36 & 23.89 $\pm$ 4.00 & 3242 $\pm$ 658 & 4.99\% & 6.5 $\pm$ 17.0 & $-$1.02\\
15:10:20.14 & $+$35:40:43.09 & 28.28 $\pm$ 0.93 & 28.15 $\pm$ 3.49 & 1005 $\pm$ 162 & 13.98\% & $-$3.5 $\pm$ 16.5 & $-$0.61\\
15:28:21.90 & $+$21:14:59.25 & 39.52 $\pm$ 1.25 & 23.44 $\pm$ 4.53 & 1686 $\pm$ 390 & 8.01\% & 27.6 $\pm$ 20.0 & \\
15:30:22.57 & $+$06:44:07.97 & 293.14 $\pm$ 8.80 & 28.49 $\pm$ 5.04 & 10289 $\pm$ 2186 & 3.35\% & 5.2 $\pm$ 5.8 & \\
15:40:43.74 & $+$46:44:48.06 & 116.36 $\pm$ 3.51 & 18.81 $\pm$ 3.35 & 6188 $\pm$ 1322 & 2.76\% & 9.9 $\pm$ 16.8 & $-$1.23\\
15:51:28.21 & $+$64:05:37.28 & 683.19 $\pm$ 20.50 & 16.82 $\pm$ 3.37 & 40612 $\pm$ 9587 & 2.95\% & $-$45.3 $\pm$ 7.6 & $-$0.35\\
15:56:48.48 & $+$55:39:05.80 & 61.22 $\pm$ 1.88 & 20.72 $\pm$ 3.09 & 2955 $\pm$ 545 & 5.85\% & 30.1 $\pm$ 16.4 & $-$0.89\\
16:11:39.63 & $+$46:18:51.12 & 47.67 $\pm$ 1.48 & 26.81 $\pm$ 3.30 & 1778 $\pm$ 281 & 7.01\% & 4.0 $\pm$ 17.0 & $-$0.83\\
16:13:34.76 & $+$45:46:54.40 & 97.08 $\pm$ 2.94 & 24.20 $\pm$ 3.38 & 4011 $\pm$ 699 & 3.16\% & 53.8 $\pm$ 15.4 & $-$0.89\\
16:14:19.24 & $+$59:59:33.77 & 273.78 $\pm$ 8.22 & 29.91 $\pm$ 3.65 & 9152 $\pm$ 1425 & 1.36\% & 21.8 $\pm$ 17.3 & $-$0.82\\
16:22:31.98 & $+$41:23:22.62 & 113.66 $\pm$ 3.43 & 29.80 $\pm$ 4.80 & 3813 $\pm$ 747 & 4.09\% & 32.8 $\pm$ 14.2 & $-$1.10\\
16:52:01.27 & $+$47:05:01.96 & 110.82 $\pm$ 3.35 & 29.56 $\pm$ 5.17 & 3749 $\pm$ 789 & 3.96\% & 2.0 $\pm$ 11.9 & $-$0.90\\
17:04:21.92 & $+$41:54:35.88 & 116.46 $\pm$ 3.52 & 28.62 $\pm$ 6.04 & 4069 $\pm$ 1008 & 6.32\% & 29.6 $\pm$ 7.9 & $-$0.97\\
17:04:32.08 & $+$54:56:42.91 & 34.96 $\pm$ 1.12 & 21.32 $\pm$ 3.87 & 1640 $\pm$ 359 & 10.23\% & 14.1 $\pm$ 15.6 & $-$0.61\\
17:16:25.50 & $+$33:05:50.64 & 176.16 $\pm$ 6.24 & 27.94 $\pm$ 5.19 & 6304 $\pm$ 1436 & 6.46\% & 24.2 $\pm$ 6.0 & $-$0.94\\
 \hline
\end{tabular}
\end{table*}

\subsection{Ancillary data}

Ancillary data were searched for in the Parkes-MIT-NRAO \citep[PMN;][]{1995ApJS...97..347G}, VLA Low-Frequency Sky Survey \citep[VLSS;][]{2008ASPC..395..370L}, {\it Spitzer} and Sloan Digital Sky Survey \citep[SDSS;][]{2012ApJS..203...21A} catalogues. Misidentification rates were estimated for these using the same procedure as outlined in \S~\ref{MisIDs}. Table~\ref{ancillary} summarises the ancillary data gathered for our sample of IFRSs from these various surveys. The ancillary data can be found within the full table, available in the electronic version. 

There are 31 IFRSs that have a counterpart in one of the following surveys from the {\it Spitzer Space Telescope}: SWIRE (9); FLS (5); {\it Spitzer} Deep Wide-Field Survey (SDWFS) (1); the `{\it Spitzer} Enhanced Imaging Products Explanatory' from the NASA/IPAC Infrared Science Archive (IRSA)\footnote{\url{http://irsa.ipac.caltech.edu/cgi-bin/Gator/nph-scan?submit=Select\&projshort=SPITZER}} (16) \citep{2003PASP..115..897L,2005ApJS..161...41L,2006AJ....131.2859F,2009ApJ...701..428A,2012AAS...21942806T}. The aperture and colour corrected 3.6$\,\mu$m flux densities from {\it Spitzer} are consistent with our 3.4$\,\mu$m flux densities from {\it WISE}, since only nine sources lay outside the 1$\sigma$ uncertainty, as expected by chance. 

There are 230 SDSS matches to our IFRSs, which yield a cross-match rate of $\sim$17 per cent. SDSS DR9 has a 95 per cent completeness for point sources, to AB magnitude limits of 22.0, 22.2, 22.2, 21.3 and 20.5, respectively in bands {\it ugriz}. The optical magnitudes quoted refer to the SDSS model magnitudes, which are measured using a weighting function as determined from the object's brightness in band {\it r}. 

SDSS DR9 was also queried for spectroscopic redshifts, which were searched for within a 2 arcsec radius of the FIRST positions of our sample. 19 spectra were returned. No flags were given in the field zWarning, and all had reduced $\chi^2$ spectral fit values of $<$ 2, apart from three which had values of 2.06, 3.19 and 7.06. All spectra were `science primary', which ensured they were the best spectra available at each location.

\begin{table*}
 \centering
  \caption{Cross-Identifications of Ancillary Data}
  \begin{tabular}{lllccccc}
  \hline
  Survey & Telescope & Reference & Mean Resolution & Astrometric Precision & Search Radius & Misidentification & Matches\\
     & & & (arcsec) & (arcsec) & (arcsec) & Rate & \\
 \hline
 Various & {\it Spitzer} & -- & 1.66 & 0.2 & 5 & $\sim$1\% & 31\\
SDSS  DR9 & Sloan & \citet{2012ApJS..203...21A} & 1.3 & 0.1 & 2 & 3\% & 230\\
 PMN & Parkes & \citet{1995ApJS...97..347G} & 300 & $\sim$10 & 120 & 10\% & 19\\
 VLSS & VLA & \citet{2008ASPC..395..370L} & 80 & $<$ 8 & 15 & 3\% & 214\\
 \hline
\label{ancillary}
\end{tabular}
\end{table*}

\vspace{14mm}

\subsubsection{Searching NED}

To search for additional spectroscopic redshifts, the NASA/IPAC Extragalactic Database \citep[NED;][]{2011AAS...21734408S} was queried for each IFRS to within 2 arcsec. Of the 1317 IFRSs, 1137 unique matches were found for 1007 IFRSs, including duplicates. For the sake of simplicity, we discarded duplicate sources and simply kept the closest match from NED.

Six of the 19 spectroscopic redshifts from SDSS were returned from NED, which contained SDSS data only as recent as DR4. Additionally, 16 photometric redshifts were found \citep{2002ARep...46..531V, 2004ApJS..155..257R, 2004AJ....128..502A, 2005AJ....129.1755A, 2005ApJS..158..161H, 2006ApJS..162...38A, 2006A&A...445..889T, 2006ApJ...649...63S, 2008MNRAS.386..697R, 2009MNRAS.396.2011E}, but we do not use these photometric redshifts in our analysis as it is unlikely that the SED templates used match those of our IFRSs.

\subsubsection{Survey fields}

A number of our IFRSs were found within deep survey fields that have a large amount of multi-wavelength coverage. Table~\ref{deepFields} lists the number of sources from our sample that were identified to be within the approximate boundaries of several deep fields. The sources located within these deep fields can be found in the full table, available in the electronic version.

\begin{table*}
 \centering
  \caption{IFRSs within deep fields}
  \begin{tabular}{llc}
  \hline
Deep Field & Reference & Matches\\
 \hline
All-Wavelength Extended Groth Strip International Survey (AEGIS)  & \citet{2007ApJ...660L...1D} & 1\\
Bootes1 & \citet{2005ApJS..161....1M} & 2\\
European Large Area ISO Survey-North 1 (ELAIS-N1)  & \citet{2000MNRAS.316..749O} & 5\\
European Large Area ISO Survey-North 2 (ELAIS-N2)  &  \citet{2000MNRAS.316..749O} & 1\\
SDSS Stripe 82 & \citet{2009ApJS..182..543A} & 12\\
VLA First Look Survey (FLS)  & \citet{2003AJ....125.2411C} & 1\\
XMM Large Scale Structure (XMM-LSS) SWIRE Boundaries & \citet{2003PASP..115..897L} & 22\\
 \hline
\label{deepFields}
\end{tabular}
\end{table*}

%**********************************************************************************************************************************************************

\section{Results and Analysis}
%**********************************************************************************************************************************************************
\label{results}
 
Our final sample consists of 1317 IFRSs in $\sim$11 000 deg$^2$, although the area where {\it WISE} is sensitive enough to find IFRSs is much smaller. This implies a lower limit to the sky density of $\sim$0.1 per deg$^2$ for ${\rm S_{20 cm} > 7.5}\,$mJy. Using more sensitive observations, \cite{Norris2011} estimated a sky density of $\sim$7 per deg$^2$ for ${\rm S_{20 cm} > 0.1}\,$mJy. Additionally, the four sources in their sample with S$_{\rm 20cm} > 7.5\,$mJy have a higher density of $\sim$0.5 per  deg$^2$. 

\subsection{WISE and NVSS detections}

\subsubsection{NVSS 20cm flux density}

\begin{figure}
 \includegraphics[width=84mm,trim=0mm 25mm 0mm 15mm]{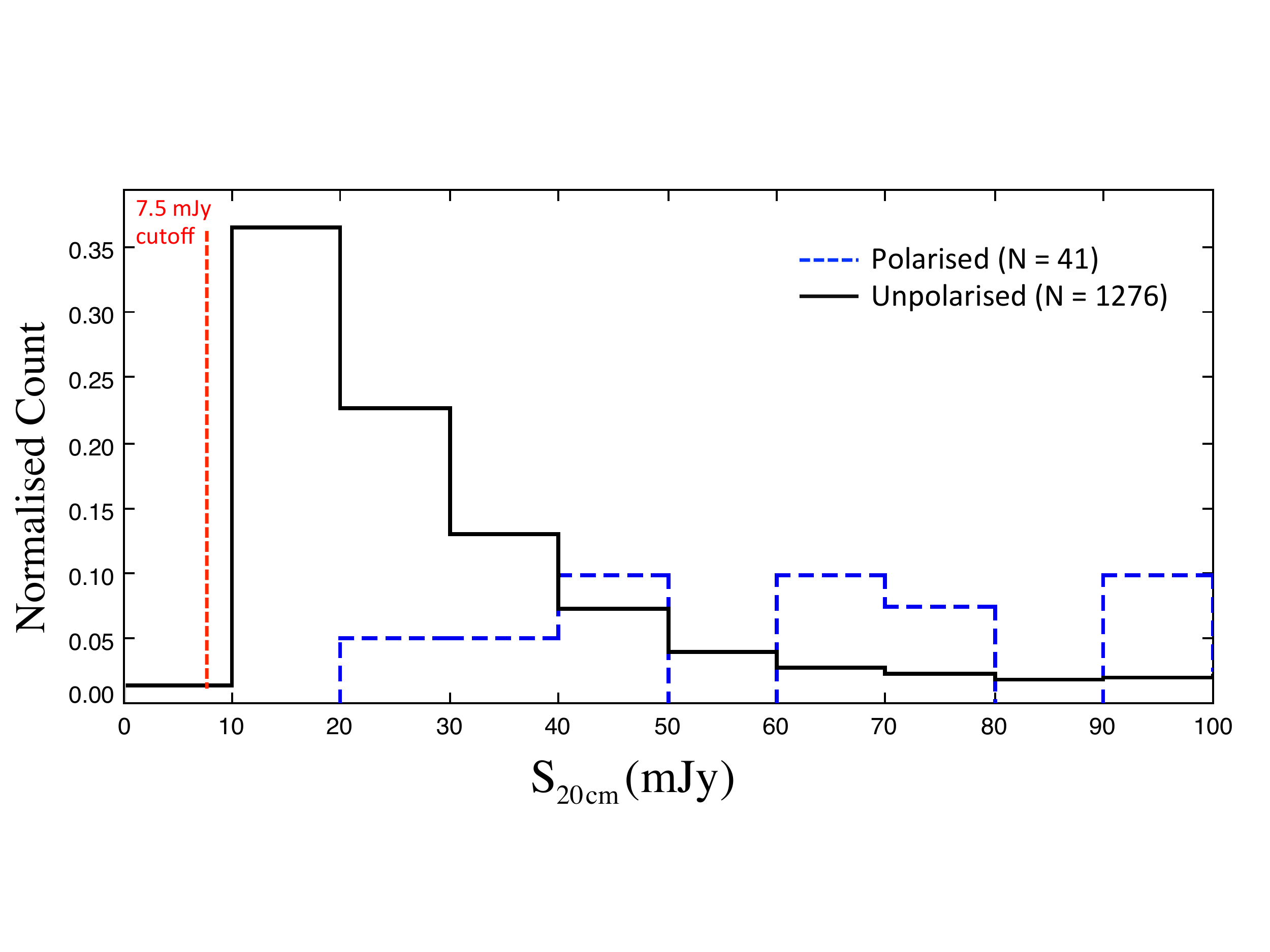}
 \caption{The normalised NVSS 20 cm flux density distribution from $0-100\,$mJy for our sample of IFRSs. The dotted red line indicates our 7.5$\,$mJy cutoff. However, since the {\it WISE} catalogue is far from complete below $\sim$40$\,\mu$Jy, and since we require $S_{\rm 3.4\,\mu m}/S_{\rm 20\,cm} > $ 500, the radio sources are highly incomplete below $\sim$20$\,$mJy. Sources polarized at levels of 8$\sigma_{\rm QU}$ and above are shown by the dotted blue line, while those with no polarization detected at this level are shown by the solid black line.}
 \label{NVSS}
\end{figure}

Because of our 7.5$\,$mJy cutoff, the 20$\,$cm flux densities found for our sample are much greater than those found for the first generation IFRSs. The distribution in the NVSS 20 cm flux density from 0 -- 100$\,$mJy for our IFRSs can be seen in Fig.~\ref{NVSS}. 

\subsubsection{NVSS 20cm polarization}

\begin{figure}
\begin{center}
\includegraphics[scale=0.5]{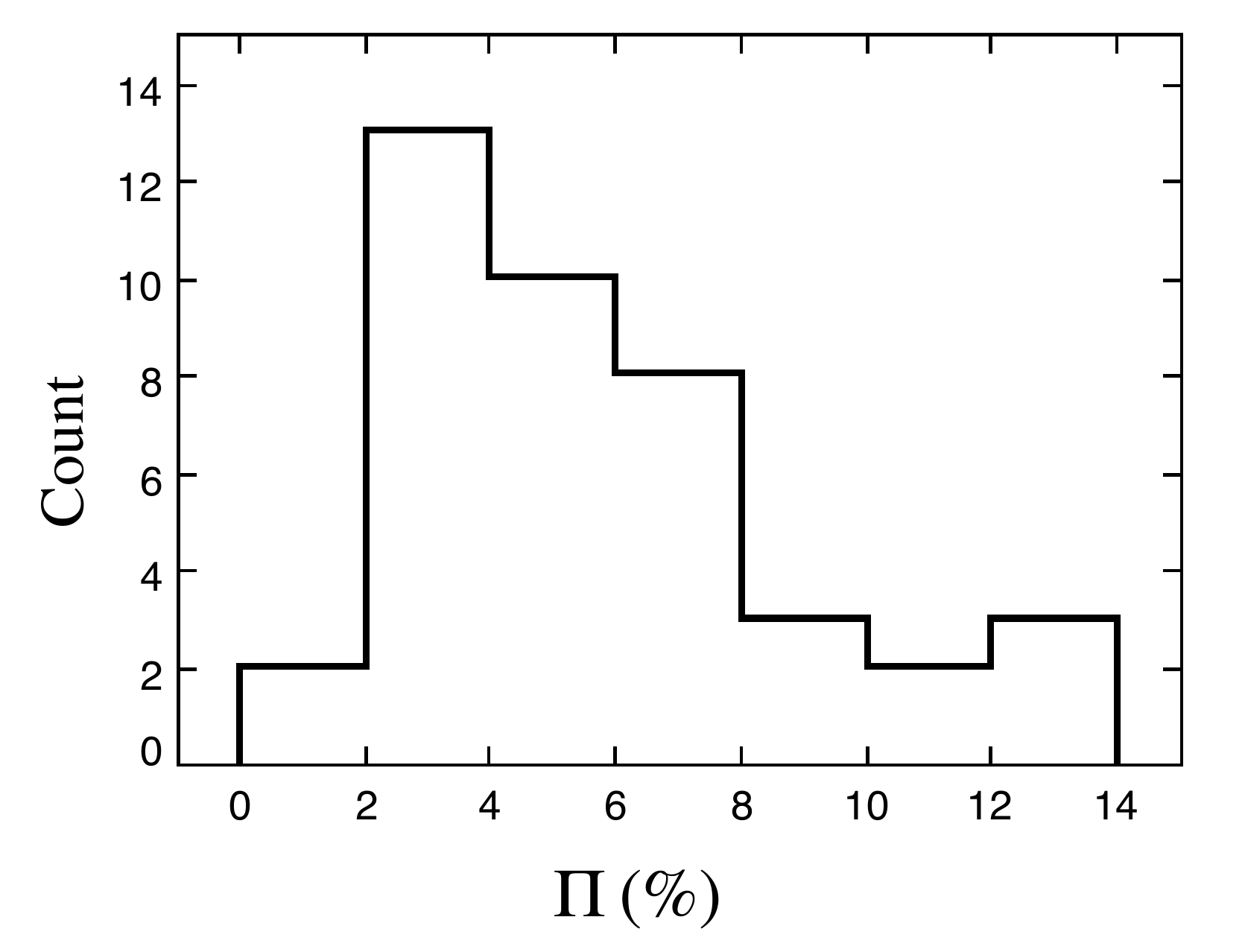}
\includegraphics[scale=0.5,trim=0mm 7mm 0mm 0mm]{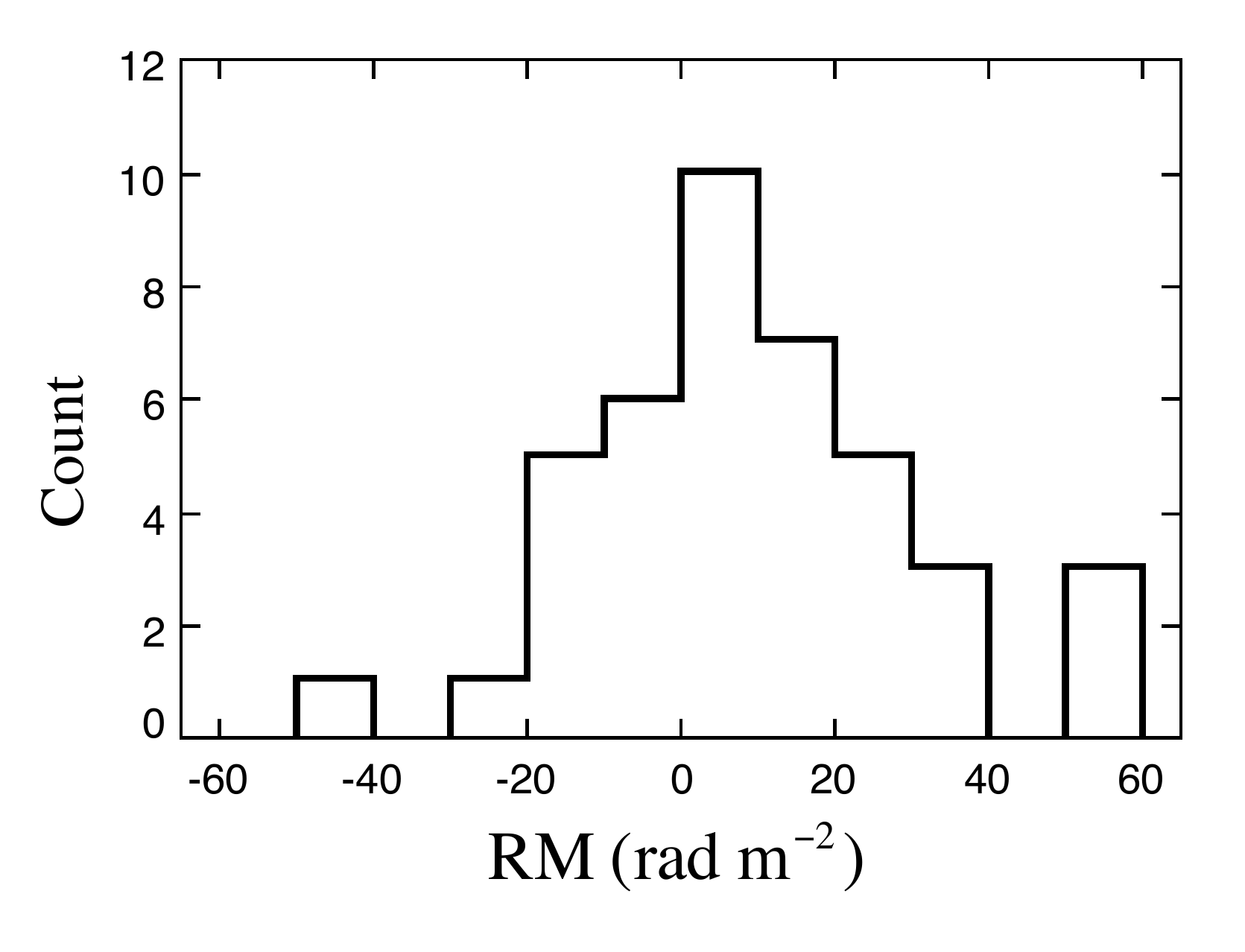}
\caption{Fractional polarization (top) and rotation measure (bottom) histograms for the 41 polarized IFRSs. The fractional polarization values range from 1 per cent $< \Pi < 14$ per cent, with the peak of 4 per cent and a median of 5.4 per cent. The RMs range from $-45.8 \le {\rm RM} \le 53.8\;$rad m$^{-2}$ and have a mean at 9.3 rad m$^{-2}$.}
\label{pHist}
\end{center}
\end{figure}

We matched our IFRS source list to the \citet{Taylor2009} RM catalogue, which lists $37~543$ polarized NVSS sources down to $8\sigma_{\rm QU}$. We accepted those sources with a match to our IFRS source list and a percentage polarization greater than 1 per cent $(\Pi = p/S \times 100$ per cent), since below this level, the instrumental on-axis polarization becomes significant. This resulted in extracting 41 polarized IFRSs, which increases the number of known IFRSs by five-fold. Fig.~\ref{pHist} shows the distribution of fractional polarization and RMs for our polarized IFRSs, which have fractional polarizations between 1 per cent and 14 per cent, and a median of 5.4 per cent. This is consistent with the findings of \cite{Middelberg2011} and \cite{Banfield2011}, who respectively found fractional polarizations ranging from $7-12$ per cent, and $\sim$$6-16$ per cent for their samples of IFRSs. \\

\begin{figure*}
 \includegraphics[width=\textwidth]{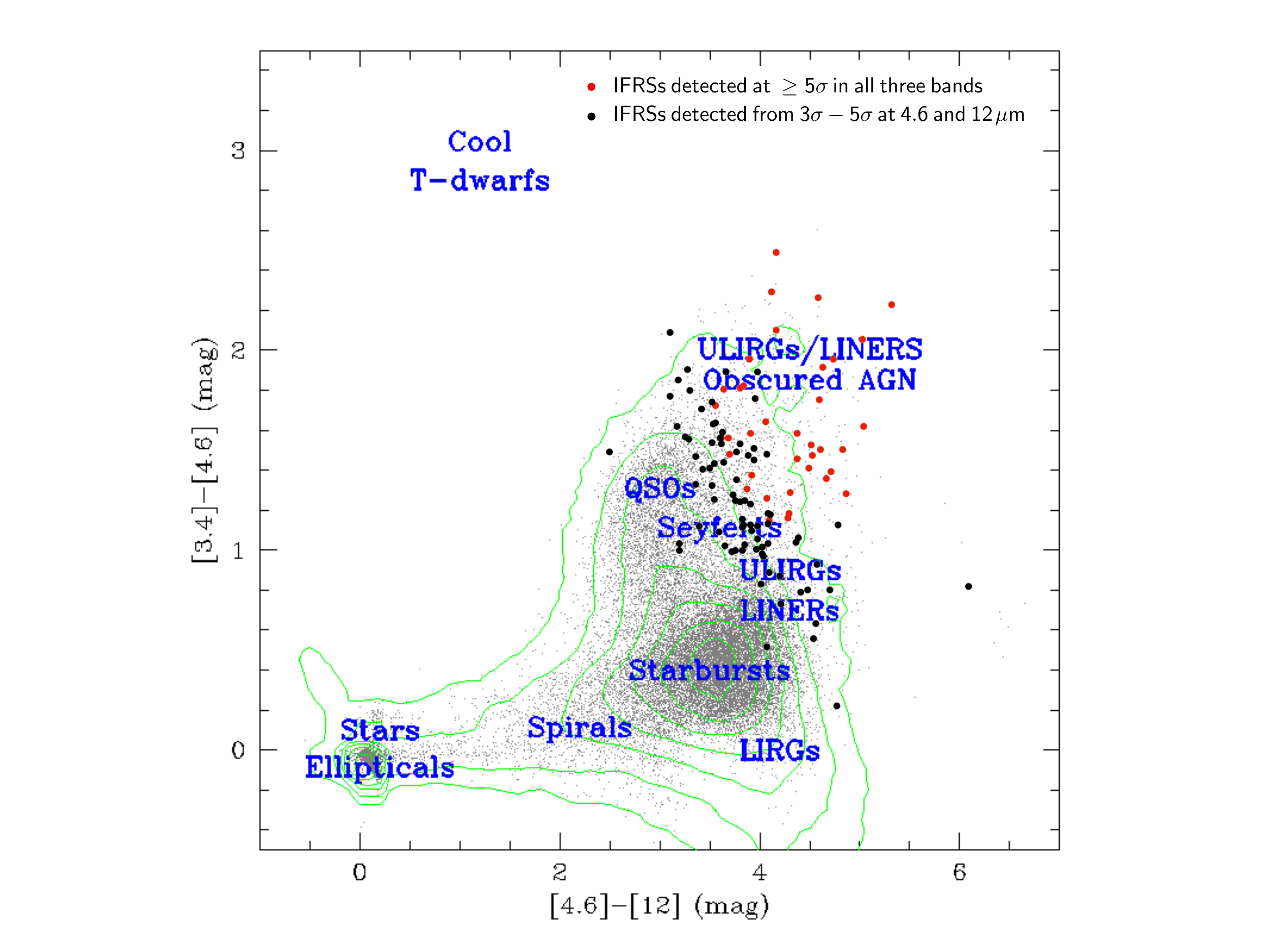}
 \caption{{\it WISE} colour-colour diagram of the 3.4, 4.6, and $12~\mu$m magnitudes for our sample of IFRSs compared to the sample from \citet{2011wise.rept....1C}. The red dots are the 37 IFRSs that were detected at $\ge5\sigma$ in all three bands, the black dots are the 107 IFRSs that were detected at $\ge5\sigma$ at 3.4$\,\mu$m and between 3$\sigma - 5\sigma$ at 4.6 or 12$\,\mu$m, and the grey points are the sources from \citet{2011wise.rept....1C}. Image adapted from \citet{2011wise.rept....1C}.}
 \label{colours}
\end{figure*}

The RMs values of our polarized IFRSs range from $-45.8 \le {\rm RM} \le 53.8\,$rad m$^{-2}$ and have a mean at 9.3 rad m$^{-2}$. If we neglect how the Galactic $\sigma$RM varies with Galactic latitude, since our IFRSs happen to be located more than 20 degrees away from the Galactic plane, then based on \citet{2010MNRAS.409L..99S}, the polarized NVSS sources will show a $\sigma$RM of 14 rad~m$^{-2}$, where $\sigma$RM $\approx$ 6 rad~m$^{-2}$ is the scatter in intrinsic RMs of polarized NVSS sources. This is close to the $\sigma$RM = 16 rad~m$^{-2}$ estimated using robust statistics from the binned RMs in Fig~\ref{pHist}. Therefore, the RMs for our polarized IFRSs are similar to those of typical polarized NVSS sources. 

However, the distribution of the properties of our polarized IFRSs cannot be effectively compared to the general population of polarized sources, since we have not corrected our sample for incompleteness. Banfield et al. (2014, in prep.) will investigate the radio polarization properties of a sample of IFRSs not suffering from incompleteness.

\subsubsection{{\it WISE} flux densities}

Fig.~\ref{colours} shows a colour-colour diagram of the IFRSs detected at 3.4, 4.6, and $12~\mu$m, as compared to those from \citet{2011wise.rept....1C}. This figure reveals that the majority of our IFRSs have {\it WISE} colours similar to those found for obscured AGN, quasars and Seyferts \citep{Jarrett2011,2013ApJ...772...26A}. 

The distribution in the {\it WISE} flux densities for our IFRS sample is listed in Table~\ref{polTable}. We show the number of IFRSs detected at $\ge5\sigma$ in the {\it WISE} bands that are polarized at 20$\,$cm compared to the IFRSs with no detectable polarization. The numbers are consistent in both groups indicating that the polarized IFRSs are similar to the unpolarized IFRSs in the infrared. However, from the $20\,$cm flux densities, we conclude that our sample of polarized IFRSs are AGN, as \citet{2010ApJ...714.1689G} determined that polarized sources with $S_{20\,{\rm cm}} \ge 1.0\;$mJy are lobe-dominated AGN. As can be seen in Table~\ref{polTable}, the number of sources detected in the infrared decreases with increasing wavelength, partly because of the lower sensitivity of {\it WISE} at longer wavelengths, and partly because the sources typically follow a power-law SED.

\begin{table*}
 \centering
  \caption{The flux density distribution of the IFRSs detected at the $5\sigma$ level or greater in the different {\it WISE} bands. The percentages in the brackets indicate the percentage of total sources detected in each band. The percentage error is given by $\frac{N_{\rm {band}}}{N_{\rm {TOT}}} / \sqrt{N_{\rm{band}}}$.}
  \label{polTable}
  \begin{tabular}{lrrrr}
  \hline
   & $N_{3.4}$ & $N_{4.6}$ & $N_{12}$ & $N_{22}$ \\
 \hline
 Unpolarized & 1276 & 458 ($35.9 \pm 1.7\%$) & 47 ($3.7 \pm 0.5\%$) & 8 ($0.6 \pm 0.2\%$)\\
 Polarized & 41 & 17 ($41.5 \pm 10.1\%$) & 1 ($2.4 \pm 2.4\%$) & 0 ($0.0 \pm 0.0\%$)\\
 All Sources & 1317 & 475 ($36.1 \pm 1.7\%$) & 48 ($3.6 \pm 0.5\%$) & 8 ($0.6 \pm 0.2\%$)\\
  \hline
\end{tabular}
\end{table*}

\subsubsection{Radio-IR flux density ratios}

\label{ratios}

\begin{figure}
 \includegraphics[width=84mm,trim=0mm 25mm 0mm 30mm,clip]{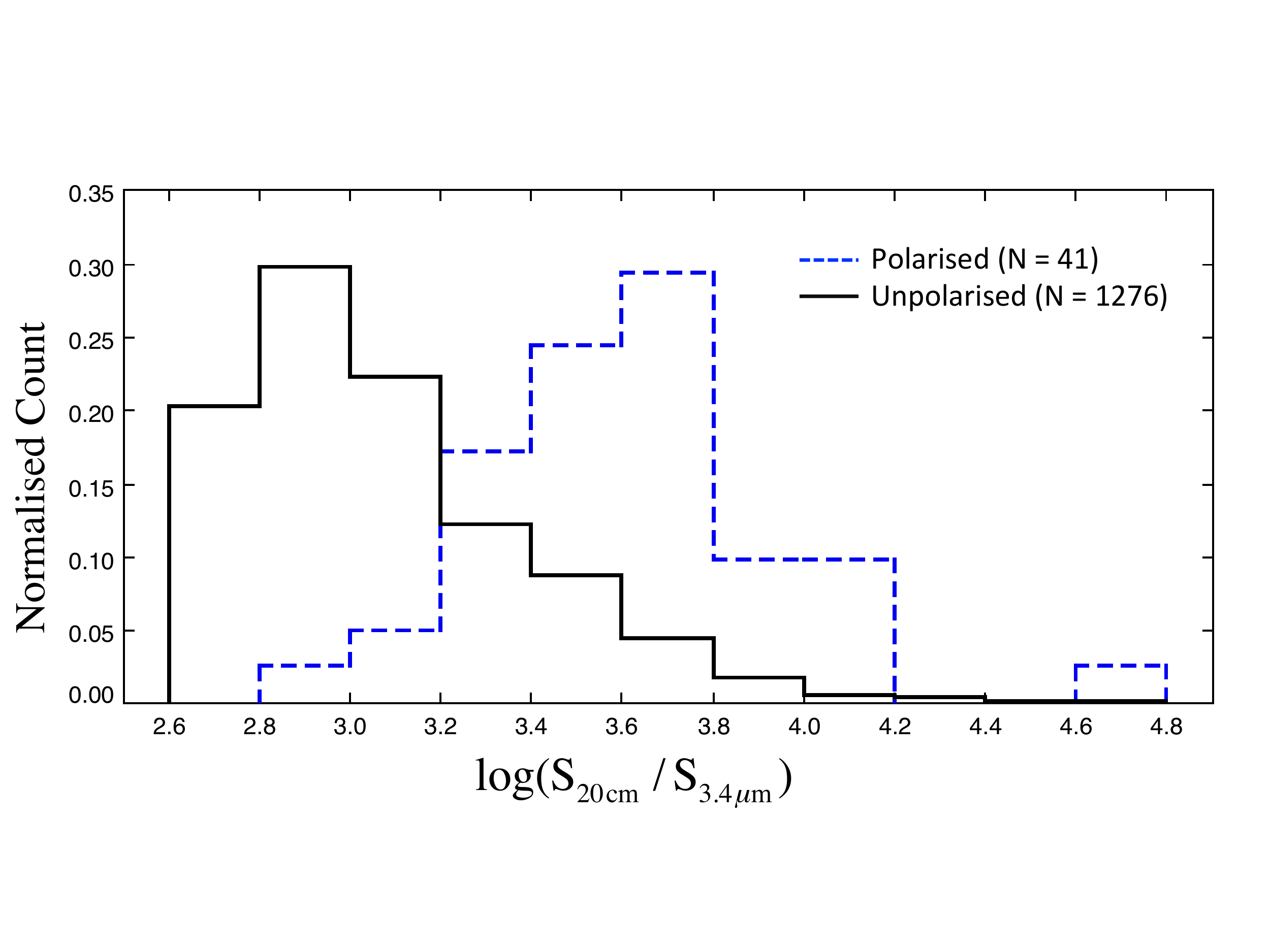}
 \caption{Normalised histogram of log(${\rm S_{20cm} / S_{3.4\mu m}}$) for sources with polarization detected $\ge 8\sigma_{\rm QU}$ (dotted blue line) and for sources with no detected polarization at this level (solid black line). The drop-off at 2.6 corresponds to our selection criterion ${\rm S_{20cm} / S_{3.4\mu m}}$ $> 500$, since $\log(500) \approx 2.7$.}
 \label{ratio}
\end{figure}

Fig.~\ref{ratio} shows the radio to infrared flux density ratio ${\rm S_{20cm} / S_{3.4\mu m}}$ for all of our sources. This is similar to the distribution in ${\rm S_{20cm} / S_{3.4\mu m}}$ for the sample from \citet{Middelberg2011}, although peaks slightly lower, since our sources are brighter in the infrared. 

We also calculate the radio to far-infrared (FIR) ratio ${\rm q_{22} = \log(S_{22\mu m}/S_{20 cm})}$, using 22 $\mu$m flux densities for 31 sources detected at 3$\sigma$ and above, and 22 $\mu$m 3$\sigma$ upper limits for the remaining 372 sources that had reliable r.m.s. noise ($\sigma$) estimates. Those detected all have values ${\rm q_{22} < -0.23}$, while those with upper limits all have values ${\rm q_{22} < -0.30}$. From the 31 sources with {\it Spitzer} counterparts, four had detected counterparts at 24 $\mu$m, for which we found ${\rm q_{24}} < -1.28$. These values of ${\rm q_{22}}$ and ${\rm q_{24}}$ are all well below those seen for SFGs, which are typically  ${\rm q_{24}} \sim 1$ \citep{2004ApJS..154..147A}, suggesting that the vast majority of our sources are AGN.

\subsection{Radio morphology}

The number of resolved and compact IFRSs from our sample was determined using two different methods: flux density ratios and visual inspection. We adopt the \citet{Kimball2008} criterion for the definition of a source that is unresolved in FIRST, which has a ratio between its peak flux $F_{\rm peak}$ and integrated flux density $F_{\rm int}$ of 

\begin{equation}
\log\frac{F_{\rm int}}{F_{\rm peak}} < 0.05 \ .\\
\label{SourceCompactness}
\end{equation}

\noindent The number of resolved and unresolved IFRSs that were found using these two methods is given in Table~\ref{morphTab}. The compactness of our sources can be used as a proxy for their projected linear size, since in $\Lambda$CDM, at $z > 1$, the angular size corresponding to a fixed linear size varies only weakly with $z$. Sources that are unresolved in FIRST have linear sizes $\lesssim 30$ kpc at any redshift $> 0.5$.

\begin{table*}
 \centering
  \caption{The total number of resolved and unresolved IFRSs (polarized IFRSs in brackets) found using the flux density ratio and visual inspection methods.}
  \label{morphTab}
  \begin{tabular}{llll}
  \hline
& Flux Density Ratio & Visual Inspection & Both\\
 \hline
Unresolved & 845 (13) & 946 (12) & 760 (5)\\
Resolved & 472 (28) & 371 (29) & 286 (21)\\
  \hline
\end{tabular}
\end{table*}

\begin{figure}
\includegraphics[scale=0.88]{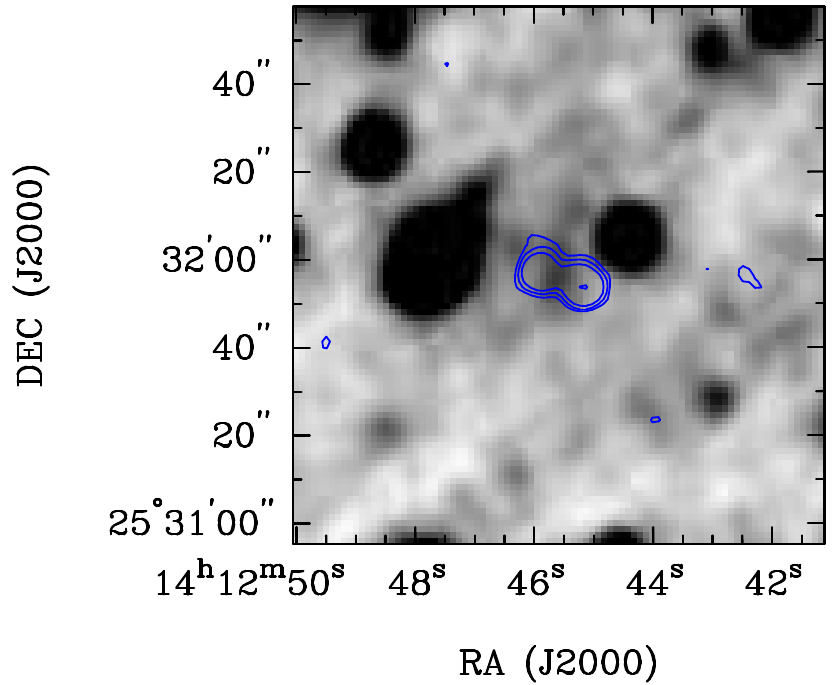}
\includegraphics[scale=0.88]{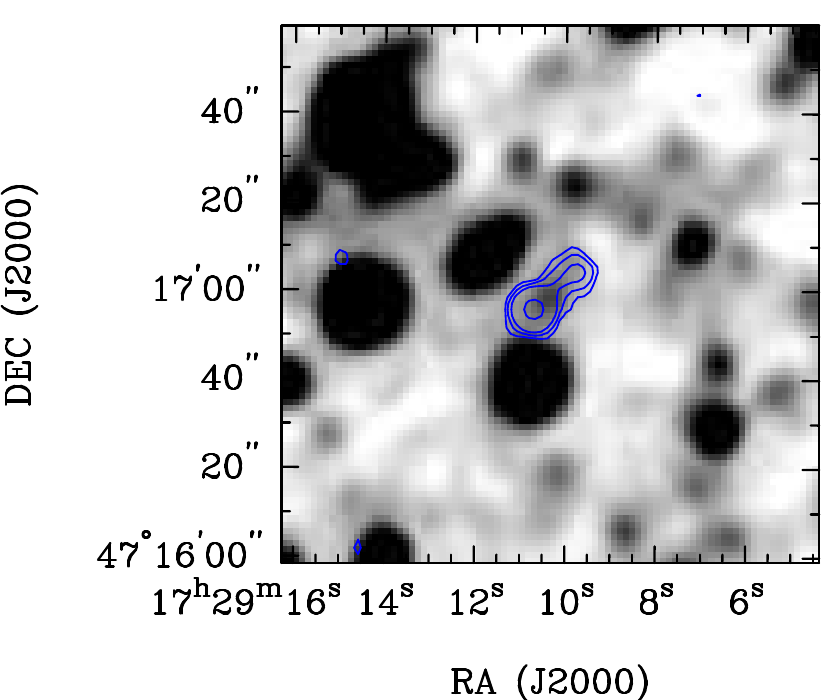}
\includegraphics[scale=0.88]{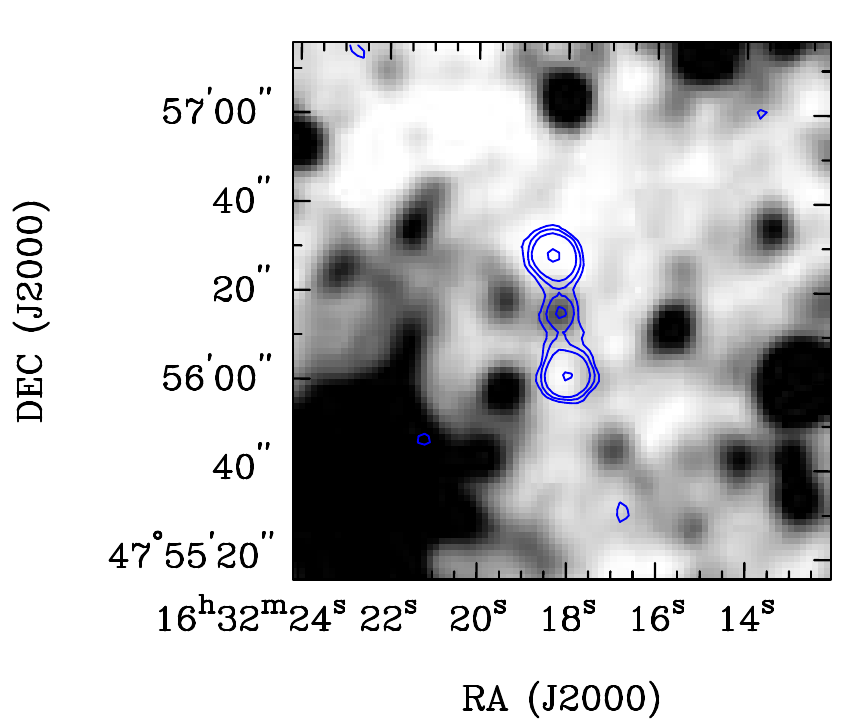}
\caption{Examples of an IFRS with a double-lobed morphology with one (top), two (middle), and three (bottom) catalogued FIRST components. The background image shows the {\it WISE} $3.4\,\mu$m detection, and the contours mark the FIRST source at 20$\,$cm. Contour levels are: 3, 6, 12, 64, 256, 1024 times the local noise level.}
\label{postage2}
\end{figure}

From the 946 IFRSs that appeared resolved from visual inspection, 157 had uncertain morphologies, and 214 had double-lobed morphologies (most likely FR~II galaxies), which included sources with a single catalogued FIRST component that could still be identified as a double-lobed galaxy (e.g. Fig.~\ref{postage2} (top)). Of the 41 polarized sources, 23 had morphologies of a double-lobed galaxy, 6 appeared resolved with unknown morphologies, while the remaining 12 sources appeared unresolved. Fig.~\ref{postage2} shows some examples of IFRSs with double-lobed morphologies with one, two and three catalogued components in FIRST. The double-lobed morphologies are consistent with previous hypotheses about IFRSs, which have found that IFRSs are well represented by more distant FR II galaxies. However, the majority of our sources are unresolved, like the IFRSs from \citet{Middelberg2011}, suggesting they may be younger radio galaxies with smaller jets.

A much larger fraction of polarized IFRSs have resolved morphologies ($\gtrsim$ 70 per cent), compared to the unpolarized IFRSs that have resolved morphologies ($\lesssim$ 35 per cent). This is consistent with that found by \citet{Banfield2011}, who found that  $\sim$80 per cent of their polarized sources and $\sim$15 per cent of their sources with no detected polarization were resolved. It is also consistent with the number of lobe-dominated AGN we expect for polarized sources with $S_{\rm 20cm} \ge 1.0$ mJy \citep{2010ApJ...714.1689G}.

\subsection{Radio spectra}
\label{specIndx}

\subsubsection{Spectral shape}

From the URC, we extracted flux densities at $\nu =$ 326, 1400 and 4850 MHz ($\lambda =$ 92, 20 and 6 cm), respectively from WENSS, NVSS, and GB6. In order to derive the radio spectra, we used the flux densities from two or three of these frequencies. Because it was not possible to match the beamsizes and epoch of observation of the various surveys, the flux density at a particular frequency was used only if the source:

\begin{enumerate}
\item was unconfused;
\item was unresolved (according to equation~\ref{SourceCompactness}, except at 6 cm);
\item was the only match within 30$\arcsec$ of NVSS source;
\item had no error flags;
\end{enumerate}

Although upper limits of the flux densities cannot be effectively used to determine the spectral index, they are effective in constraining the shape of the radio spectra. Therefore, we measure the shape of the radio spectrum for our IFRSs using all the flux densities available ($S_{20\rm cm}$ always available) and where this is unavailable, upper limits. Where the detection is not significant, a 5$\sigma$ upper limit is attributed to the source. Where no detection is made within the WENSS or GB6 footprint (i.e. when the source does not appear in the WENSS or GB6 catalogue, but is still in its footprint), an upper limit of 18 mJy is attributed to the source, since this is the sensitivity limit of both the WENSS and GB6 surveys. To identify GPS sources, we use the flux densities and limits from all three frequencies, since at least three points are needed to identify a peak. We define a GPS source as a source that:

\begin{enumerate}
\item is unresolved in FIRST; 
\item contains a positive spectral index between 92 and 20$\,$cm and a negative spectral index between 20 and 6$\,$cm; 
\item has a minimum 20$\,$cm flux density of ${\rm S_{20cm} - 3\Delta S_{20cm}}$ greater than the expected value at 20$\,$cm as extrapolated from the spectral index fit between the two points ${\rm S_{92cm} + \Delta S_{92cm}}$ and ${\rm S_{6cm} + \Delta S_{6cm}}$, where $\Delta$S is the 1$\sigma$ error in the flux density. 
\end{enumerate}

\noindent The last of these criteria ensures that the peak lies outside the uncertainty in the spectral index, separating the GPS sources from the flat-spectrum sources. We also use this criterion to identify curvature in other radio spectra. We define a CSS source as a source that:

\begin{enumerate}
\item is unresolved in FIRST; 
\item has a spectral index of $\alpha < -0.8$.
\end{enumerate}

To identify CSS sources, we use flux densities from two or three frequencies, and upper limits on the flux density only at $20\,{\rm cm}$ and $6\,{\rm cm}$, since we expect all CSS sources with ${\mbox S_{\rm 20 cm} > 7.5}$~mJy to be detected in WENSS. 

\begin{figure}
\begin{center}
\includegraphics[scale=0.23]{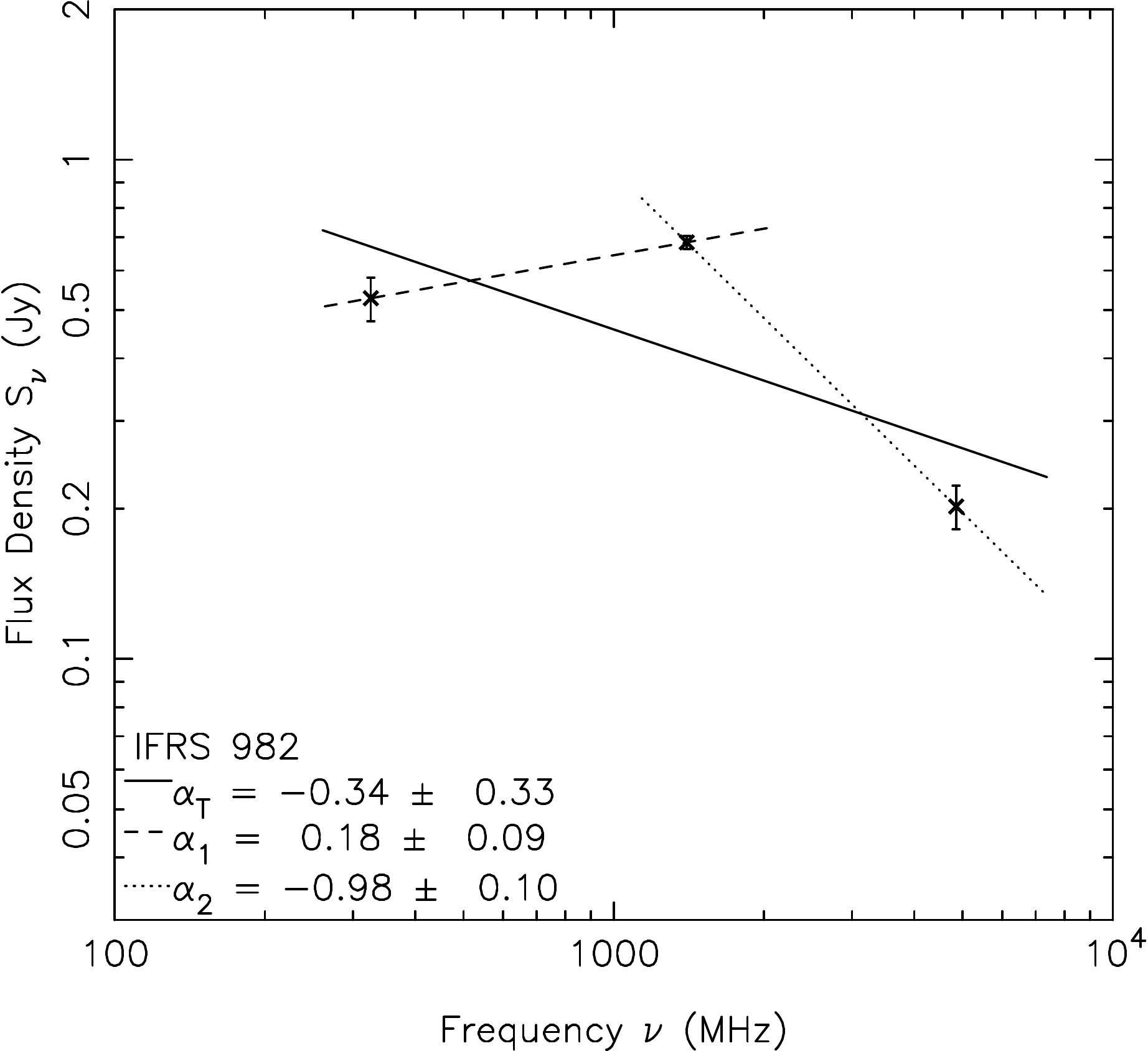}
\includegraphics[scale=0.23]{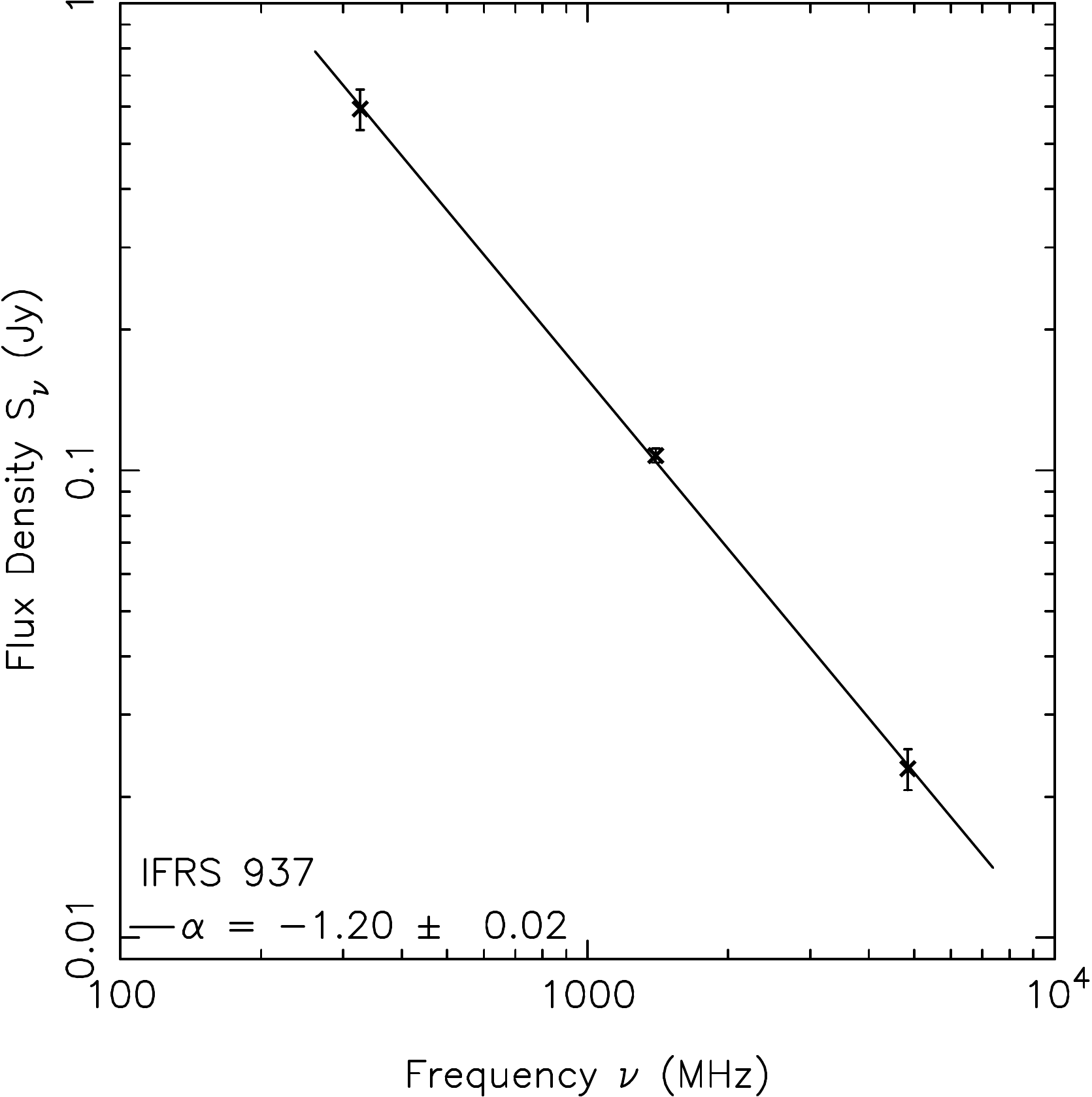}
\caption{The radio spectra of a GPS (left) and a CSS (right) source from our sample of IFRSs.}
\label{specIndices2}
\end{center}
\end{figure}

We find that 208 of our IFRSs have a steep spectrum of $\alpha < -0.8$, while 32 have a gigahertz-peaked spectrum. All of these GPS candidates and 124 of the CSS candidates are considered compact in FIRST according to equation~\ref{SourceCompactness}, and we therefore respectively define them as GPS and CSS sources. Fig.~\ref{specIndices2} shows the radio spectra of a GPS and a CSS source from our sample. These findings are consistent with \cite{MiddelbergVLBI} and \cite{Middelberg2011}, who found that their IFRS sample consisted of GPS and CSS sources, based on finding compact AGN cores and curvature in the radio spectra. This implies that a substantial fraction of our IFRSs are young and evolving AGN, with jets that have not broken out far beyond the galaxy. 

Of the 13 polarized sources that are compact according to equation~\ref{SourceCompactness}, six are CSS sources, three are USS sources, and one is a GPS source.

\subsubsection{Spectral indices}

We derived spectral indices using a least squares fit in the log-log domain for all sources that: (1) had a signal-to-noise ratio $\ge5$; (2) did not show a GPS spectrum. Given the 18 mJy sensitivity limits at 92 and 6$\,$cm, sources close to the median 20 cm flux density of 25.5 mJy will not be detected at 92 cm if they are shallower than $\alpha = 0.24$, nor at 6 cm if they are steeper than $\alpha = -0.28$. Hence there is a large bias on our spectral indices, with the 6 cm data tending to give shallower spectral indices, and the 92 cm tending to give steeper spectral indices. 

\begin{figure}
\includegraphics[scale=0.33,trim=0mm -10mm 0mm 0mm]{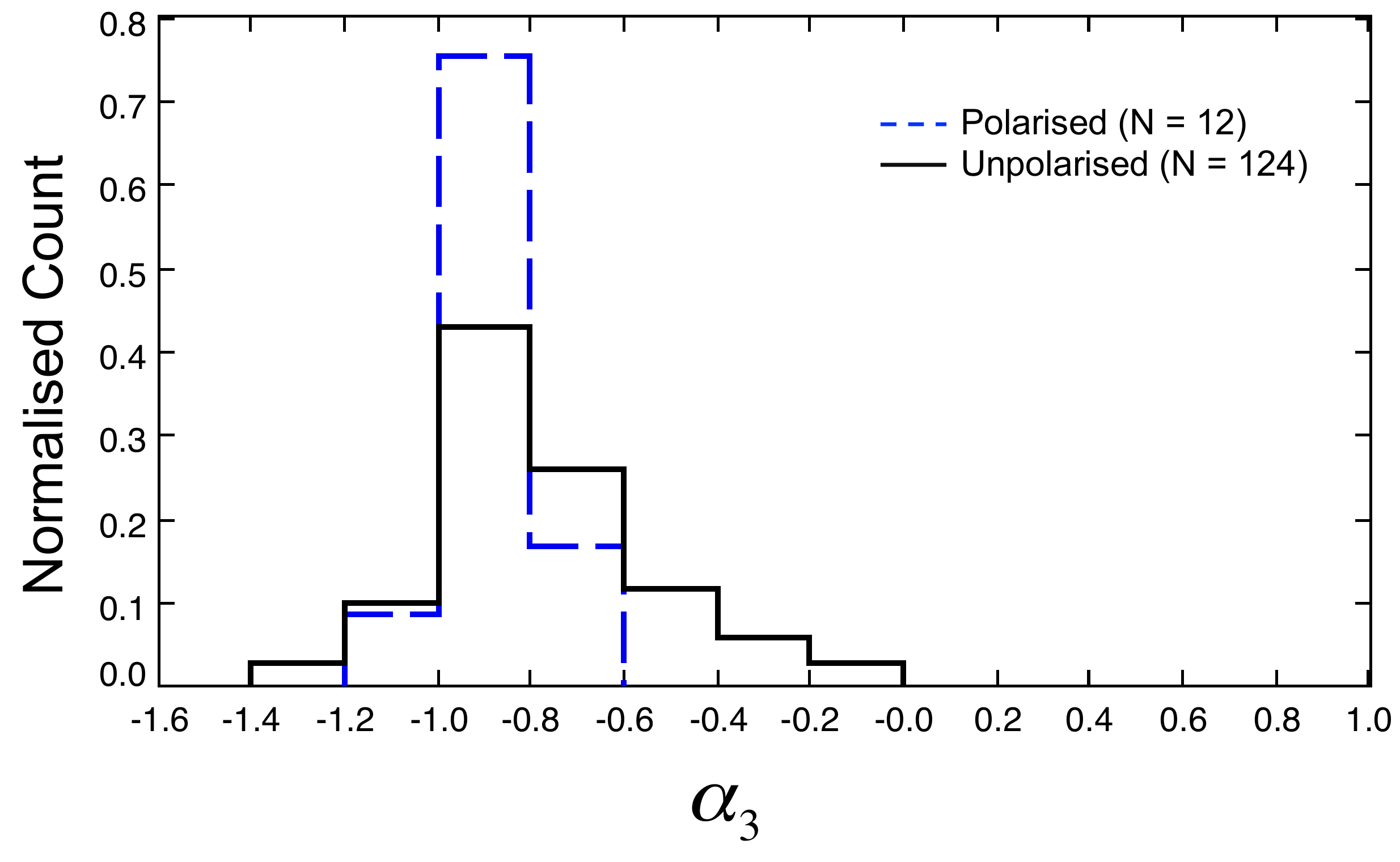}
\includegraphics[scale=0.33,,trim=0mm -10mm 0mm 0mm]{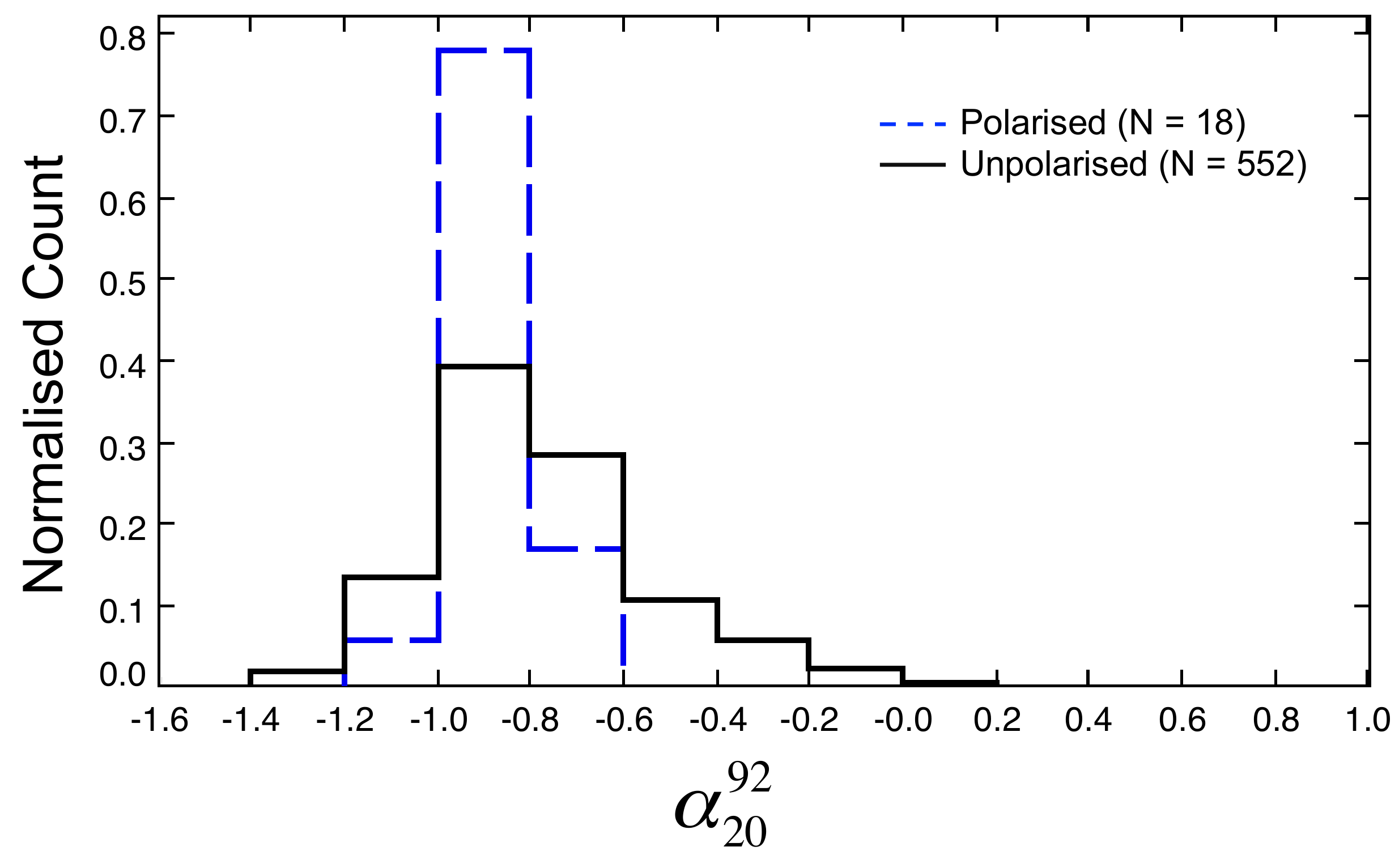}
\includegraphics[scale=0.33,,trim=0mm -10mm 0mm 0mm]{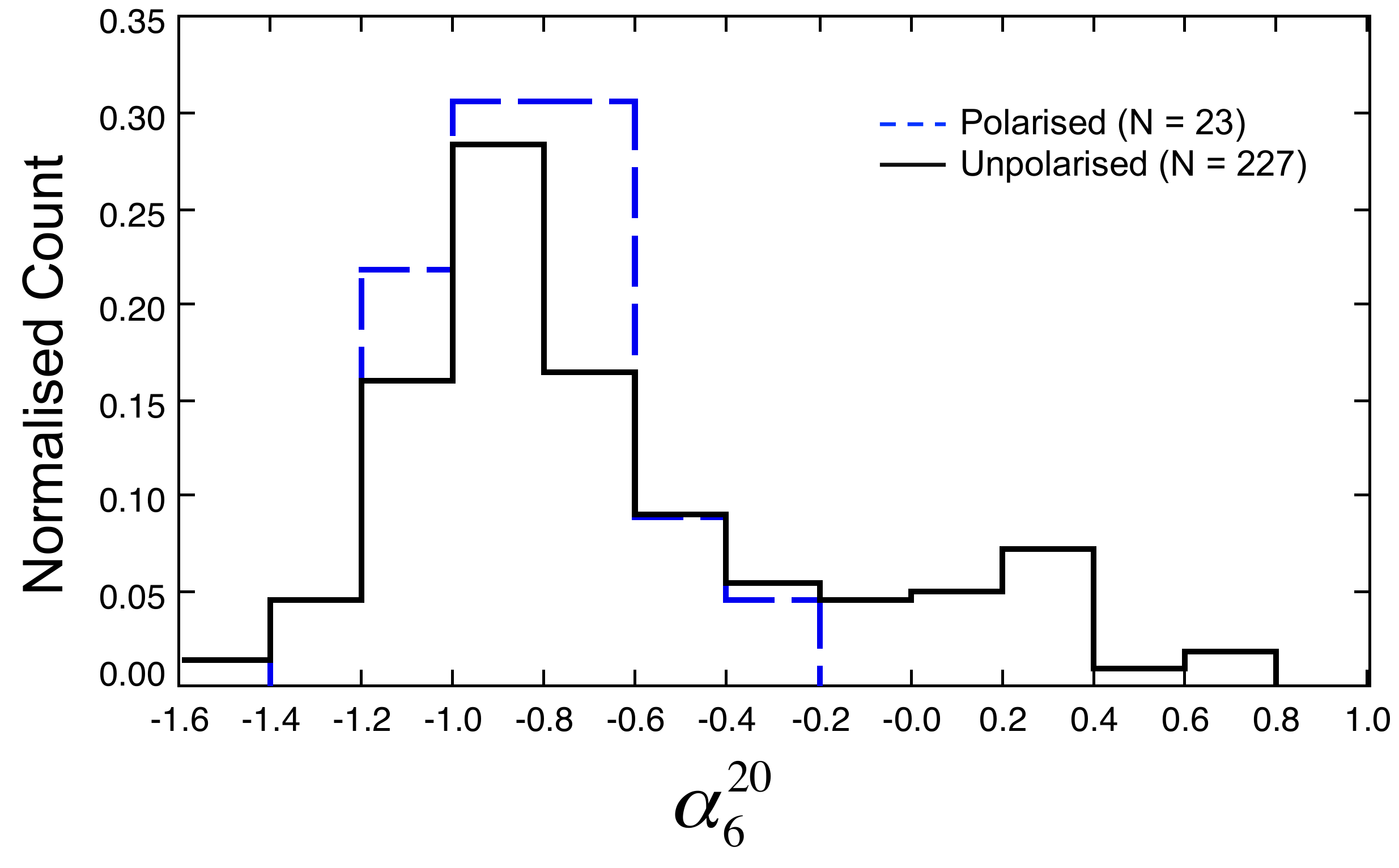}
\caption{Normalised histogram of the spectral indices of the compact non-GPS IFRSs as found respectively between frequencies one (92$\,$cm), two (20$\,$cm) and three (6$\,$cm) ($\alpha_3$: top), frequencies one and two ($\alpha_{20}^{92}$: middle) and frequencies two and three ($\alpha_{6}^{20}$: bottom). Sources polarized at $8\sigma_{\rm QU}$ and above are shown by the blue dotted line, and sources with no detectable polarization at this level are shown by the solid black line.}
\label{SI2}
\end{figure}

\begin{figure}
\includegraphics[scale=0.24]{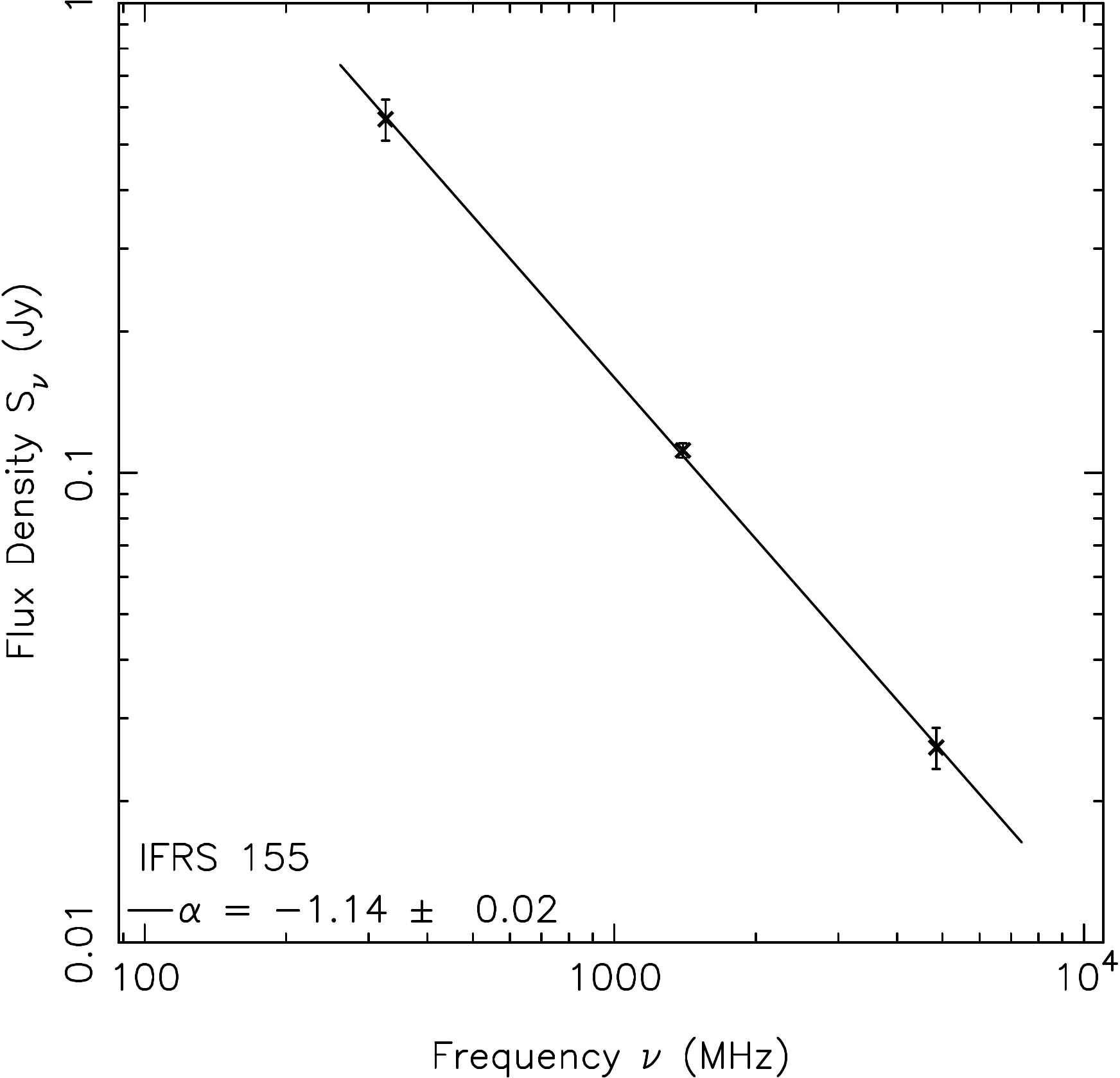}
\includegraphics[scale=0.24]{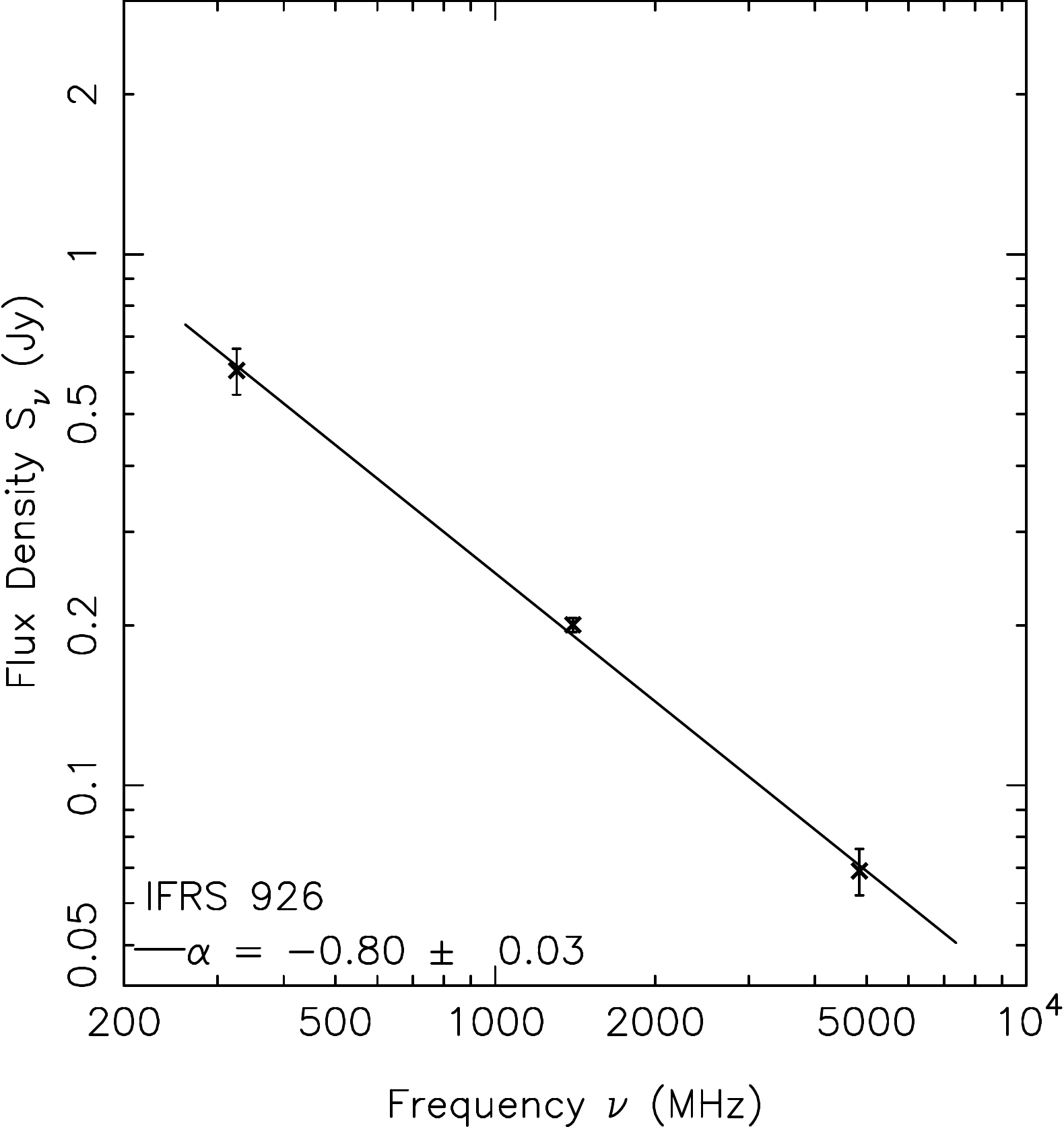}
\includegraphics[scale=0.25]{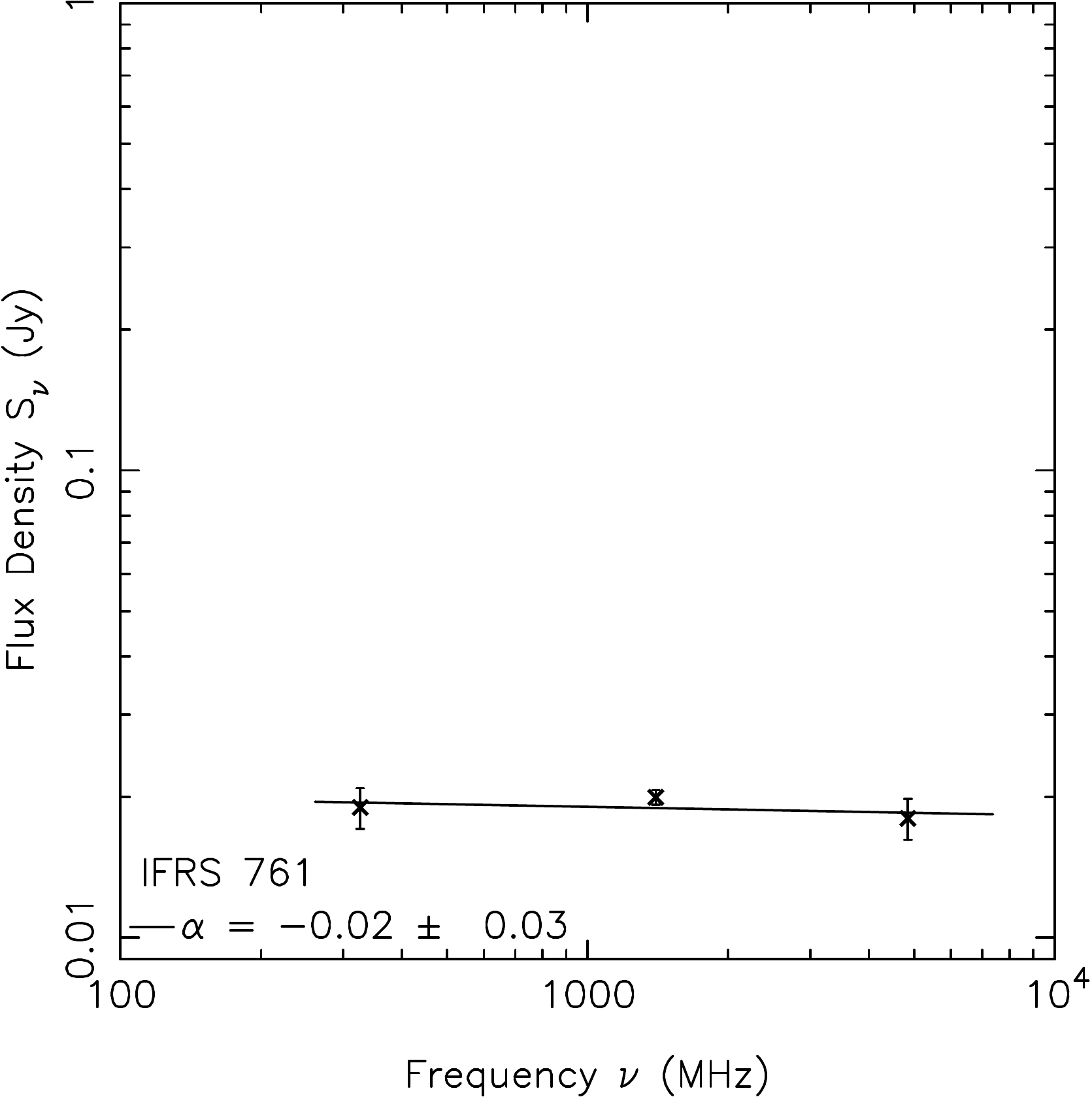}
\includegraphics[scale=0.25]{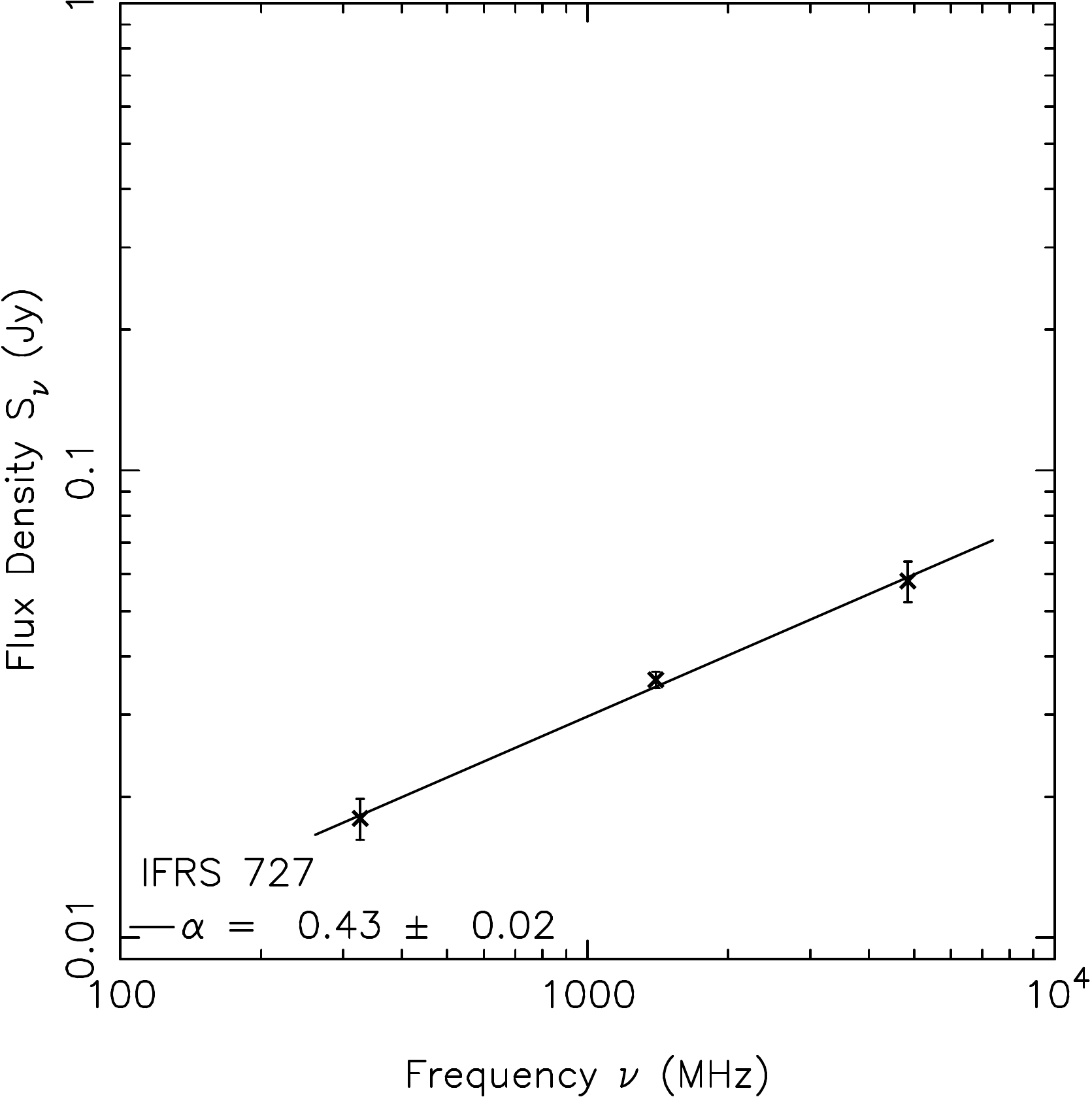}
\caption{The radio spectra of some IFRSs from our sample with a USS (top left), steep (top right), flat (bottom left) and inverted (bottom right) spectral index.}
\label{specIndices}
\end{figure}

The median values of the spectral index for three different groupings of frequencies are listed in Table~\ref{specIndxTab}, and their distributions are shown in Fig.~\ref{SI2}. We find that our sample is made up of USS, steep, flat and inverted sources, as shown in Fig.~\ref{specIndices}. \cite{Banfield2011} and \cite{Middelberg2011} found much steeper median spectral indices of $-1.07$ and $-1.01$ (with the majority of sources having $\alpha < -1.0$) when they studied the ELAIS-N1 and ATLAS fields respectively. Those from \citet{Banfield2011} were measured from the flux densities at 92 and 20$\,$cm. However, the spectral indices from \cite{Middelberg2011} were calculated primarily from the flux densities at 20, 13, 6 and 3$\,$cm, and in some cases, also from the 36$\,$cm flux densities. Therefore a GPS source will appear to be steeper when measured at the frequencies used by \cite{Middelberg2011} than at the frequencies measured in this paper.

Inverted sources with $\alpha > 0$ are possibly very young GPS sources that peak above 4850 MHz, like those seen by \citet{2009AN....330..180H}. These could also be variable sources with varying flux densities across the different epochs, or Blazars.

Objects with steep radio spectra are more likely to be found at high redshifts than objects with flatter radio spectra, and the USS criterion ($\alpha \lesssim -1.0$) has been employed to discover most of the known radio galaxies at $z > 3.5$ \citep[][and references therein]{2013PASA}. Most of our IFRSs have steep spectral indices (see Table~\ref{specIndxTab}), and many are USS, suggesting that a significant fraction of them are likely to be located at high redshifts. 

\begin{table*}
 \centering
  \caption{The median spectral indices of our IFRSs. The first row contains sources with available flux densities at 92, 20 and 6$\,$cm, while the second row has only flux densities at 92 and 20$\,$cm, and the third row only at 20 and 6$\,$cm. Sources in the first row are also in the second and third row. $N_{\rm USS}$ signifies the number of USS ($\alpha \lesssim -1.0$) sources.}
  \label{specIndxTab}
  \begin{tabular}{cccccccc}
  \hline
  Wavelengths used to fit $\alpha$ & $N_{\rm total}$ & $\alpha_{\rm total}$ & $N_{\rm polarized}$ & $\alpha_{\rm polarized}$ & $N_{\rm unpolarized}$ & $\alpha_{\rm unpolarized}$ & $N_{\rm USS}$\\
 \hline
92, 20 \& 6$\,$cm ($\alpha_3$) & 136 & $-0.82$ & 12 & $-0.90$ & 124 & $-0.82$ & 16 \vspace{1mm}\\
92 \& 20$\,$cm ($\alpha^{92}_{20}$) & 570 & $-0.82$ & 18 & $-0.88$ & 552 & $-0.81$ & 84\vspace{1mm}\\
20 \& 6$\,$cm ($\alpha^{20}_{6}$) & 249 & $-0.80$ & 23 & $-0.91$ & 227 & $-0.79$ & 55\\
  \hline
\end{tabular}
\end{table*}

\subsection{Optical matches}

Most of our IFRSs are undetected in SDSS to the limiting magnitude of {\it r} $=22.2$, which is consistent with previous studies of IFRSs, in which few optical counterparts were found. The distribution of the SDSS {\it r} magnitudes for the 230 detected sources is shown in Fig.~\ref{fluxComp} (bottom), which shows an increase up to a magnitude of 22.2, beyond which SDSS becomes highly incomplete. This shows that at least a fraction of our IFRSs are also optically brighter than first generation IFRSs. The {\it WISE} sources studied by \citet{2013AJ....145...55Y} peak at an SDSS {\it r} magnitude of 20, and these sources lie almost entirely at $z < 1$. This suggests that most of our IFRSs are likely to lie at redshifts of $z > 1$, but could also be at $z < 1$ and suffer from more obscuration than typical {\it WISE} galaxies.

\subsection{Redshifts}

\label{Redshifts}

\begin{table*}
\centering
\caption{Sources from our sample with matching spectroscopic redshifts from SDSS DR9. Listed is the SDSS RA and DEC, spectroscopic redshift and uncertainty, the NVSS flux density at 20 cm, the luminosity at 20 cm, the {\it WISE} flux density and luminosity at 3.4 $\mu$m, the flux density ratio between 20$\,$cm and 3.4$\,\mu$m, the sky separation between the FIRST and SDSS positions, and the class (and subclass) of the spectra as labelled by SDSS. Notes: (1) `BL' stands for broadline. (2) The luminosities are given at the observed frequency and have not been K-corrected.
}
\label{specz}
\begin{tabular}{ccccccccccc}
\hline\hline
RA & DEC & Spectroscopic & S$_{20}$ & L$_{20}$ & S$_{3.4}$ & L$_{3.4}$ & S$_{20}$/S$_{3.4}$ & Sky Separation & Class \\
     \multicolumn{2}{c}{J2000.0} & redshift & [mJy] & [W Hz$^{-1}$] & [$\mu$Jy] & [W Hz$^{-1}$]  & & [arcsec] &\\
\hline
  02:06:22.45 & $-$00:43:20.68 & 2.3767 $\pm$ 0.0005 & 19.9 $\pm$ 0.7 & 8.8 $\times 10^{26}$ & 28.3 $\pm$ 5.8 & 1.2 $\times 10^{24}$ & 703 $\pm$ 175 & 0.124 & BL QSO\\
  11:35:41.11 & $+$45:40:42.85 & 2.1420 $\pm$ 0.0007 & 17.3 $\pm$ 0.7 & 5.9 $\times 10^{26}$ & 29.1 $\pm$ 5.5 & 9.9 $\times 10^{23}$ & 594 $\pm$ 140 & 0.797 & BL QSO\\
  11:36:34.93 & $+$61:06:20.53 & 2.0296 $\pm$ 0.0005 & 18.3 $\pm$ 0.7 & 5.5 $\times 10^{26}$ & 27.3 $\pm$ 5.5 & 8.2 $\times 10^{23}$ & 670 $\pm$ 164 & 1.360 & BL QSO\\
  12:27:43.49 & $+$36:42:55.74 & 2.1151 $\pm$ 0.0015 & 24.1 $\pm$ 0.8 & 8.0 $\times 10^{26}$ & 28.8 $\pm$ 6.2 & 9.5 $\times 10^{23}$ & 836 $\pm$ 216 & 0.205 & QSO \\
  13:29:22.21 & $+$05:20:14.38 & 2.9943 $\pm$ 0.0004 & 44.5 $\pm$ 1.4 & 3.4 $\times 10^{27}$ & 29.9 $\pm$ 6.4 & 2.3 $\times 10^{24}$ & 1487 $\pm$ 374 & 0.251 & QSO \\
  13:55:04.58 & $+$36:38:02.00 & 2.2817 $\pm$ 0.0006 & 15.9 $\pm$ 0.6 & 6.4 $\times 10^{26}$ & 28.8 $\pm$ 5.6 & 1.1 $\times 10^{24}$ & 555 $\pm$ 133 & 0.184 & BL QSO\\
  14:08:55.02 & $+$55:52:17.96 & 2.5526 $\pm$ 0.0002 & 62.9 $\pm$ 1.9 & 3.3 $\times 10^{27}$ & 26.4 $\pm$ 5.0 & 1.4 $\times 10^{24}$ & 2380 $\pm$ 537 & 0.393 & BL QSO\\
  14:19:18.81 & $+$39:40:35.87 & 0.0196 $\pm$ 0.0000 & 18.5 $\pm$ 1.0 & 1.6 $\times 10^{22}$ & 26.7 $\pm$ 4.4 & 2.3 $\times 10^{19}$ & 695 $\pm$ 159 & 0.500 & SFG\\
  14:29:48.64 & $-$02:59:21.28 & 2.6837 $\pm$ 0.0005 & 18.7 $\pm$ 0.7 & 1.1 $\times 10^{27}$ & 27.4 $\pm$ 5.8 & 1.6 $\times 10^{24}$ & 682 $\pm$ 175 & 0.115 & BL QSO\\
  14:38:21.80 & $+$34:40:00.94 & 2.3452 $\pm$ 0.0010 & 10.7 $\pm$ 0.5 & 4.6 $\times 10^{26}$ & 20.4 $\pm$ 4.3 & 8.7 $\times 10^{23}$ & 526 $\pm$ 141 & 0.324 & QSO \\
  14:52:51.72 & $+$52:39:56.05 & 2.3372 $\pm$ 0.0005 & 126.1 $\pm$ 3.8 & 5.3 $\times 10^{27}$ & 22.7 $\pm$ 4.7 & 9.6 $\times 10^{23}$ & 5551 $\pm$ 1349 & 0.131 & BL QSO\\
  14:55:06.54 & $+$06:40:18.92 & 2.2183 $\pm$ 0.0003 & 51.5 $\pm$ 1.6 & 1.9 $\times 10^{27}$ & 23.0 $\pm$ 5.0 & 8.6 $\times 10^{23}$ & 2237 $\pm$ 574 & 0.081 & BL QSO\\
  15:16:09.85 & $+$22:25:07.80 & 2.7756 $\pm$ 0.0008 & 20.1 $\pm$ 0.7 & 1.3 $\times 10^{27}$ & 23.1 $\pm$ 4.5 & 1.5 $\times 10^{24}$ & 871 $\pm$ 207 & 0.036 & QSO \\
  15:17:03.80 & $+$24:01:27.51 & 2.9306 $\pm$ 0.0005 & 18.1 $\pm$ 0.7 & 1.3 $\times 10^{27}$ & 29.2 $\pm$ 4.4 & 2.1 $\times 10^{24}$ & 620 $\pm$ 120 & 0.260 & BL QSO\\
  15:20:44.37 & $+$27:06:36.38 & 2.7324 $\pm$ 0.0003 & 46.8 $\pm$ 1.5 & 2.9 $\times 10^{27}$ & 25.9 $\pm$ 4.7 & 1.6 $\times 10^{24}$ & 1803 $\pm$ 393 & 0.218 & BL QSO\\
  15:33:17.30 & $+$12:18:00.89 & 2.7974 $\pm$ 0.0016 & 23.3 $\pm$ 0.8 & 1.5 $\times 10^{27}$ & 25.0 $\pm$ 4.3 & 1.6 $\times 10^{24}$ & 932 $\pm$ 197 & 0.261 & BL QSO\\
  15:38:26.90 & $+$14:55:05.29 & 2.6189 $\pm$ 0.0002 & 15.1 $\pm$ 0.6 & 8.4 $\times 10^{26}$ & 27.1 $\pm$ 5.2 & 1.5 $\times 10^{24}$ & 558 $\pm$ 134 & 1.428 & BL QSO\\
  15:43:14.72 & $+$32:51:38.19 & 2.2652 $\pm$ 0.0003 & 50.3 $\pm$ 1.6 & 2.0 $\times 10^{27}$ & 20.2 $\pm$ 4.2 & 7.9 $\times 10^{23}$ & 2495 $\pm$ 617 & 0.759 & BL QSO\\
  17:26:16.51 & $+$32:16:20.01 & 2.6405 $\pm$ 0.0008 & 28.4 $\pm$ 0.9 & 1.6 $\times 10^{27}$ & 24.2 $\pm$ 4.5 & 1.4 $\times 10^{24}$ & 1175 $\pm$ 267 & 0.203 & BL QSO\\
\hline
\label{allZs}
\end{tabular}
\end{table*}

Table~\ref{allZs} lists the 19 sources which have spectroscopic redshifts in SDSS DR9. Their spectra and postage stamps are shown in Fig.~\ref{Spectra} in Appendix A. 18 of these sources are identified as quasars in the range $2 < z < 3$, 14 of which contain broad emission lines. Their strong radio emission means that they are necessarily radio-loud quasars. 

One source is identified as a SFG at a redshift of $z \approx 0.02$, which clearly hosts an AGN, because of its high flux density ratio S$_{\rm 20cm} / $S$_{\rm 3.4\mu m}$ and its GPS spectrum. This source is probably either a misidentification, or a composite galaxy with a radio-loud AGN embedded within a SFG, similar to F00183-7112 \citep{2012MNRAS.422.1453N}. 

\begin{figure}
\begin{center}
\includegraphics[scale=0.55]{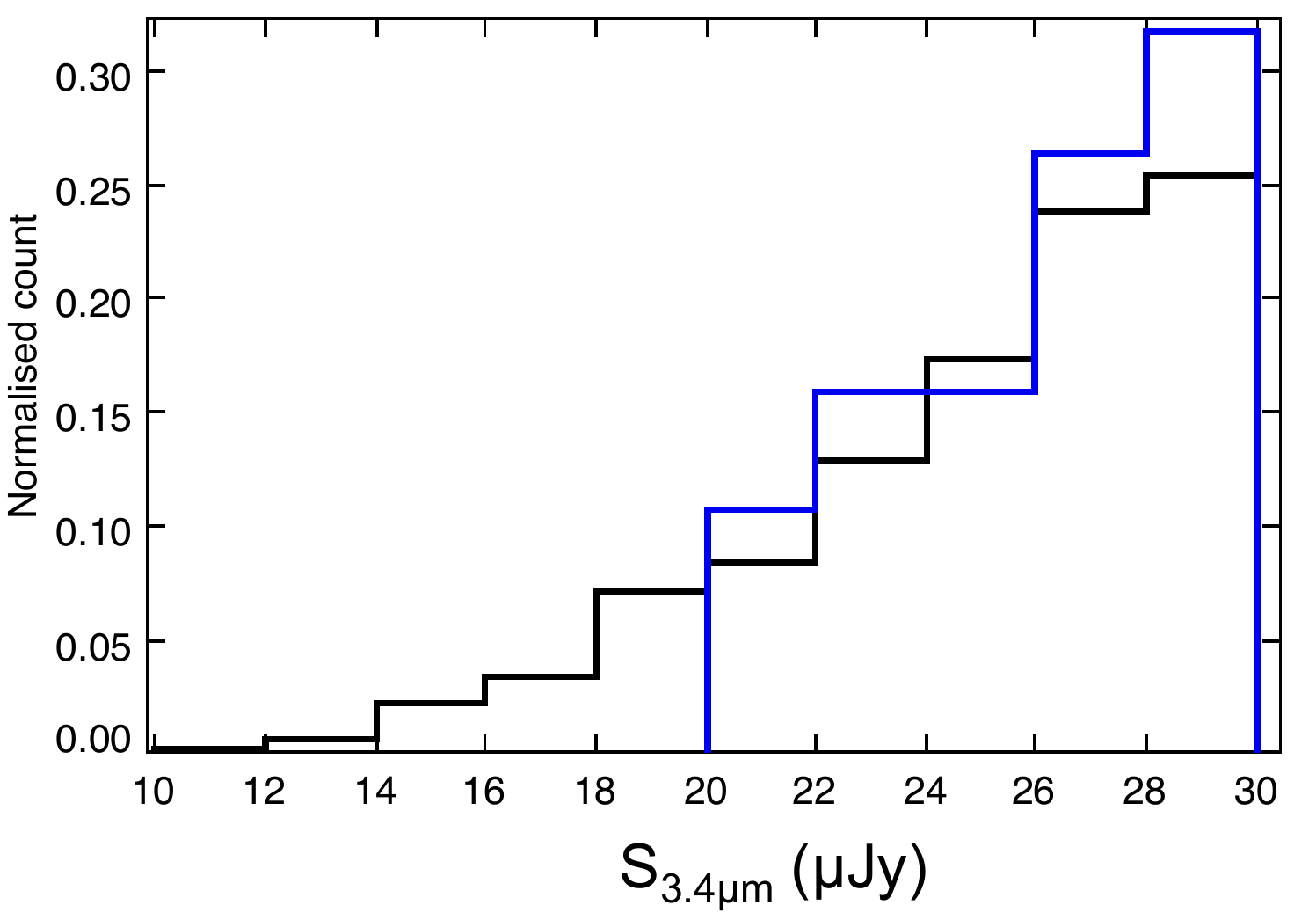}
\includegraphics[scale=0.65]{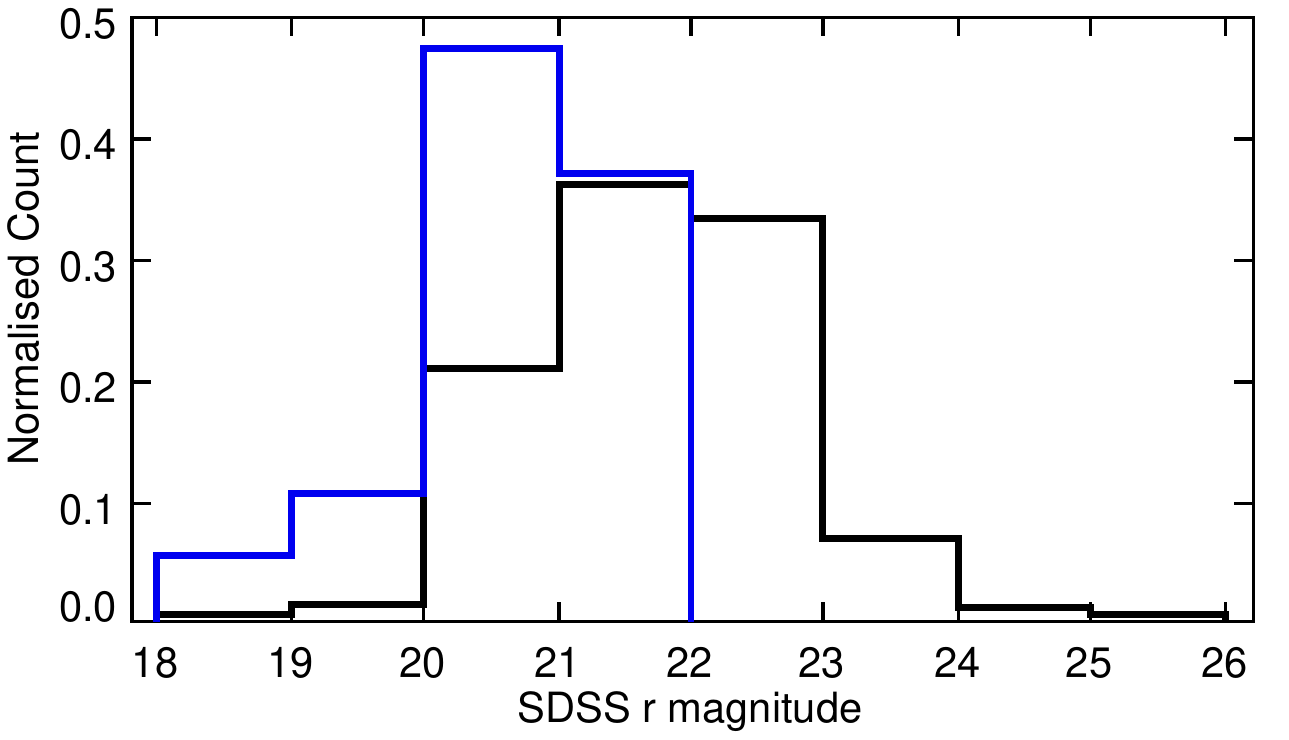}
\caption{Normalised histogram of the 3.4 $\mu$m flux density (top) and the SDSS {\it r} magnitudes (bottom) for the 19 IFRSs with spectroscopic redshifts (blue) and the IFRSs without redshifts (black). It can be seen that the sources in blue are more representative of the bright end of the distribution at both wavelengths. The distribution of {\it r} magnitudes becomes highly incomplete above the SDSS limiting magnitude of 22.2.}
\label{fluxComp}
\end{center}
\end{figure}

Because these sources with spectroscopy are much brighter than first generation IFRSs, they probably represent the closer and brighter tail end of the IFRS population. This is also suggested by the distribution in 3.4$\,\mu$m flux density for these sources (Fig.~\ref{fluxComp}), which ranges from $20-30\,\mu$Jy and has a higher median of 26.7$\,\mu$Jy, as compared to the rest of the sample, which ranges from $10-30\,\mu$Jy and has a median of 25.9$\,\mu$Jy. Fig.~\ref{fluxComp} demonstrates that the sources with spectroscopic redshifts are taken from the upper half of the infrared and optical brightness distribution, implying that the fainter half is located at higher redshift. Despite being brighter in the optical and infrared, these 18 sources are still classified as IFRSs and share the same region in ${\rm S_{20cm} / S_{3.4\mu m}}$ space as HzRGs (Fig.~\ref{Olay1}).

Fig.~\ref{Olay2} shows where our {\it WISE} IFRSs with spectroscopy appear in the diagram adapted from \cite{Norris2011}. Our sources are very densely concentrated around $S_{\rm 3.4\mu m} \approx 30 \mu$Jy in this figure and do not have sufficient range in IR flux density to show the observed $S_{\rm 3.4\mu m} - z$ relation, although they are consistent with it. From this we conclude that distance, rather than dust obscuration, is likely to be the main mechanism for the infrared-faintness of our sample of IFRSs. Furthermore, Fig.~\ref{Olay1} and \ref{Olay2} confirm our hypothesis that at least a fraction of IFRSs are indeed high-redshift radio-loud AGN.

\subsection{X-ray data}

\citet{2009MNRAS.396.2011E} identified an X-ray counterpart in the {\it Chandra} XBo\"otes survey for one of the IFRSs from our sample (IFRS ID 545 in full catalogue). The source respectively has a soft (0.5$-$2 keV), hard (2$-$7 keV) and full-band (0.5$-$7 keV) X-ray flux of 0.44, 0.46 and 0.99~$\times 10^{-15}$ erg s$^{-1}$ m $^{-2}$. This gives a hardness ratio (HR)\footnote{The HR is defined as HR $= h - s/h + s$, where $h$ and $s$ are respectively the number of counts detected in the $2-7$ keV and $0.5-2$ keV bands.} of $-0.60^{+0.23}_{-0.29}$, from which \citet{2009MNRAS.396.2011E} concludes that the source is a type 1 (unobscured) AGN. The IFRS has an X-ray to optical flux ratio of $\log(f_{\rm X}/f_{\rm opt}) = -0.29$, which \citet{2009MNRAS.396.2011E} states is well within the expected locus for a typical AGN (0 $\pm$ 1), rather than that for a SFG or a low luminosity AGN, which typically have $\log(f_{\rm X}/f_{\rm opt}) \le -1$. The source is fitted with a photometric redshift of $z = 0.605 ^{+0.727}_{-0.499}$. However, this is an incorrect photometric redshift, since we find an SDSS source 0.16 arcsec away with a spectroscopic redshift of $z = 2.3452 \pm 0.0010$ (row 10 in Table~\ref{allZs}). At this distance, the source has a full-band X-ray luminosity of $L_{\rm X} = 4.22 \times 10^{36}$ W. 

Further cross-matches from the {\it Chandra} and {\it XMM-Newton} X-ray Telescopes were searched for within a 5\arcsec~radius of the FIRST radio positions for our entire sample using the {\it Chandra} Data Archive \citep{2000ASPC..216..184R}\footnote{\url{http://cda.harvard.edu/chaser/}} and the {\it XMM-Newton} Science Archive \citep[XSA;][]{2002astro.ph..6412A}\footnote{\url{http://xmm.esac.esa.int/xsa/index.shtml}}. No additional matches were found.

The {\it ROSAT} All-Sky Survey (RASS) bright and faint source catalogues \citep{1999yCat.9010....0V,2000yCat.9029....0V} were searched for matches within 10\arcsec~of the FIRST radio positions, and only one match from the faint source catalogue was returned, which had RXS designation J144102.9+534040.

\section{Discussion}

\subsection{How our sample relates to the original IFRSs}

Our sample consists of much brighter IFRSs than the original first generation IFRSs discovered in ATLAS, since the radio and infrared flux densities are much larger. We suggest that this is due to them being lower-redshift counterparts of the first generation IFRSs. Both generations of IFRSs have ratios of $S_{\rm 20cm} / S_{\rm 3.4\mu m} > 500$ and flux densities $S_{\rm 3.4\mu m} < 30\,\mu$Jy, so they are likely to be from the same parent population. Our IFRSs are brighter, and have lower $S_{\rm 20cm} / S_{\rm 3.4\mu m}$ ratios. Our sample has a lower sky density than the first generation IFRSs, which is consistent with them being brighter (and probably closer) versions of the same object.

We find a higher fraction of resolved IFRSs compared to \citet{Middelberg2011}, who found no resolved IFRSs. AGN may appear compact in the radio either because (1) their jets are small, suggesting they are quite young \citep{1998PASP..110..493O}, or (2) they are oriented with their lobes pointed along the line-of-sight of the observer. We find a number of GPS and CSS sources, which is consistent with the first scenario. This suggests that the more compact first generation IFRSs may be more representative of younger radio galaxies than our IFRSs, possibly located at higher redshift in the younger Universe.

The optical magnitudes of our sample are also consistent with this interpretation. \cite{Norris2011} found a median 3.6 $\mu$m flux density of $\sim$0.2 $\mu$Jy, while \cite{2008MNRAS.391.1000G} found an upper limit on the median 3.6 $\mu$m flux density of $\sim$3.1$\mu$Jy, which is approximately 10 times brighter. Our median 3.4 $\mu$m flux density is $\sim$26 $\mu$Jy, approximately 10 times brighter still. If this difference is simply due to cosmic distance, rather than obscuration, we expect the optical brightness to scale in the same way. If the \cite{2010ApJ...710..698H} optical magnitudes are representative of the population as a whole, first generation (non-detected) IFRSs have typical optical magnitudes of about $z{\rm_{AB}} = 26$, while the brighter \cite{2008MNRAS.391.1000G} IFRSs have typical magnitudes of $R_{\rm AB} = 24.4$. Our median SDSS magnitude of $z{\rm_{AB}} = 21$ is 100 times brighter than those from \cite{2010ApJ...710..698H}, the same factor brighter as for the infrared emission. This suggests that obscuration is not the dominant effect in reducing the infrared flux density for these IFRSs.

Furthermore, the spectroscopic redshifts from our sample suggest that the brighter IFRSs are found at $2 < z < 3$ and form a continuous population with the faintest IFRSs which are believed to be at $z > 3$. The distribution of redshifts we find is consistent with the results from \cite{2010ApJ...710..698H}, who found that IR-detected IFRSs cannot be explained easily at $z < 2$, but can be modelled at $z \ge 2$. \cite{2010ApJ...710..698H} find that the radio-loud quasar 3C 273 at $z \ge 2$ accurately models the SEDs of the IFRSs from their sample. Similarly, all our IFRSs in Table~\ref{allZs} are radio-loud quasars. 

\citet{Herzog2013} present Very Large Telescope spectra of infrared-detected IFRSs which satisfy the \citet{Zinn2011} criteria, and are similarly brighter IFRSs. The spectra reveal that their IFRSs are located at 2 $\lesssim z \lesssim 3$, have IR and radio luminosities in the range from which HzRGs are selected, and follow their ${\rm S_{3.6\mu m}} - z$ relation, in agreement with our findings for brighter IFRSs.

The place occupied by our IFRSs in ${\rm S_{20cm} / S_{3.4\mu m}}$ space (Fig.~\ref{Olay1}) suggests that HzRGs, first generation IFRSs and {\it WISE} IFRSs are all from the same parent population of radio galaxies.

\subsection{Are IFRSs misidentifications?}

{\it WISE} sources are dominated by low-redshift, low-luminosity objects at $z < 1$ \citep{2013AJ....145...55Y}. So objects mistakenly identified with {\it WISE} sources should have low redshifts, whereas all of our objects with spectroscopy except one have $z > 2$. This result cannot therefore be attributed to misidentifications.
Furthermore, our misidentification rate is estimated to be 0.24 $\pm$ 0.27 per cent, so $\sim$3 of our sources are false-positives. 

\subsection{Are IFRSs hotspots or lobes?}

The VLBI detections from \citet{NorrisVLBI} and \citet{MiddelbergVLBI} suggest that at least a third of IFRSs are not radio lobes. Additionally, the majority of IFRSs are unresolved at high resolution. \cite{Middelberg2011} find the vast majority of sources are unresolved on scales of $\sim$2 arcsec. From our sample, 845 (64 per cent) of our sources are unresolved at the 5 arcsec FIRST resolution, which puts an upper limit of $\sim30$ kpc on their projected linear size at $z > 0.5$.

\citet{2012ApJ...759...86W} found {\it Spitzer} observations of hotspots in the radio lobes of FR II galaxies, ranging in IR flux density from $< 1\,\mu$Jy up to $\sim$70$\,\mu$Jy. It is therefore possible that our sample consists of a number of hotspots. However, 213 of our sources are identified as double-lobed galaxies which have hotspots that are not coincident with the corresponding faint IR source. Additionally, careful inspection was carried out to ensure that the faint {\it WISE} sources did not coincide with an identifiable lobe or hotspot. Furthermore, if they were hotspots, the redshifts shown in Table~\ref{allZs} would reflect the overall distribution of the radio galaxy population, very few of which are at $z > 2$.
We therefore conclude that they are not hotspots.

\subsection{Are IFRSs nearby AGN?}

We find 18 reliable spectroscopic redshifts from our sample at $z \ge 2$. This shows that the brightest members of our sample are not nearby AGN. Additionally, we find steep radio spectral indices for most of our sample, as well as many USS sources. This shows that our IFRSs are more likely to be taken from a higher redshift population, particularly in the case of the USS sources. However, we cannot rule out the possibility that some small fraction of our sample is located at low redshift and is suffering from significant dust extinction. 

\subsection{The nature of IFRSs}

In \S~\ref{ratios}, we showed that most of our IFRSs are not SFGs. We have also ruled out the hypotheses that their majority is made up of misidentifications, hotspots, lobes or nearby AGN. The radio spectra, {\it WISE} colours and many other properties of our sample are consistent with the IFRSs being high-redshift radio-loud AGN. Above all, we have shown that the brightest IFRSs have spectroscopic redshifts $> 2$, and so we conclude that the IFRSs are most likely high-$z$ radio-loud AGN. 

Previous studies of IFRSs have been almost entirely unsuccessful in detecting their emission in the infrared and optical bands, resulting in speculations about their nature based only on their radio detections and upper limits in the infrared and the optical. We have shown that there exists a significant population of IFRSs that can be detected in the infrared and optical, which show similar properties to these first generation IFRSs. IFRSs seem to span a continuous population of high-redshift AGN which are from the same parent population of AGN from which HzRGs and their lower redshift versions come.

If IFRSs are high-redshift AGN and follow the relation between redshift and 3.4 $\mu$m flux density, where those with lower infrared flux densities are found at even higher redshifts, then we have found a very effective way to find HzRGs using their 3.4$\,\mu$m emission, a technique parallel to using the K-z diagram  \citep{2003MNRAS.339..173W}. Therefore {\it WISE} provides a great new all-sky method to find many HzRGs, which will be valuable in studying cosmic AGN evolution.

%**********************************************************************************************************************************************************
%**********************************************************************************************************************************************************

\section{Conclusions}

We have compiled the first detectable sample of IFRSs, consisting of 1317 sources generated by cross-correlating the NVSS, FIRST and {\it WISE} surveys. Below we summarise and discuss our results.

\begin{itemize}
\item Our sample is brighter and has a higher sky density than first generation IFRSs, while retaining the same values of distance-independent measures such ${\rm S_{20cm} / S_{3.4\mu m}}$. This suggests that our sample consists of a lower-redshift, brighter population of IFRSs.
\item The 403 sources that have reliable measurements or upper limits of the 22 $\mu$m flux density all have q$_{22} < -0.23$, which is strong evidence that they are AGN, rather than SFGs. 
\item The {\it WISE} colours of our sources significantly detected at 3.4, 4.6 and 12$\,\mu$m are similar to those of obscured AGN, QSOs and Seyferts. 
\item 41 IFRSs are polarized at levels $\ge 8\sigma_{\rm QU}$, with fractional polarizations ranging between 1 $ < \Pi <$ 14 per cent, and RMs ranging from $-45$ to 54 rad m$^{-2}$.
\item There are 213 sources that contain double-lobed radio morphologies.
The majority of our sources are unresolved, suggesting that most of them are either beamed radio sources or young radio sources with small jets.
\item We derive radio spectra for a significant fraction of our sample and find that the majority have steep spectral indices, and many have USS, suggesting that they are more likely to be located at high redshift than low redshift. We also find 32 GPS sources and 124 CSS sources, which is consistent with some of our IFRSs being young radio sources with small jets. 
\item Amongst our brightest IFRSs, we find 18 spectroscopic redshifts from SDSS DR9, which reveal that these are quasars located at 2 $< z <$ 3, giving strong evidence that our IFRSs are high-redshift radio-loud AGN.
\item One of these sources at $z = 2.3452$ has an X-ray detection, at which redshift it has a luminosity of $L_{\rm X} = 4.22 \times 10^{36}$ W.
The X-ray hardness ratio suggests that the source is an unobscured AGN.
\item The properties of our large sample of IFRSs imply that they are radio-loud AGN at $z \ge 2$.
We suggest that the fainter IFRSs that are undetectable in the infrared represent a population of radio sources at even higher redshift. 
\item Searching for IFRSs is an effective all-sky method for finding HzRGs and investigating cosmic AGN evolution.
\end{itemize}

\section*{Acknowledgments}
This research has made use of the NASA/IPAC Infrared Science Archive and the NASA/IPAC Extragalactic Database (NED) which are operated by the Jet Propulsion Laboratory, California Institute of Technology, under contract with the National Aeronautics and Space Administration. We gratefully acknowledge the people and institutes that contributed to the NVSS and FIRST surveys. This publication makes use of data products from the {\it Wide-field Infrared Survey Explorer}, which is a joint project of the University of California, Los Angeles, and the Jet Propulsion Laboratory/California Institute of Technology, funded by the National Aeronautics and Space Administration. Funding for SDSS-III has been provided by the Alfred P. Sloan Foundation, the Participating Institutions, the National Science Foundation, and the U.S. Department of Energy Office of Science. The SDSS-III web site is http://www.sdss3.org/. The National Radio Astronomy Observatory is a facility of the National Science Foundation operated under cooperative agreement by Associated Universities, Inc.

%**********************************************************************************************************************************************************

%**********************************************************************************************************************************************************

%**********************************************************************************************************************************************************

%**********************************************************************************************************************************************************

\bibliographystyle{mn2e}
\bibliography{references}

\section*{Appendix A}

\begin{figure*}
\includegraphics[scale=0.195,trim=0mm 0mm 30mm 10mm]{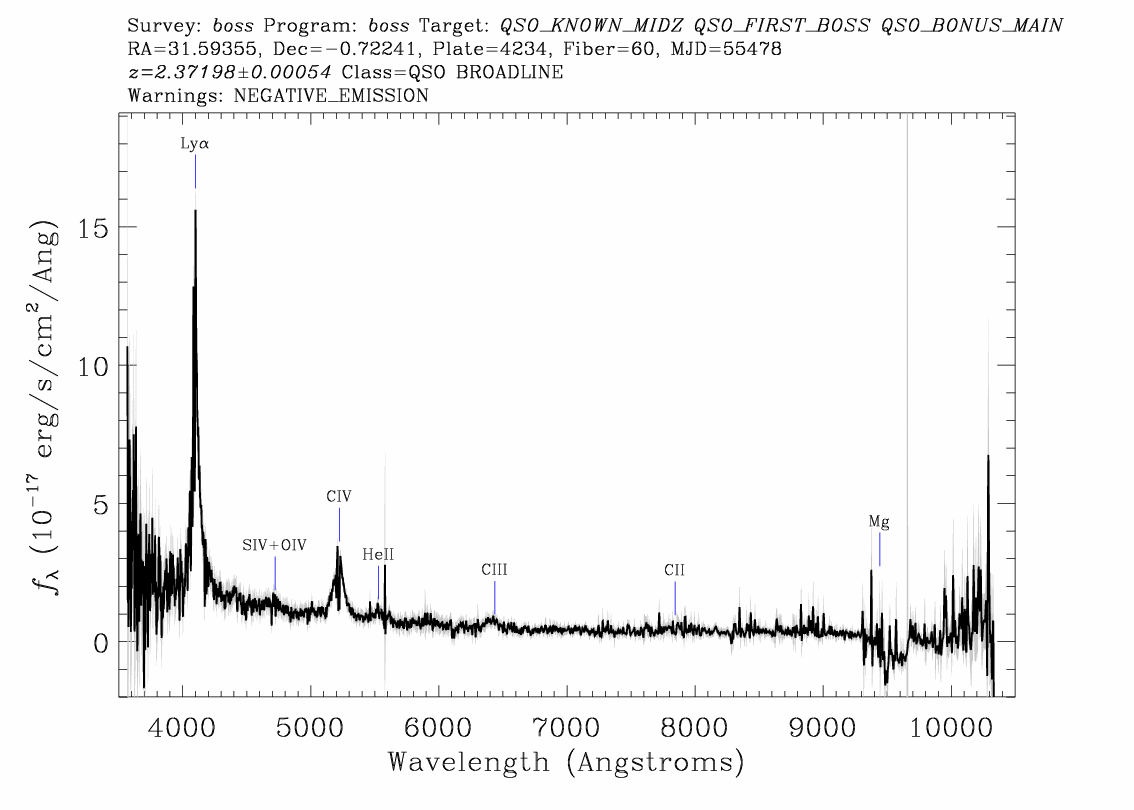}
\includegraphics[scale=0.295]{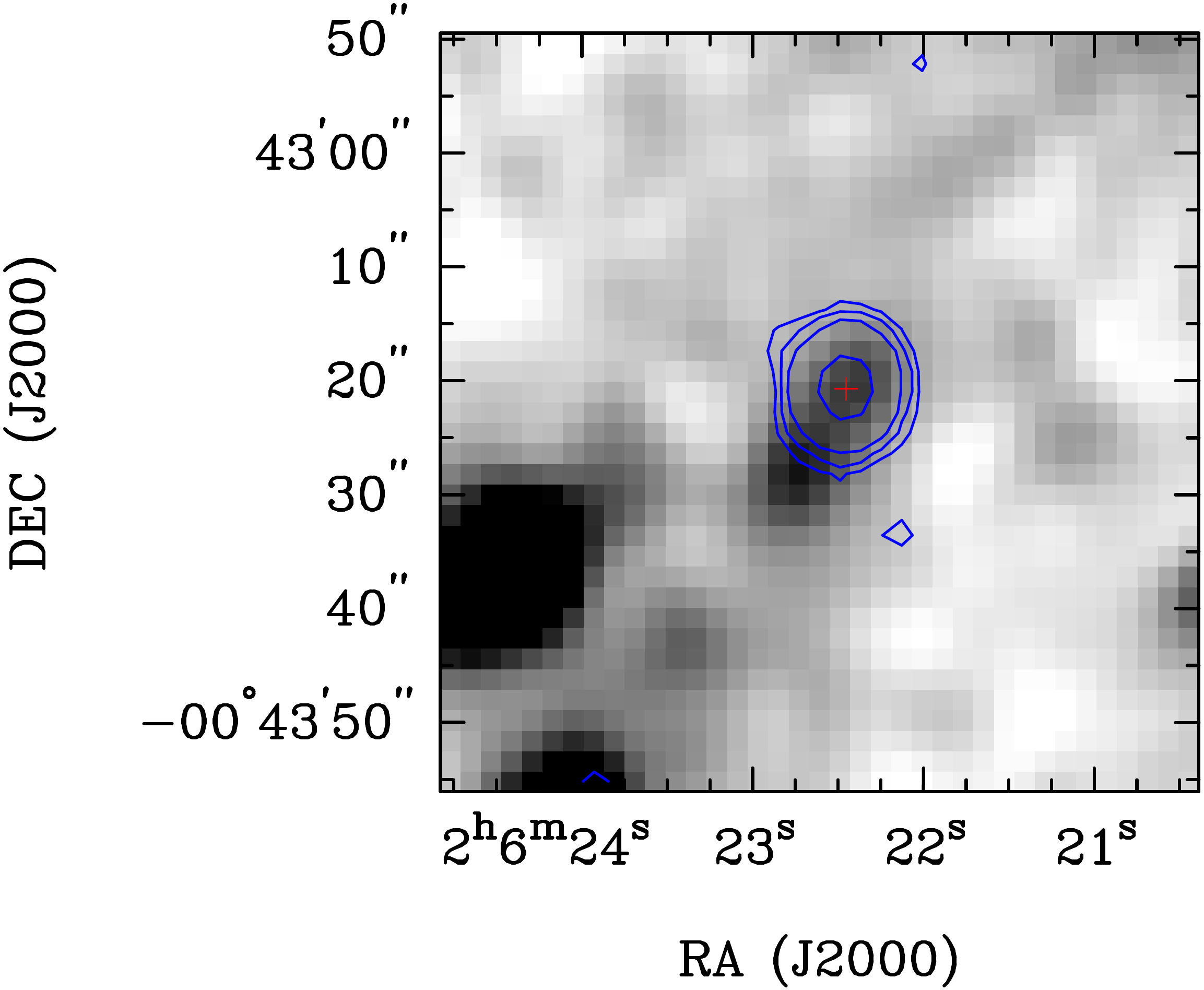}
\includegraphics[scale=0.195,trim=0mm 0mm 30mm 10mm]{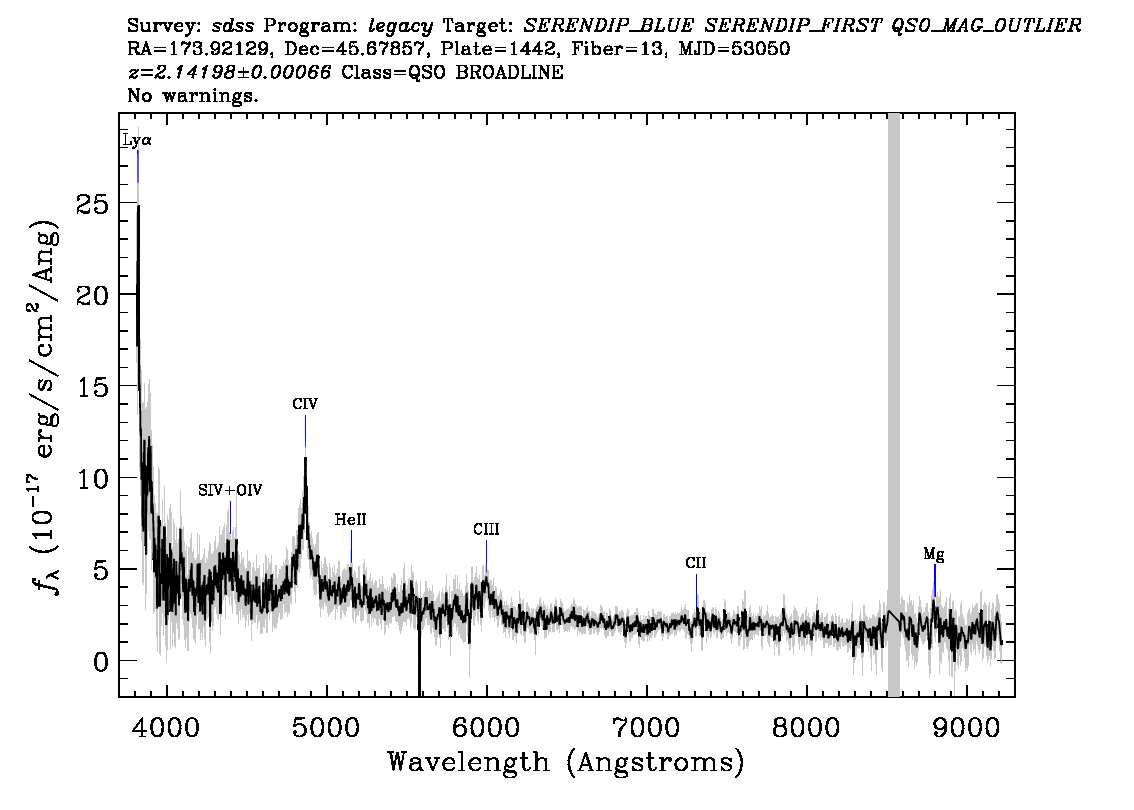}
\includegraphics[scale=0.295]{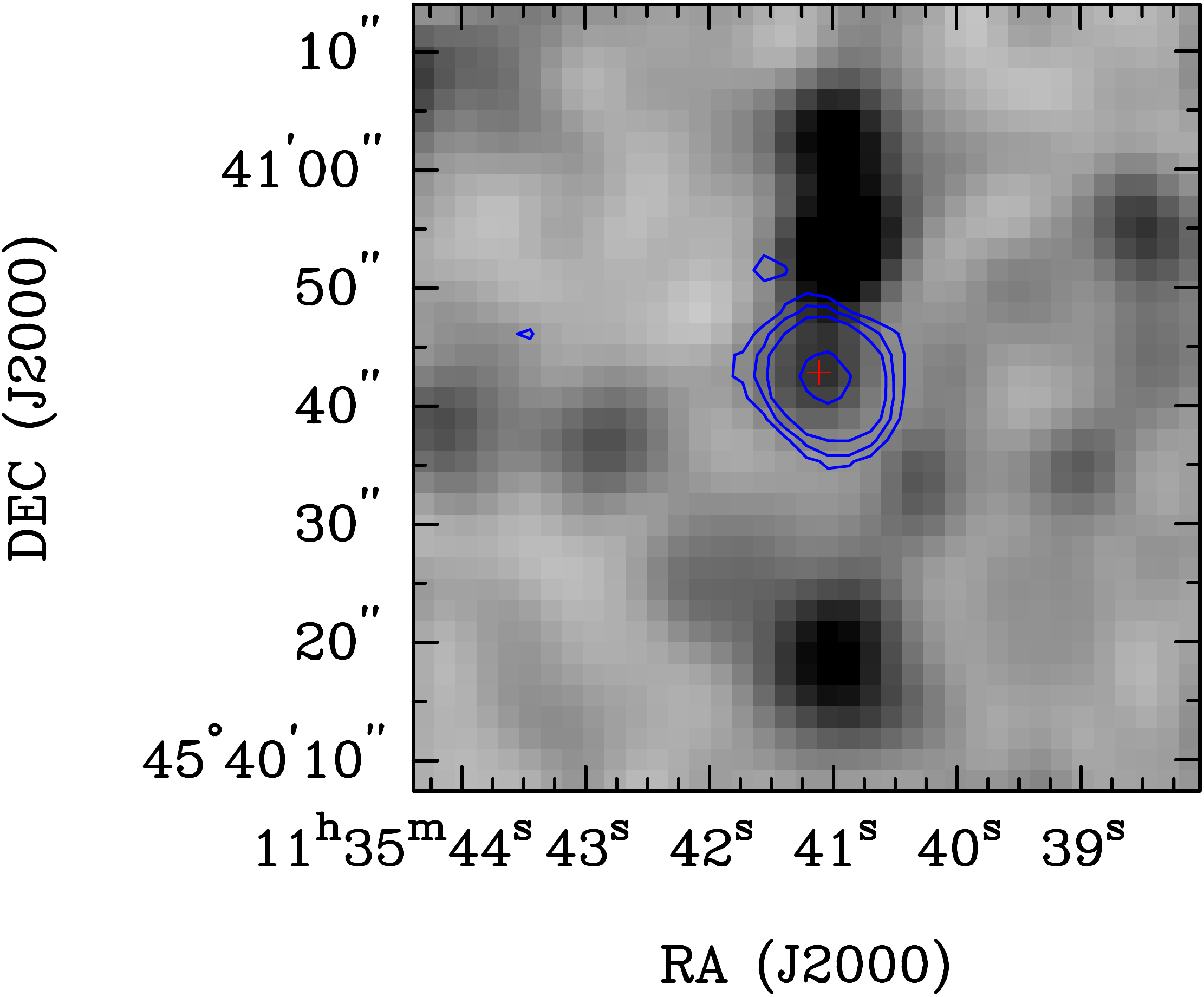}
\includegraphics[scale=0.195,trim=0mm 0mm 30mm 10mm]{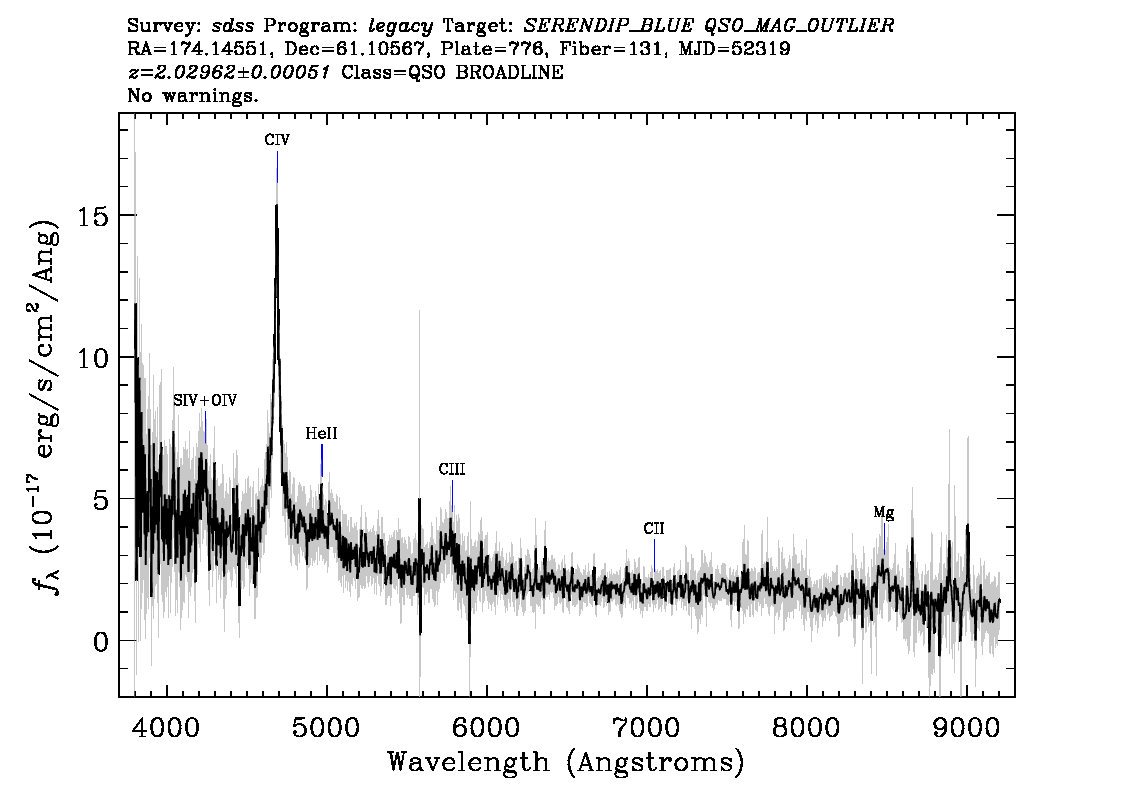}
\includegraphics[scale=0.295]{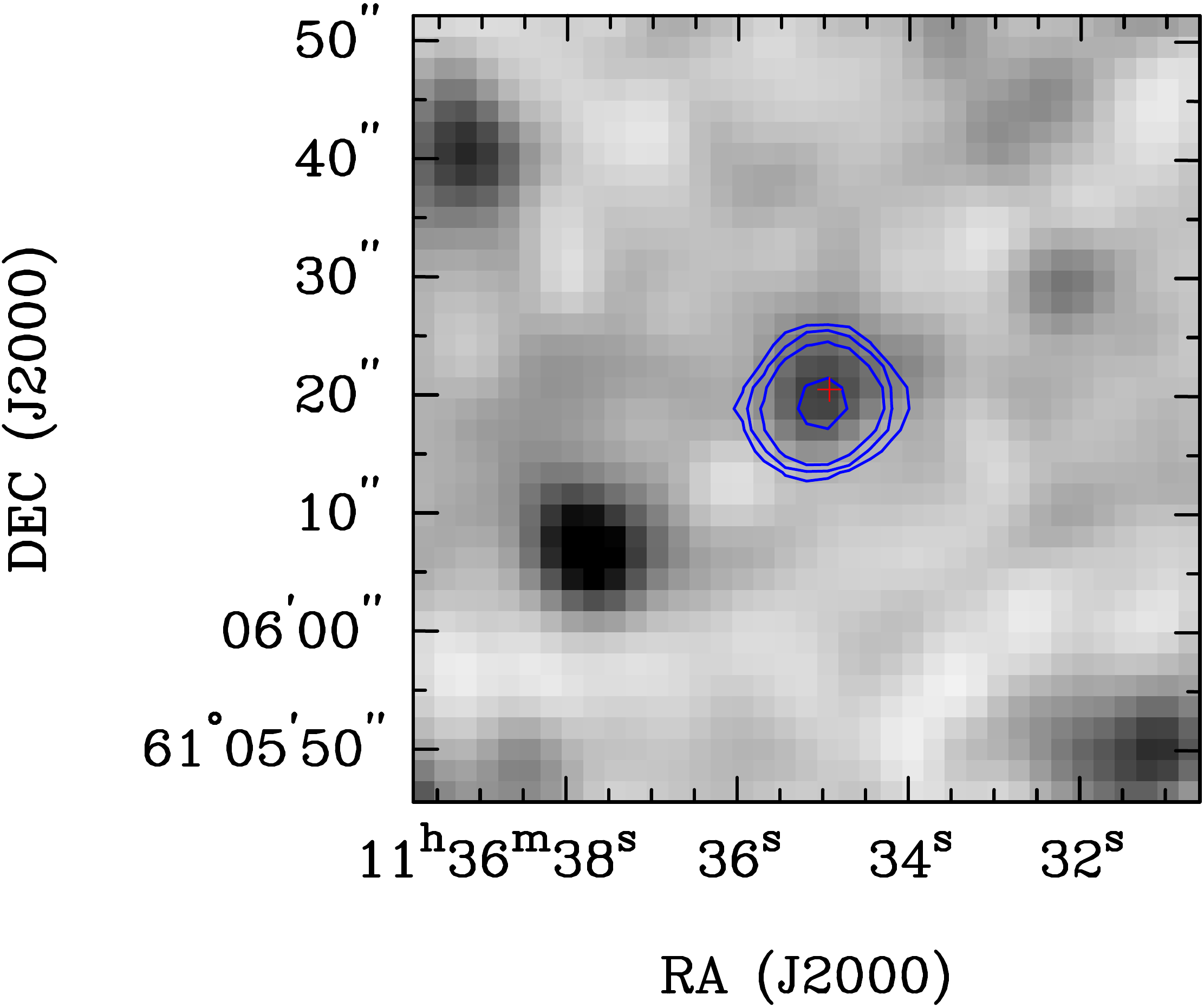}
\includegraphics[scale=0.195,trim=0mm 0mm 30mm 10mm]{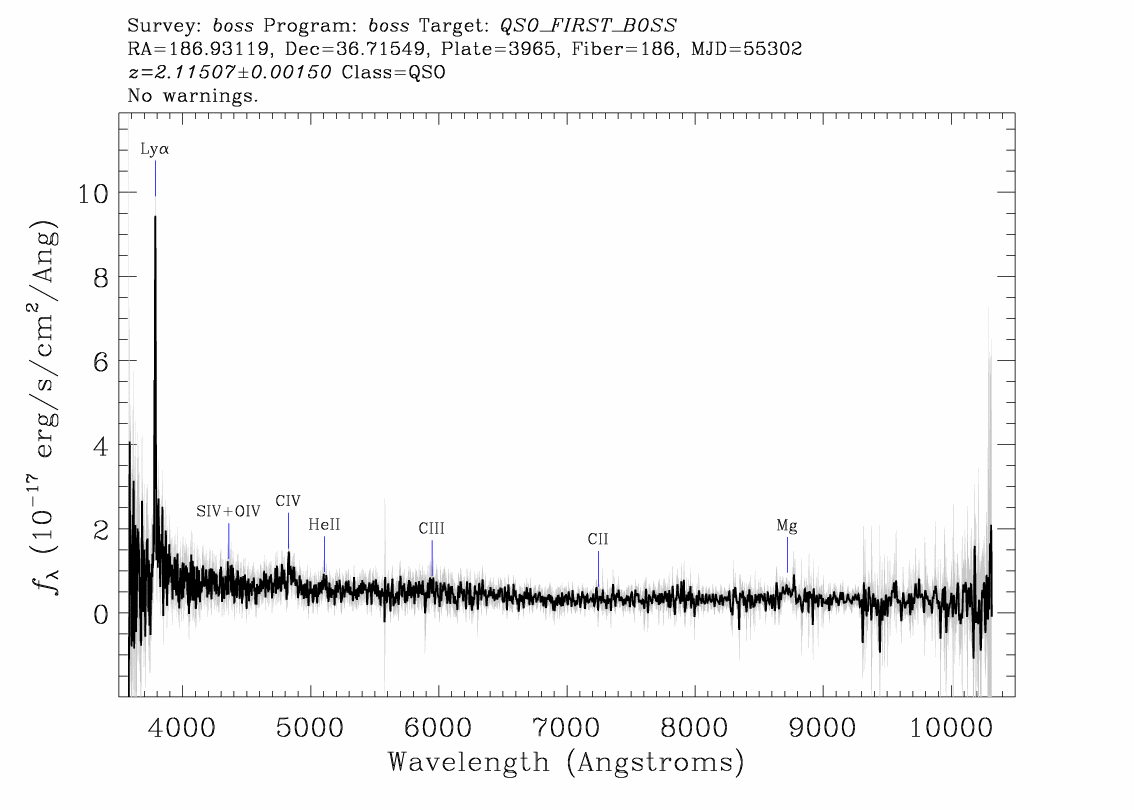}
\includegraphics[scale=0.295]{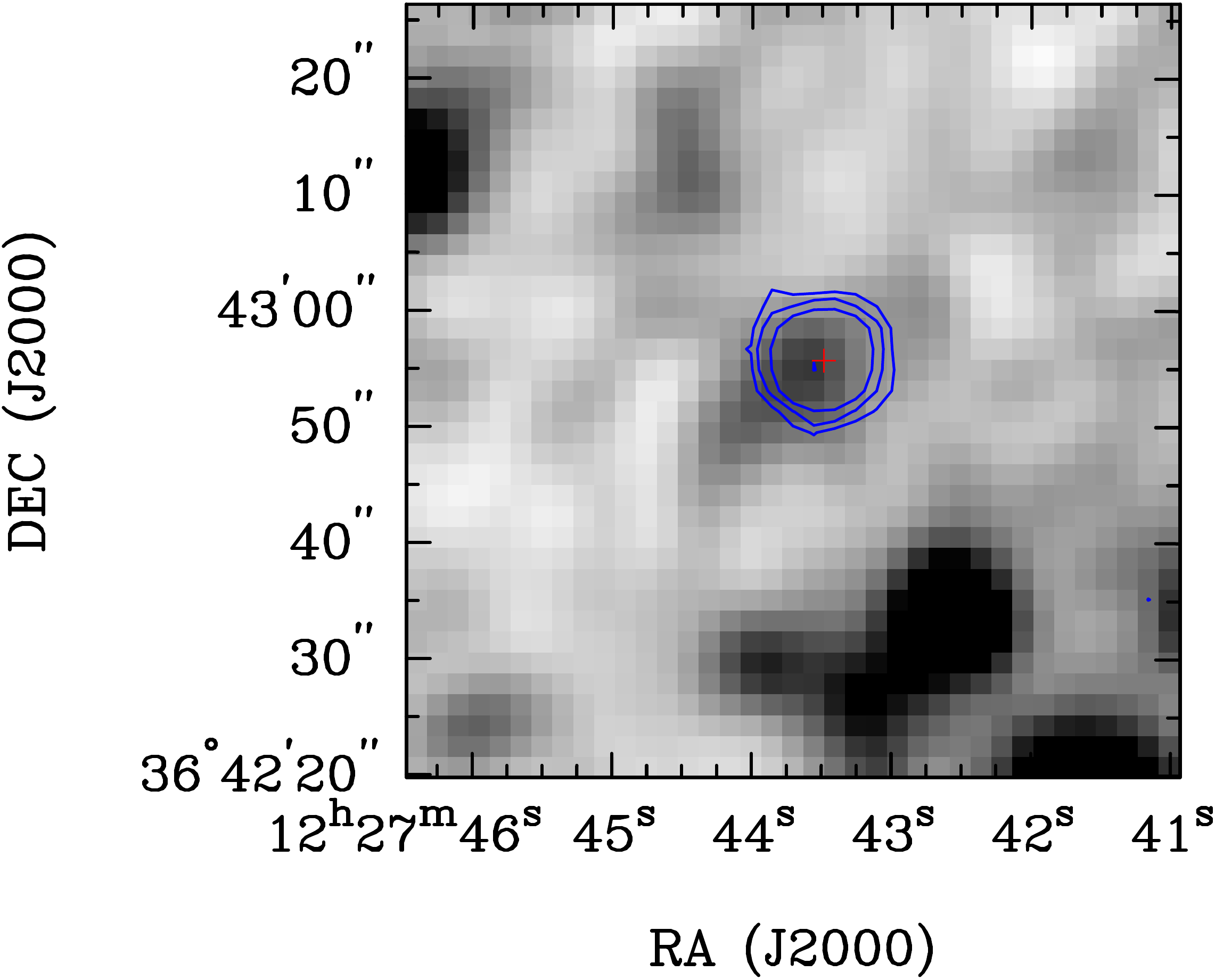}
\caption{Spectra and postage stamps of the 19 IFRSs with spectroscopy in SDSS DR9, appearing in the same order as in Table~\ref{allZs}. The background image shows the {\it WISE} $3.4\,\mu$m detection, the red cross represents the position of the SDSS spectroscopy, and the contours mark the FIRST source at 20$\,$cm, at levels of 3, 6, 12, 64, 256 and 1024 times the local noise level.}
\label{Spectra}
\end{figure*}

\begin{figure*}
\includegraphics[scale=0.215,trim=0mm 0mm 30mm 10mm]{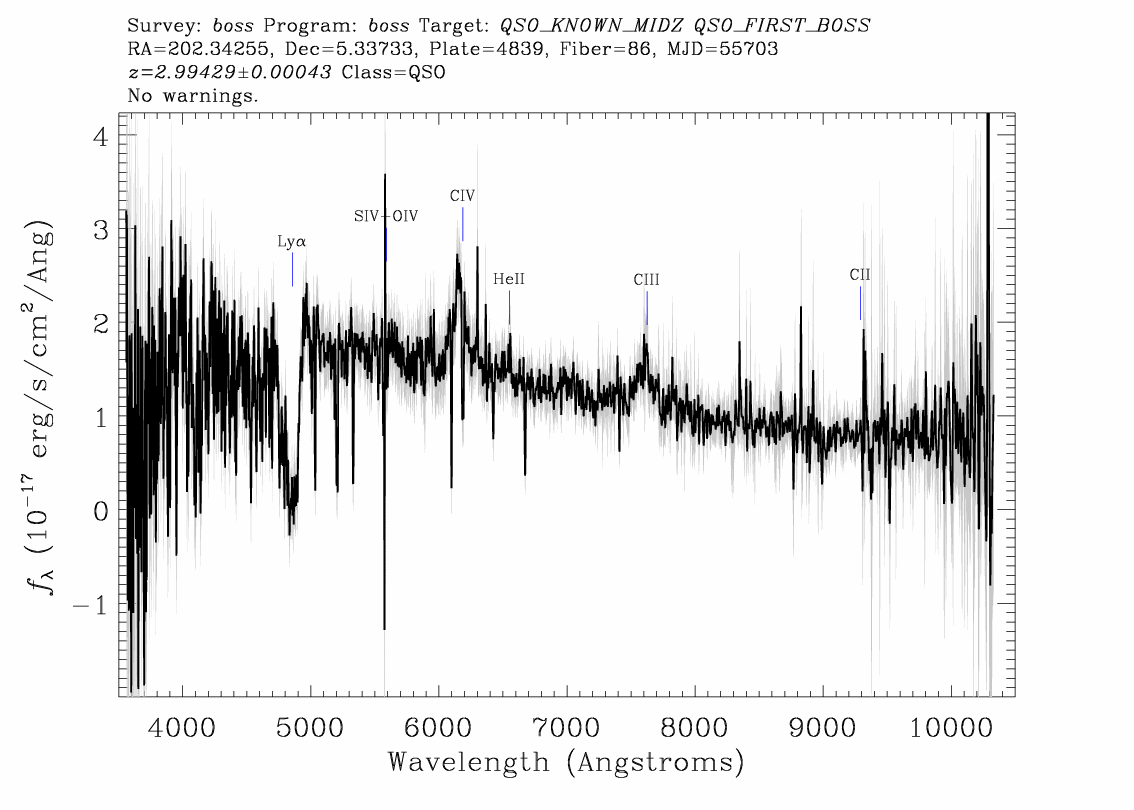}
\includegraphics[scale=0.3]{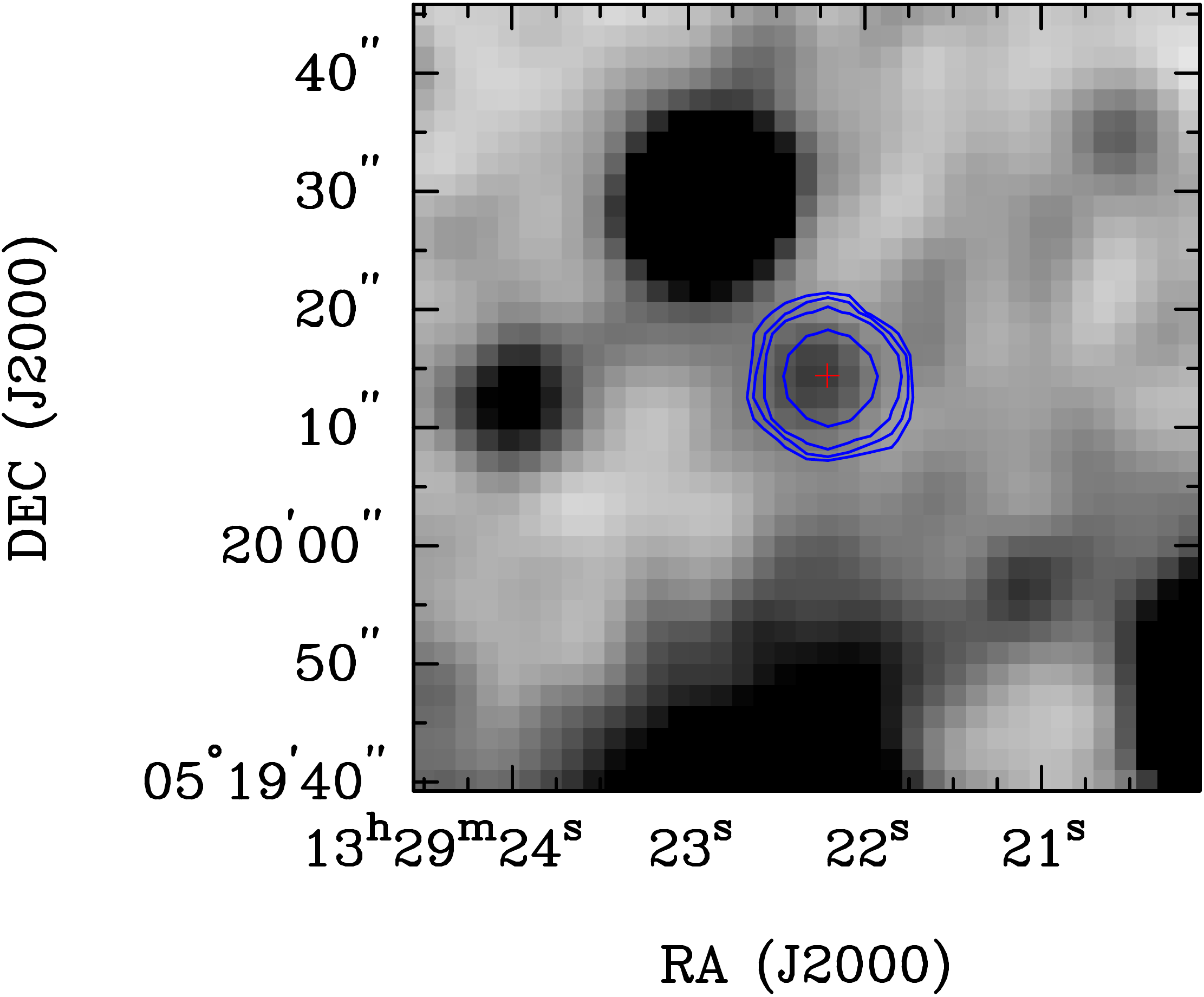}
\includegraphics[scale=0.215,trim=0mm 0mm 30mm 10mm]{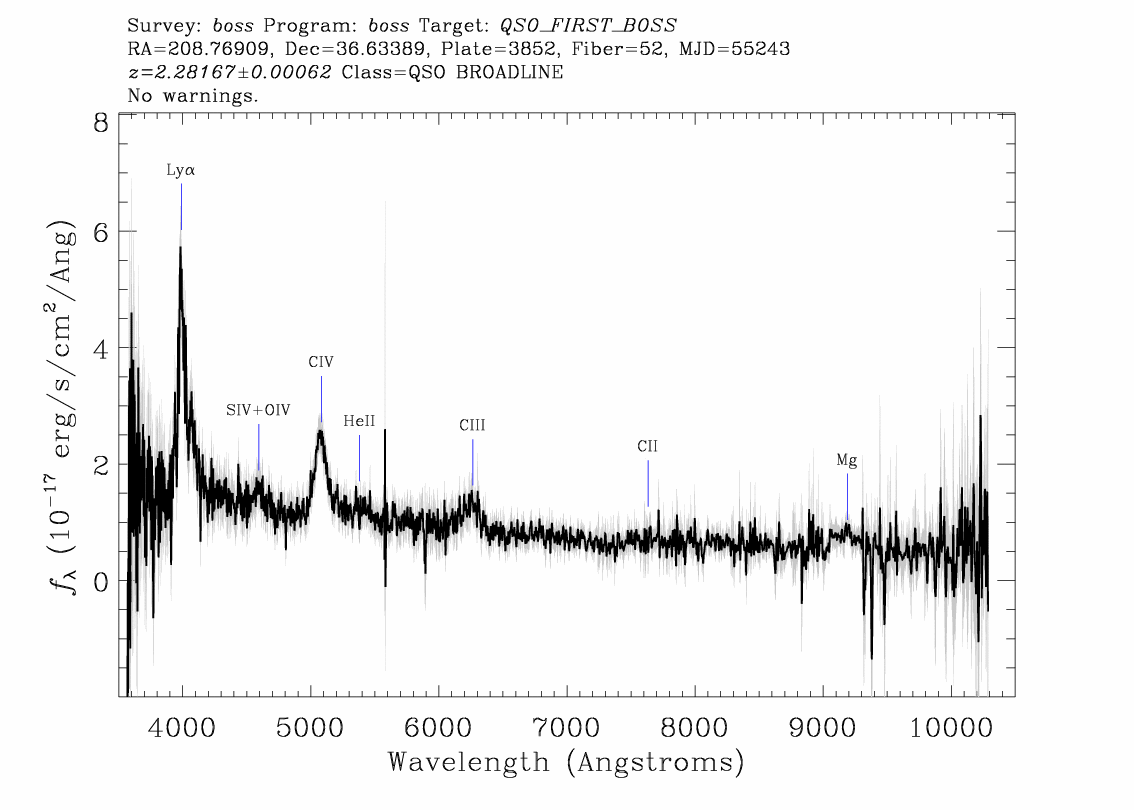}
\includegraphics[scale=0.3]{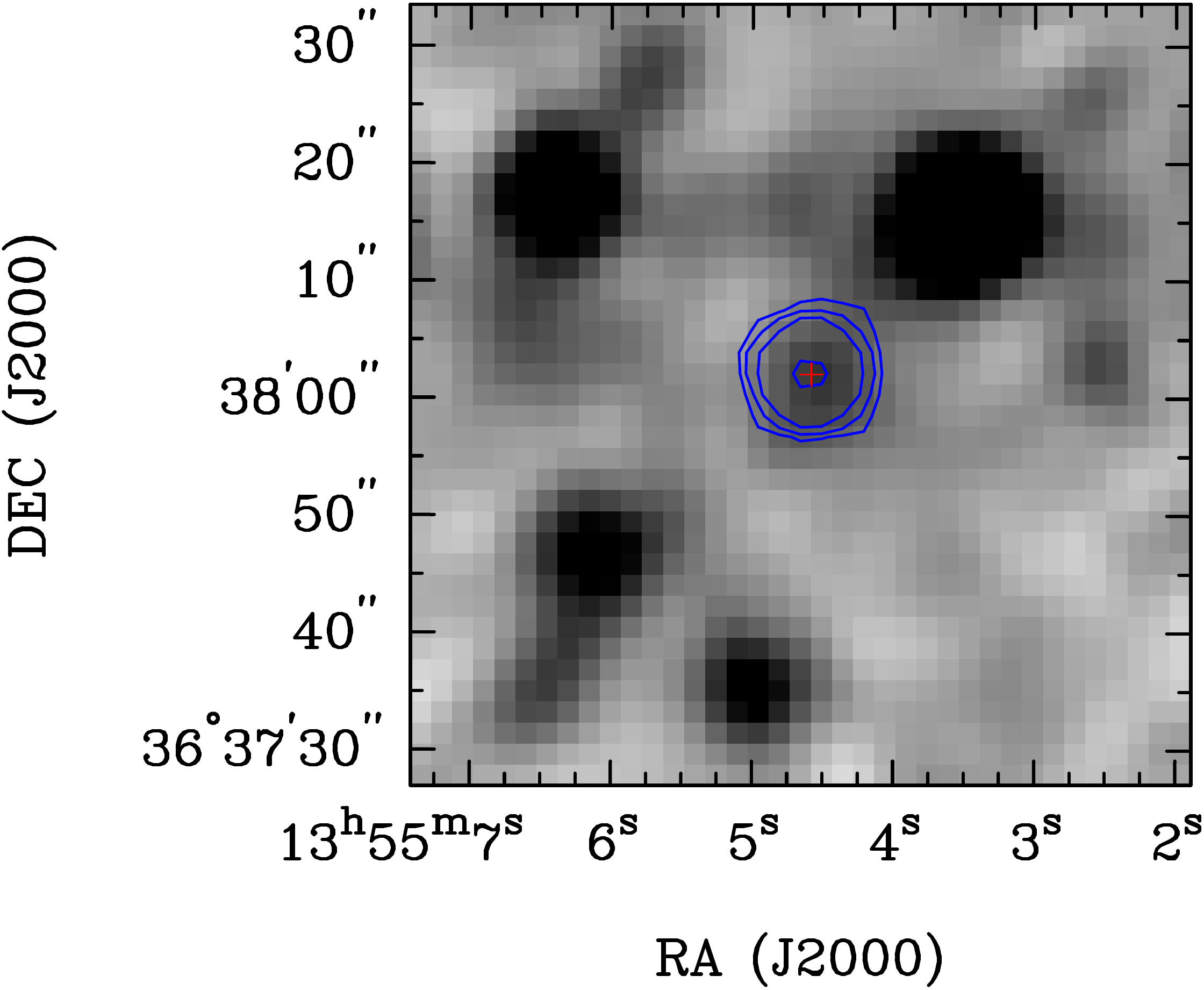}
\includegraphics[scale=0.215,trim=0mm 0mm 30mm 10mm]{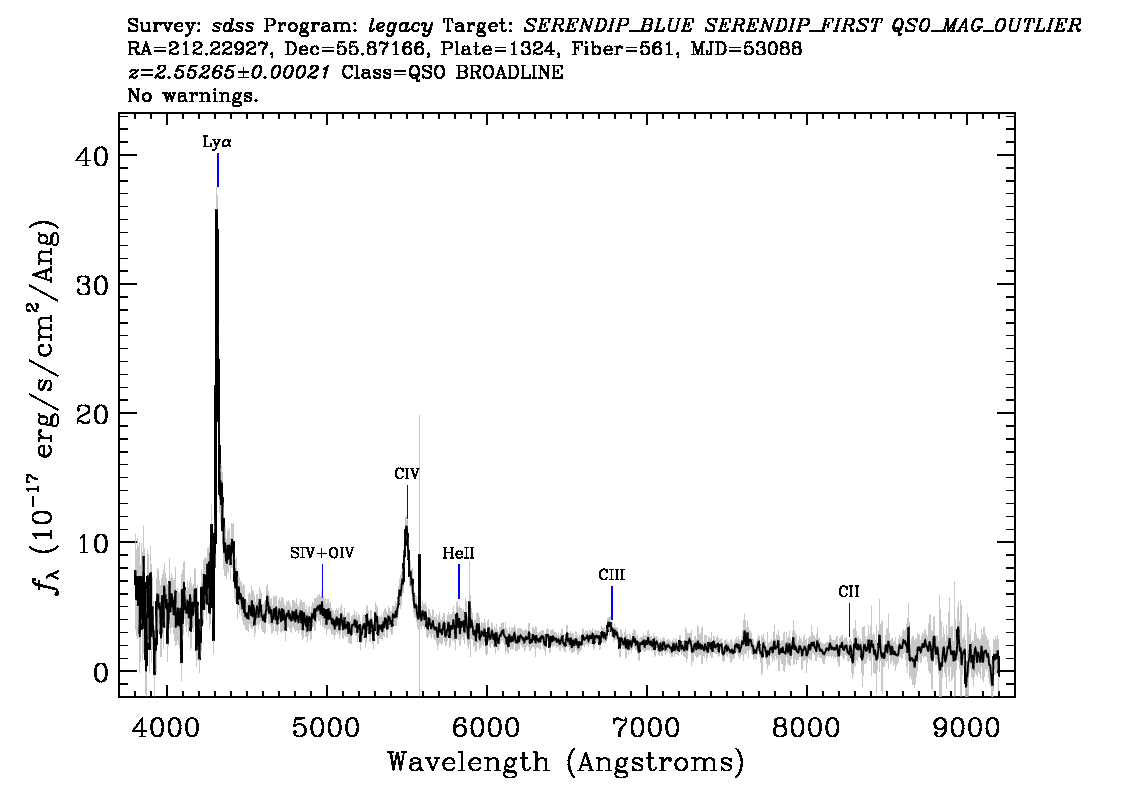}
\includegraphics[scale=0.3]{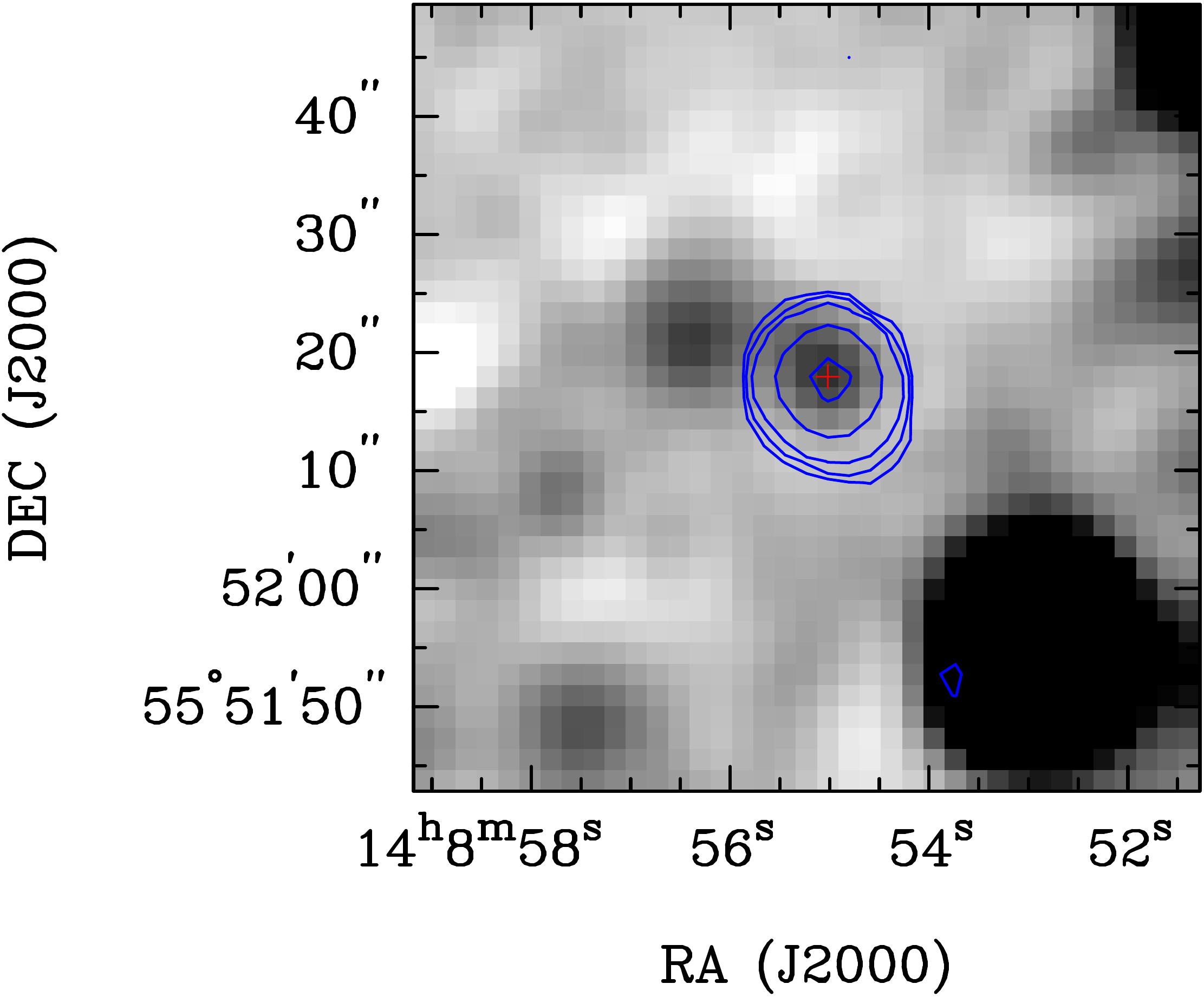}
\includegraphics[scale=0.215,trim=0mm 10mm 30mm 10mm]{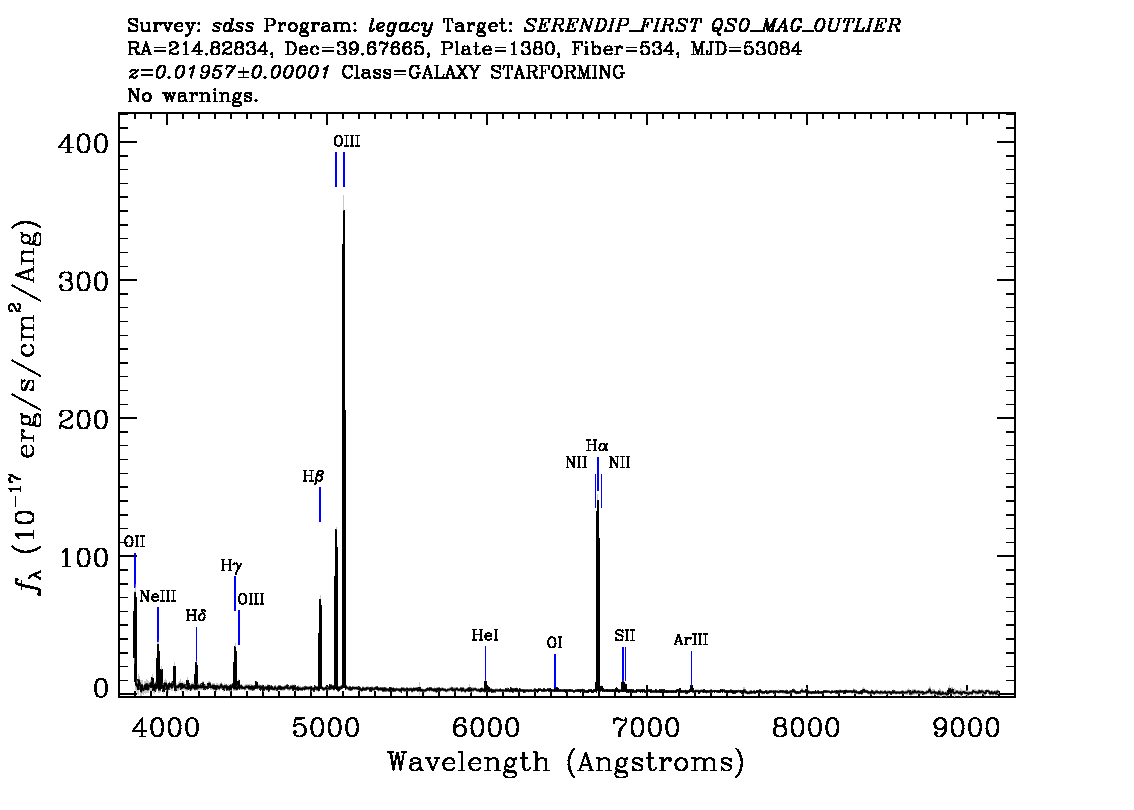}
\label{SFG}
\includegraphics[scale=0.3]{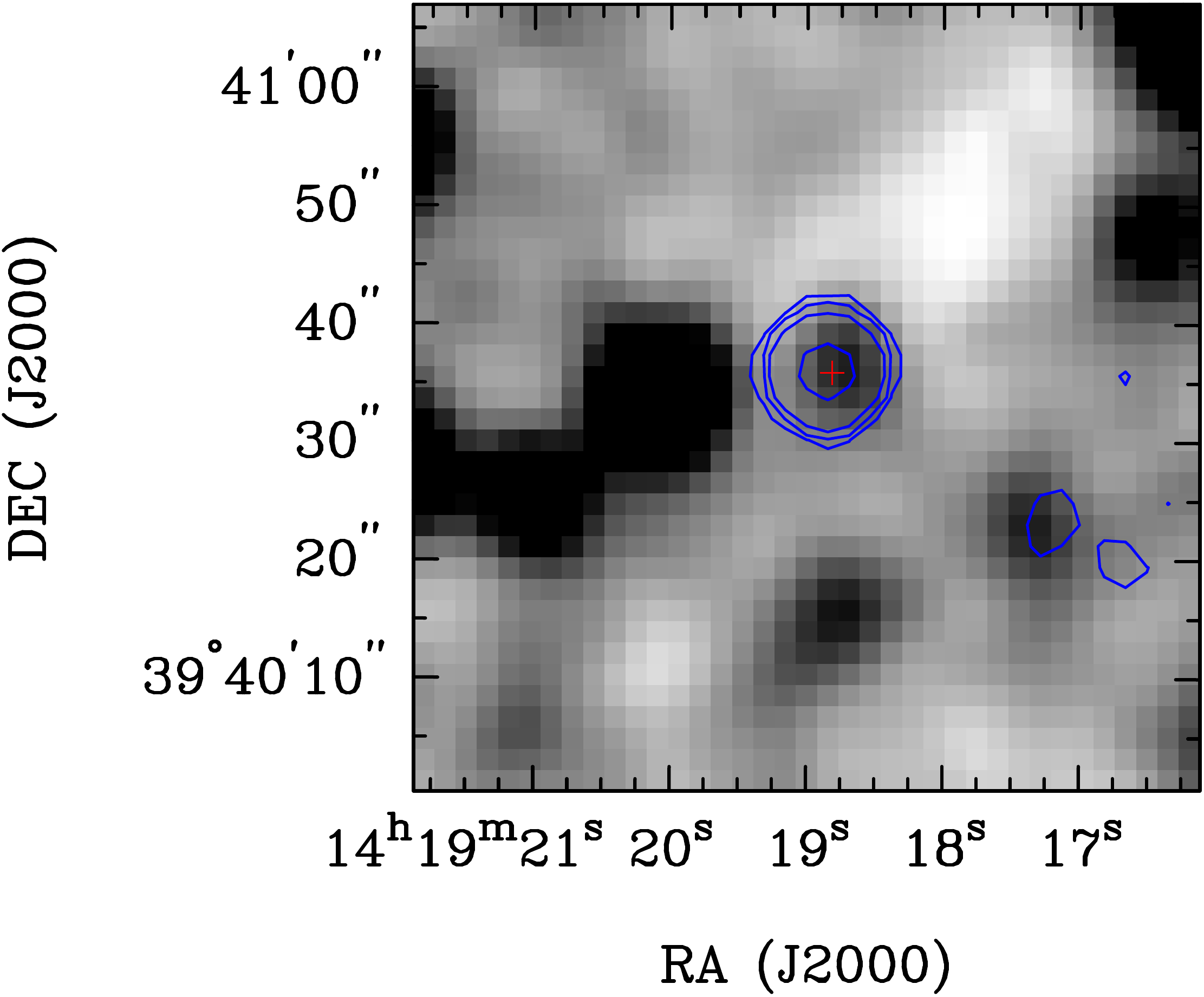}\\
\vspace{4mm}
{{\bf Fig.~\ref{Spectra}.} (continued)}
\end{figure*}

\begin{figure*}
\includegraphics[scale=0.215,trim=0mm 0mm 30mm 10mm]{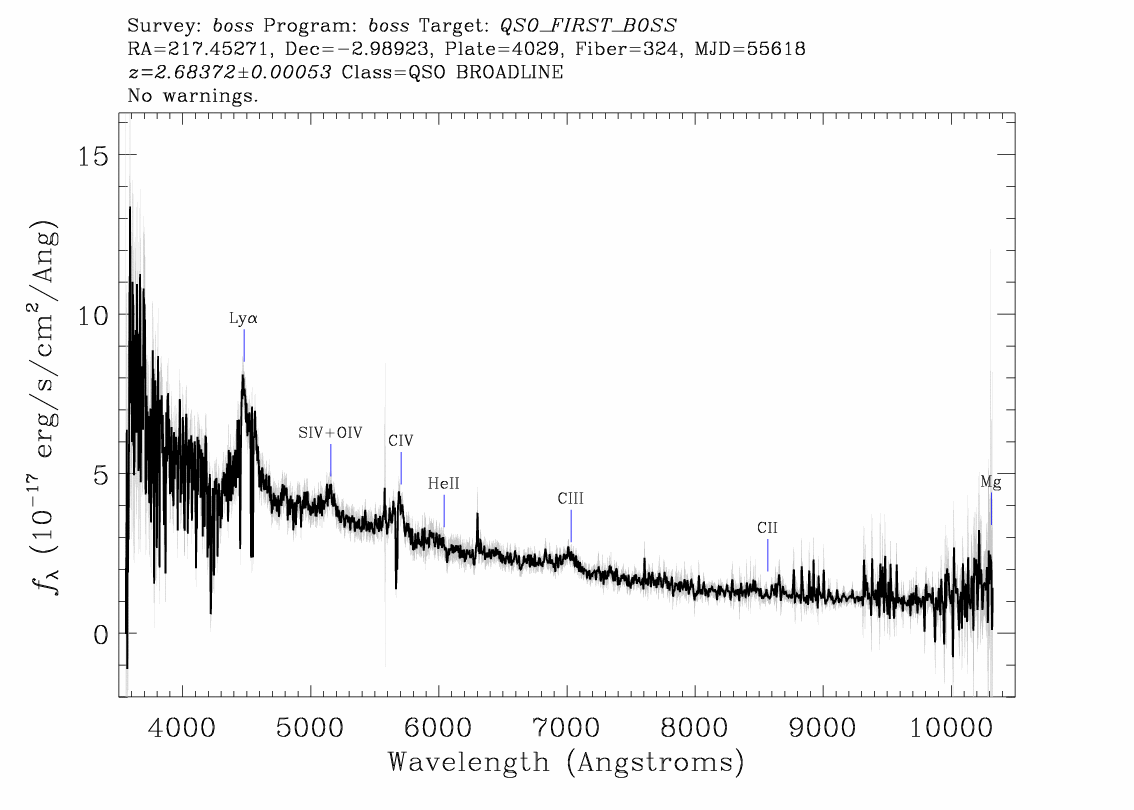}
\includegraphics[scale=0.3]{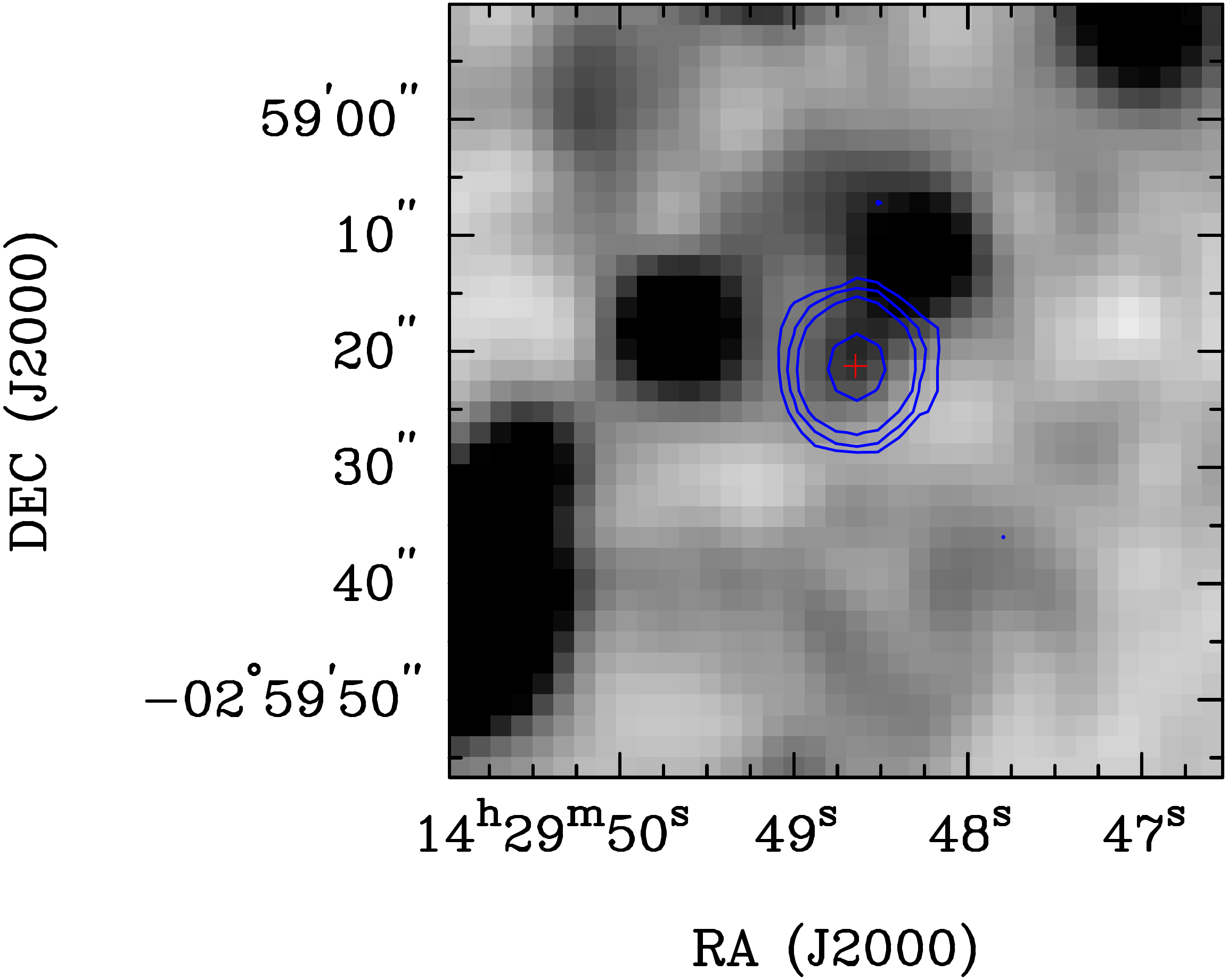}
\includegraphics[scale=0.215,trim=0mm 0mm 30mm 10mm]{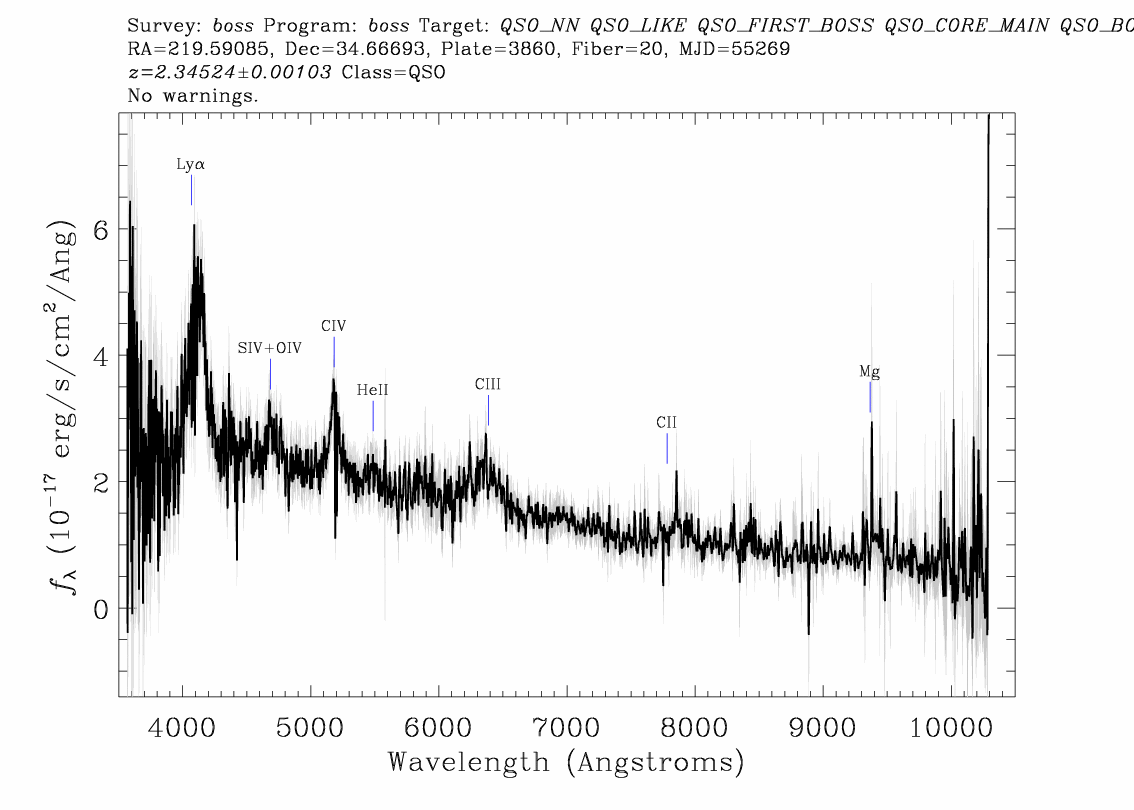}
\includegraphics[scale=0.3]{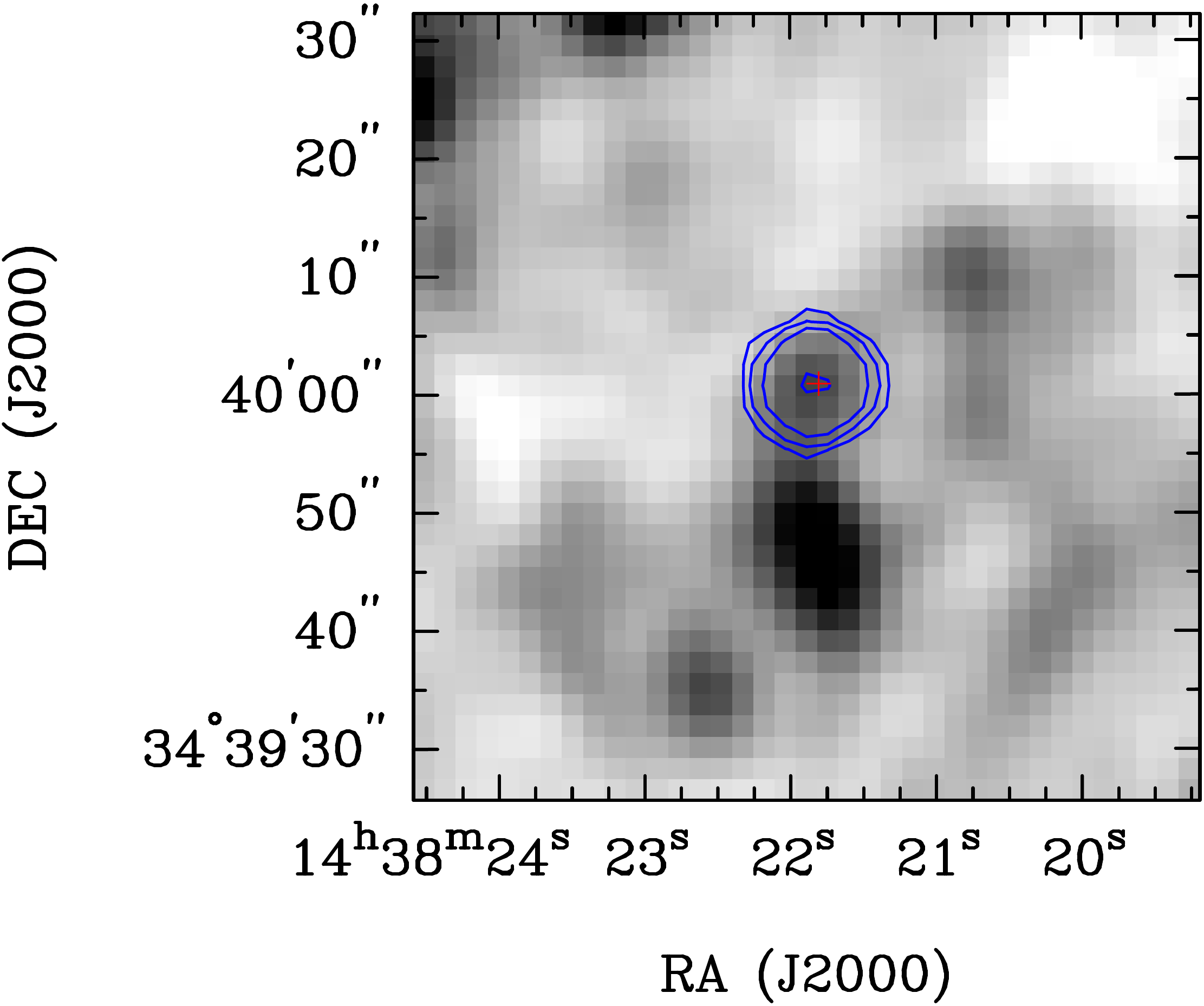}
\includegraphics[scale=0.215,trim=0mm 0mm 30mm 10mm]{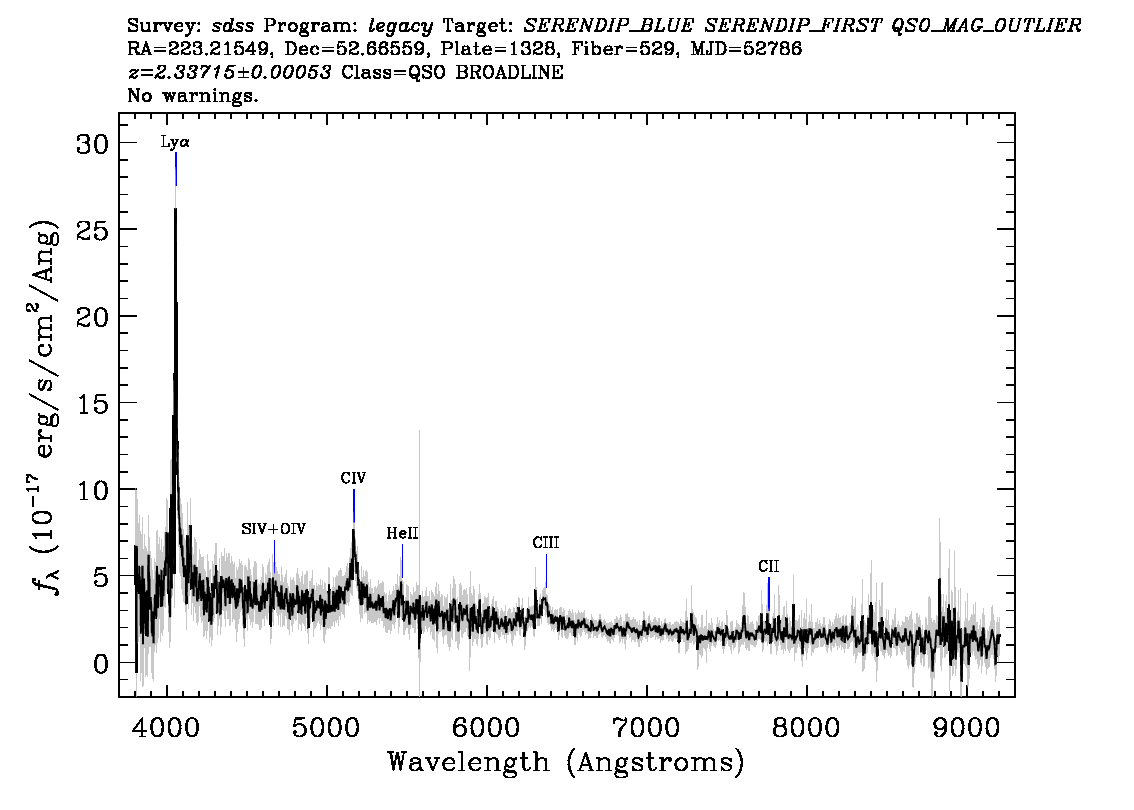}
\includegraphics[scale=0.3]{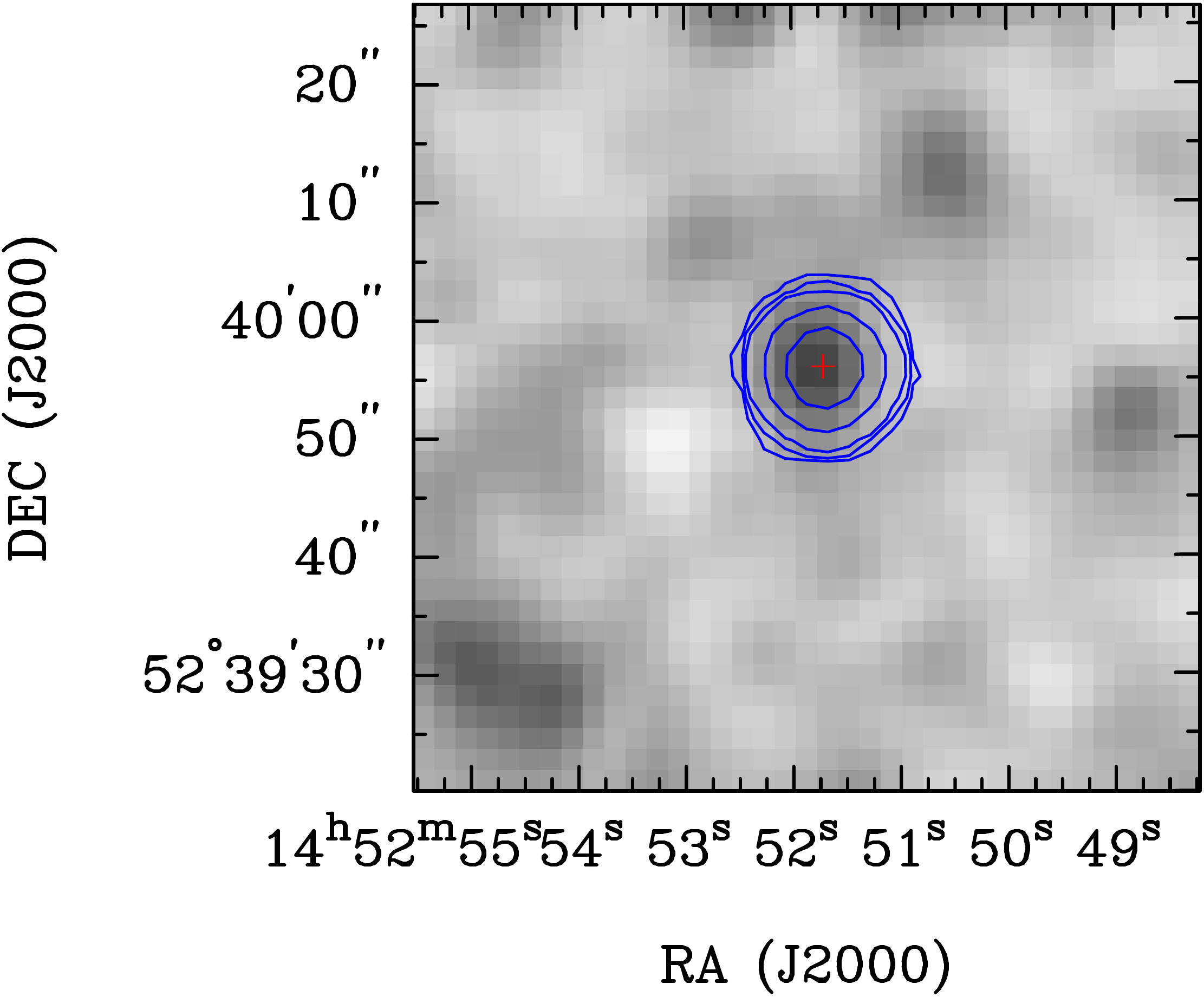}
\includegraphics[scale=0.215,trim=0mm 0mm 30mm 10mm]{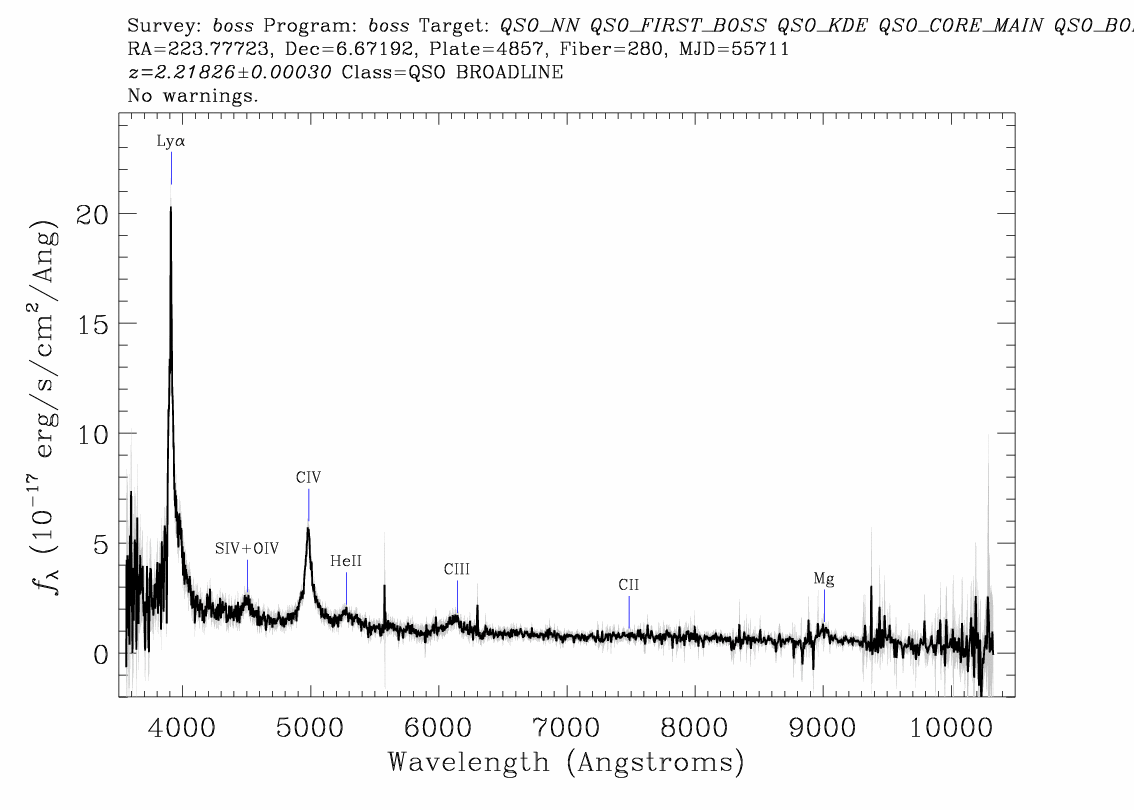}
\includegraphics[scale=0.3]{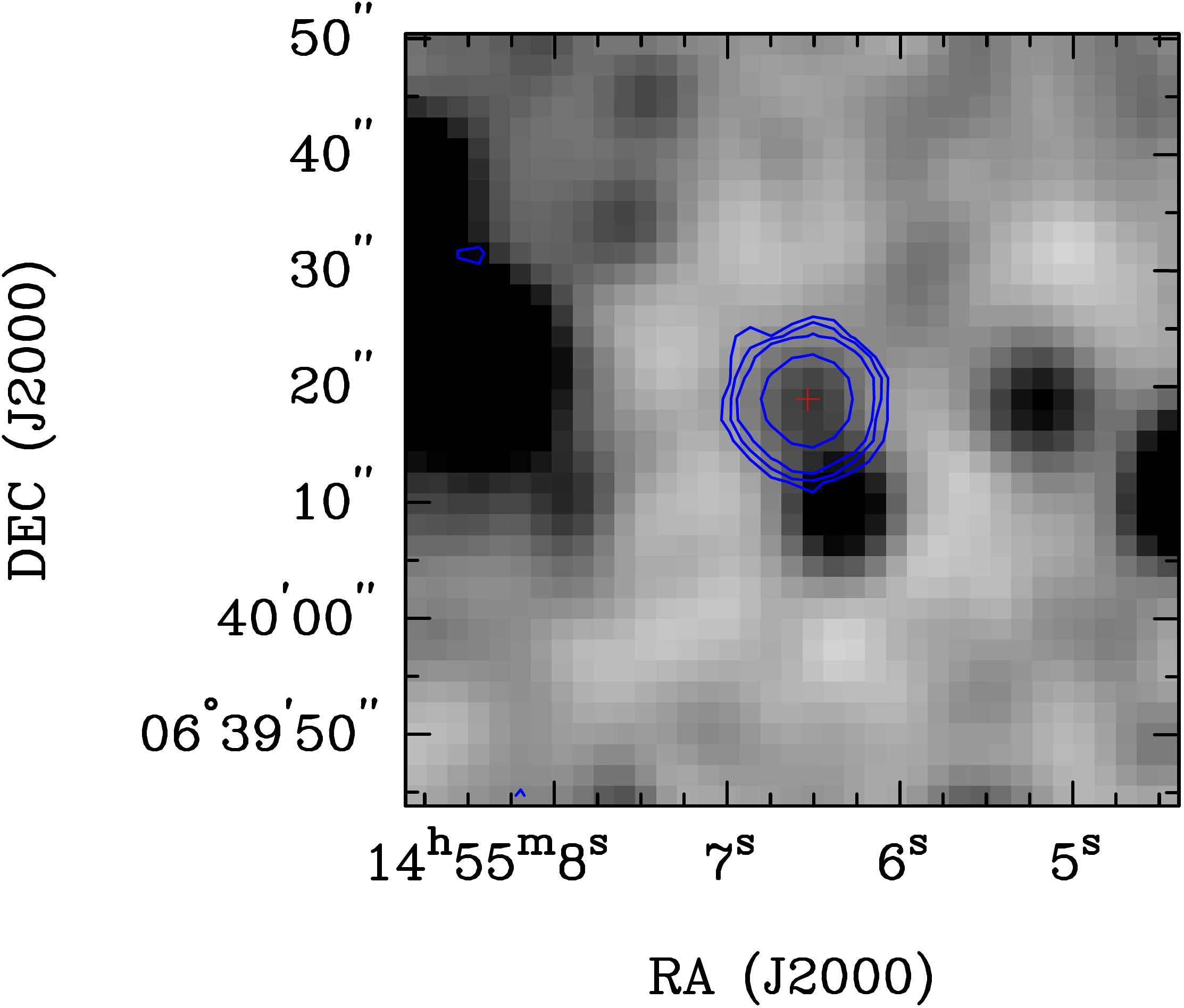}\\
\vspace{4mm}
{{\bf Fig.~\ref{Spectra}.} (continued)}
\end{figure*}

\begin{figure*}
\includegraphics[scale=0.215,trim=0mm 0mm 30mm 10mm]{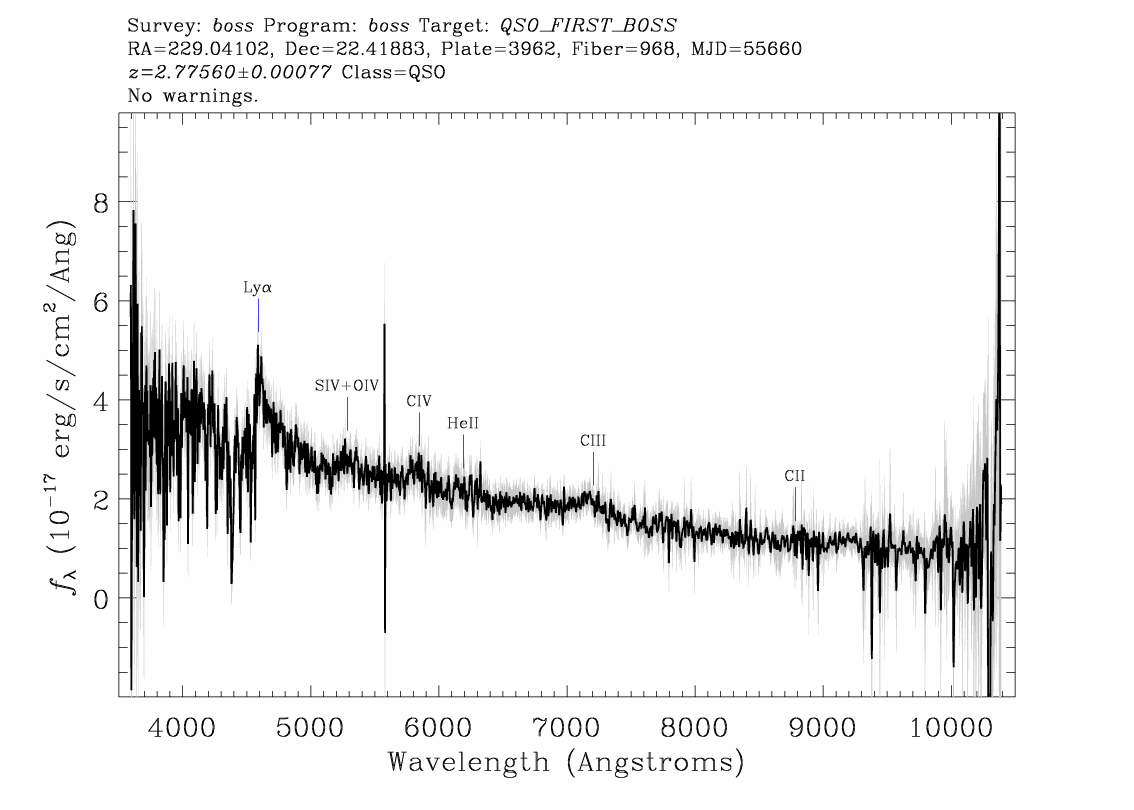}
\includegraphics[scale=0.3]{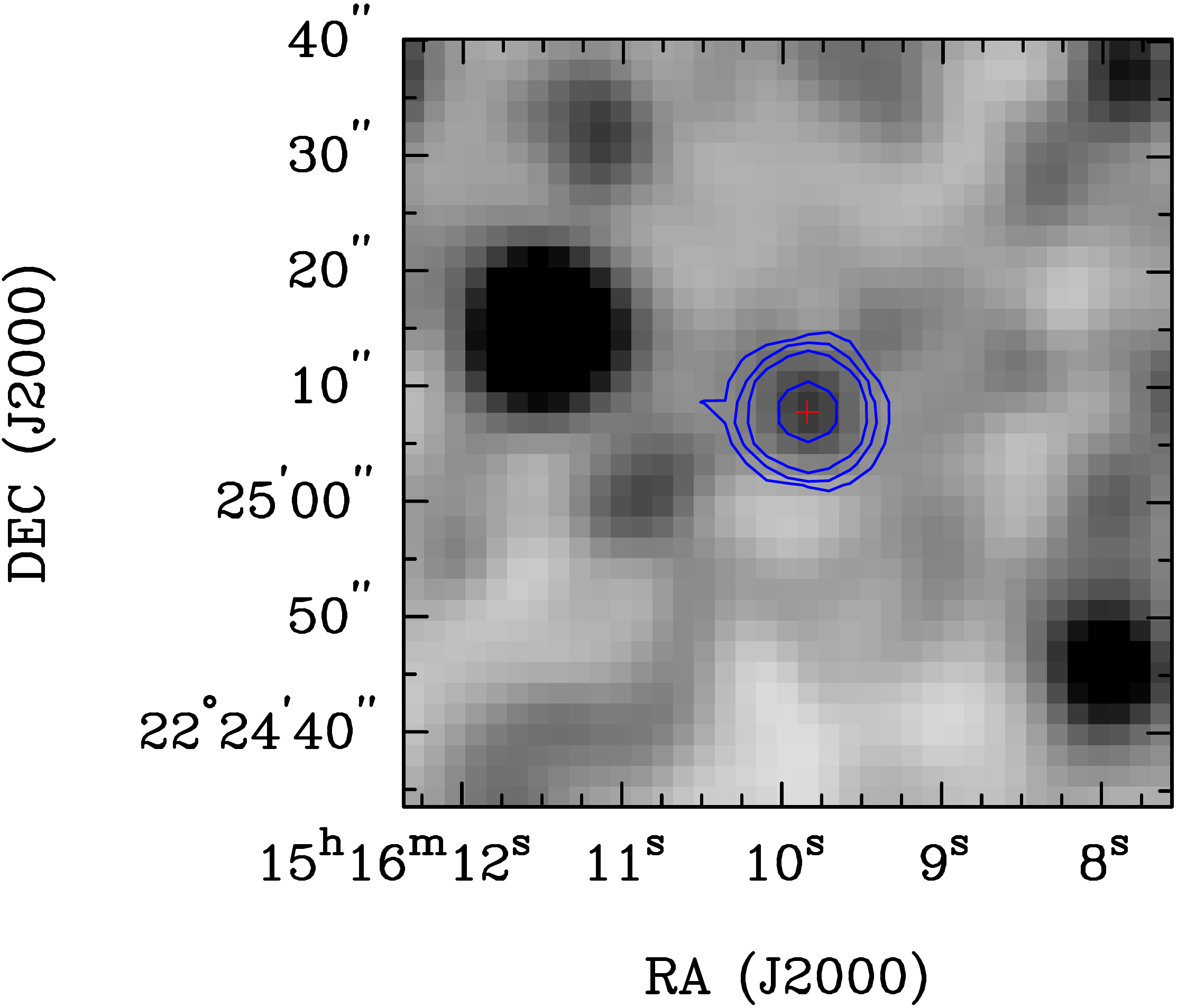}
\includegraphics[scale=0.215,trim=0mm 0mm 30mm 10mm]{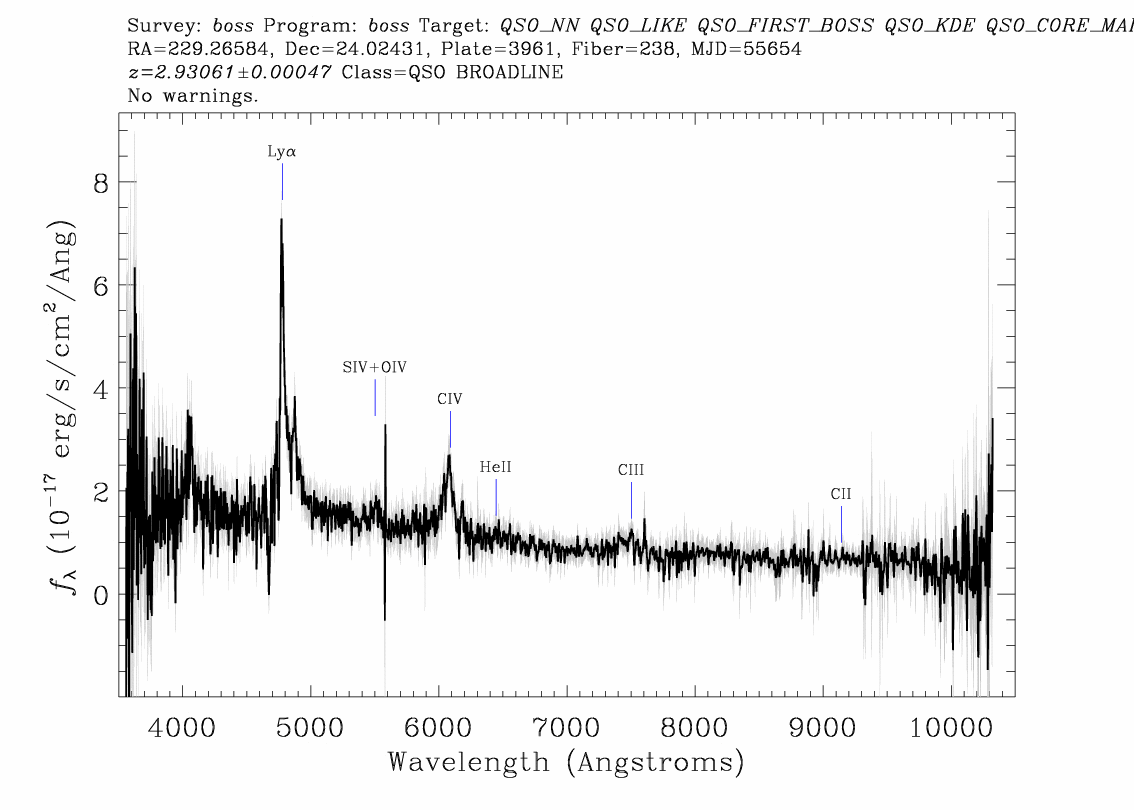}
\includegraphics[scale=0.3]{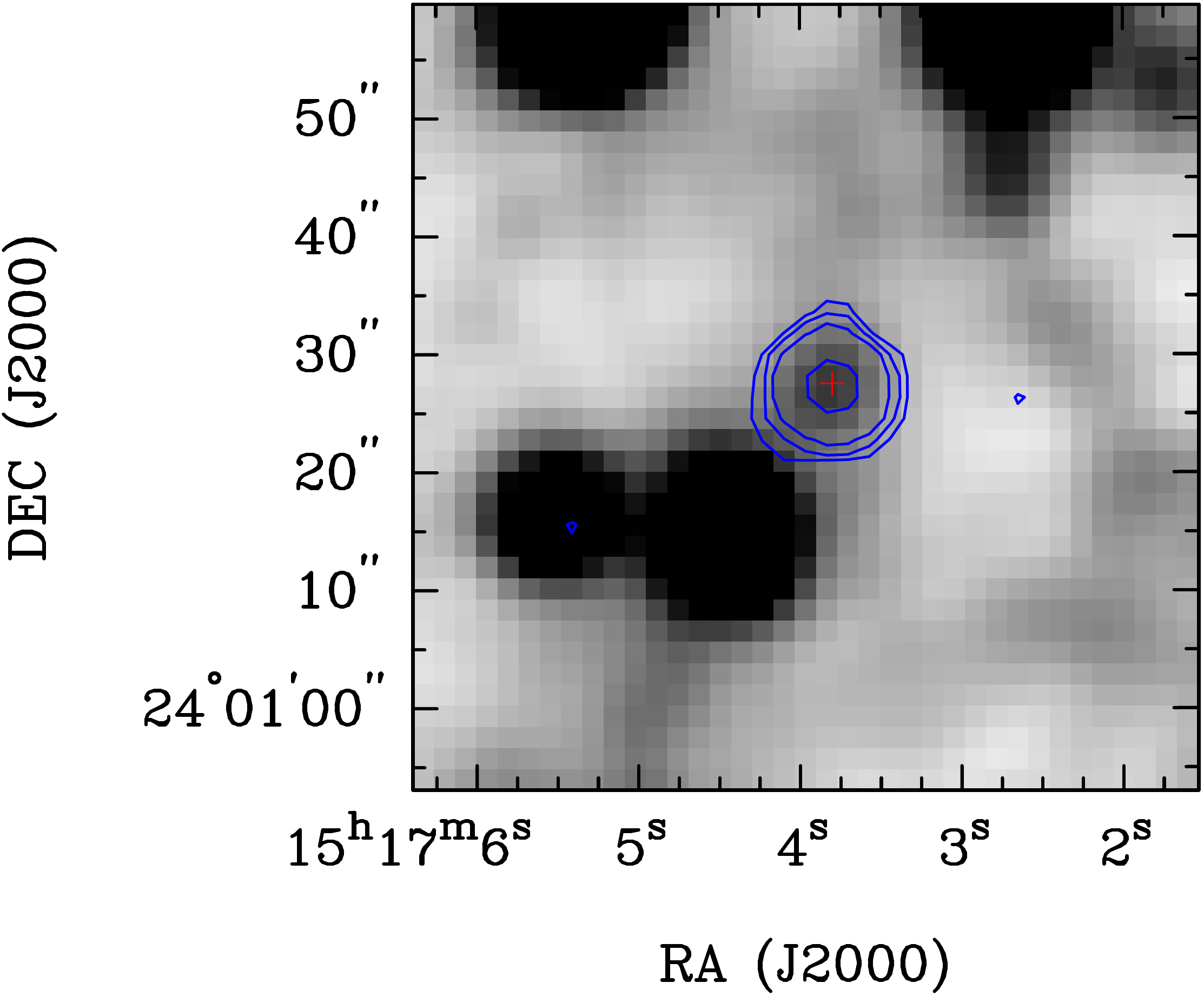}
\includegraphics[scale=0.215,trim=0mm 0mm 30mm 10mm]{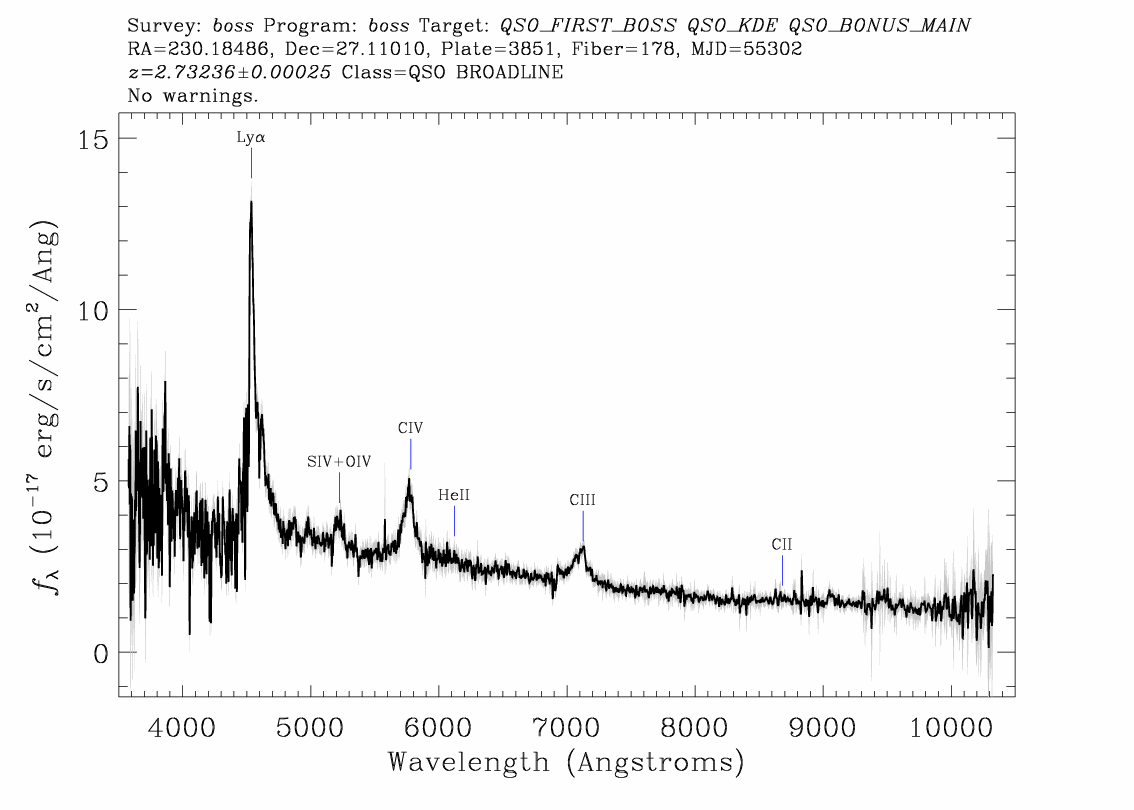}
\includegraphics[scale=0.3]{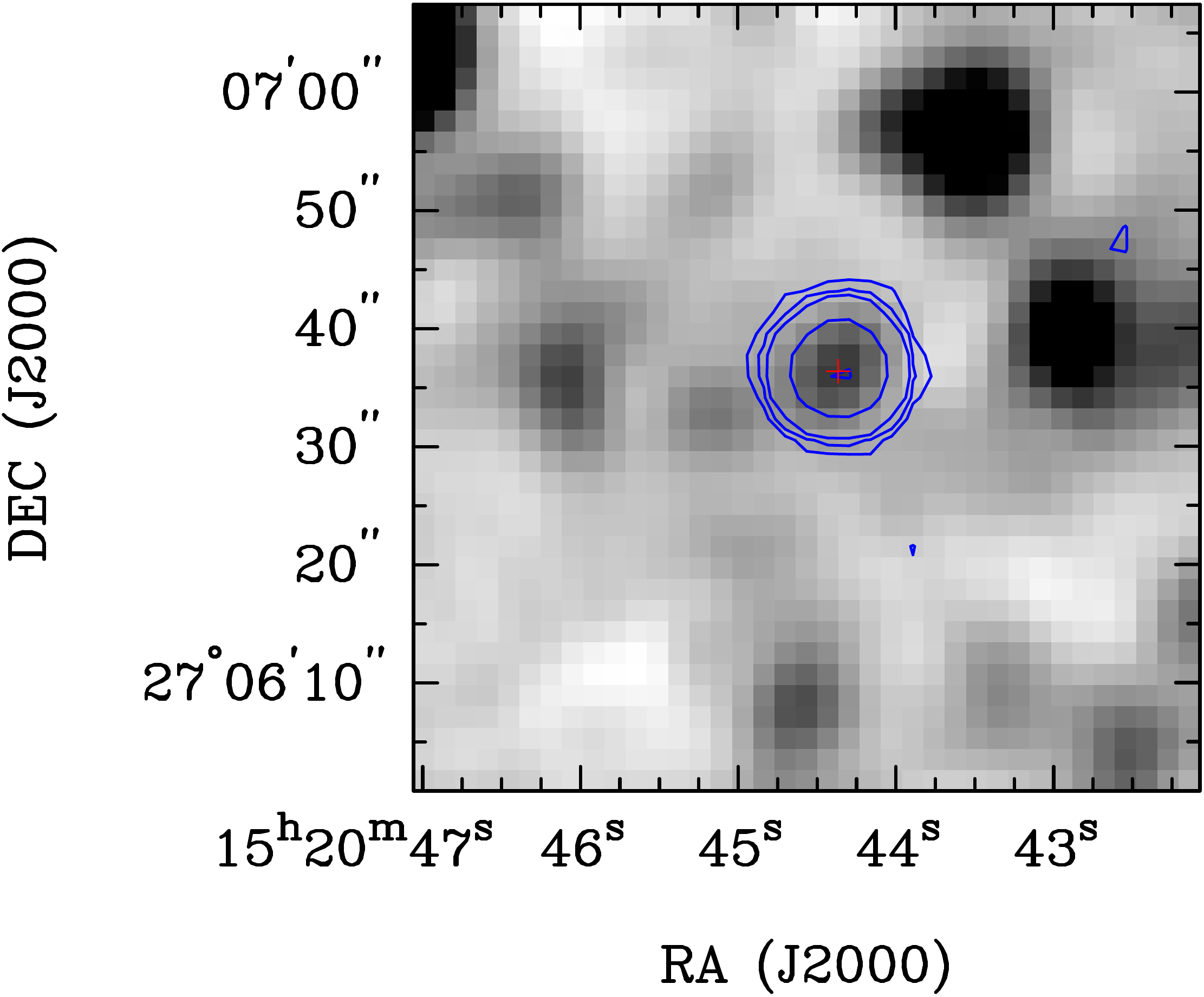}
\includegraphics[scale=0.215,trim=0mm 0mm 30mm 10mm]{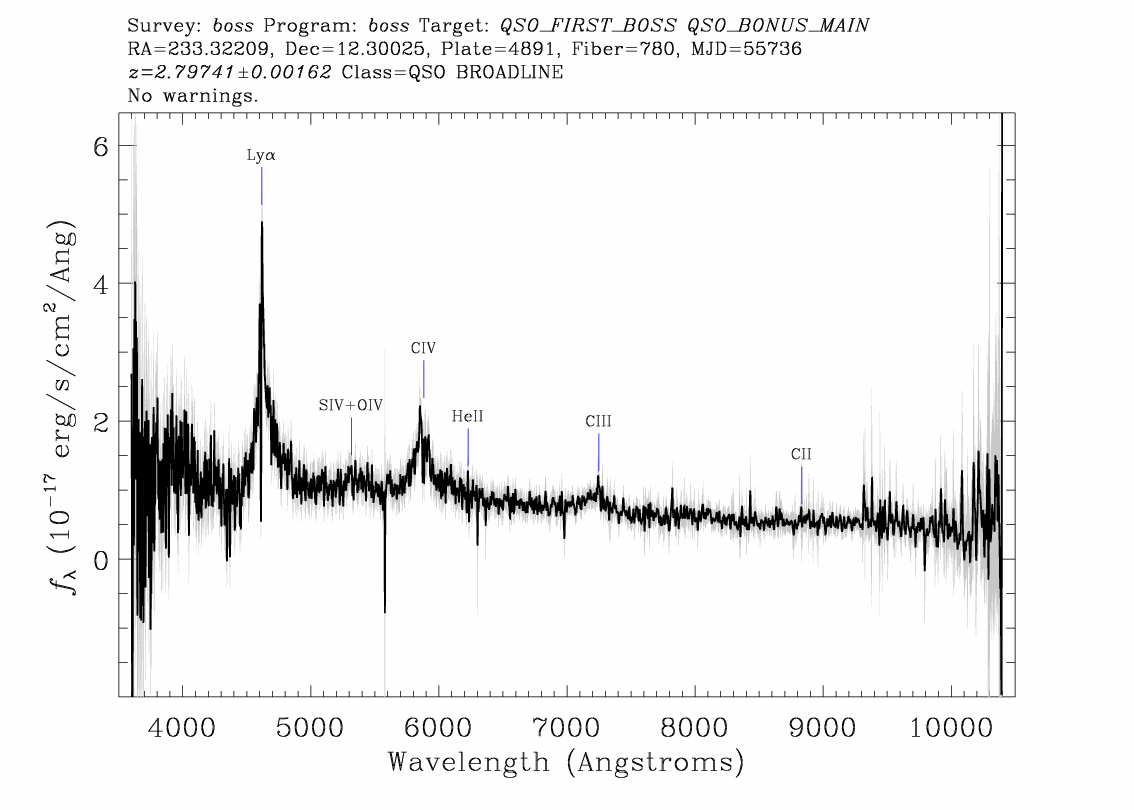}
\includegraphics[scale=0.3]{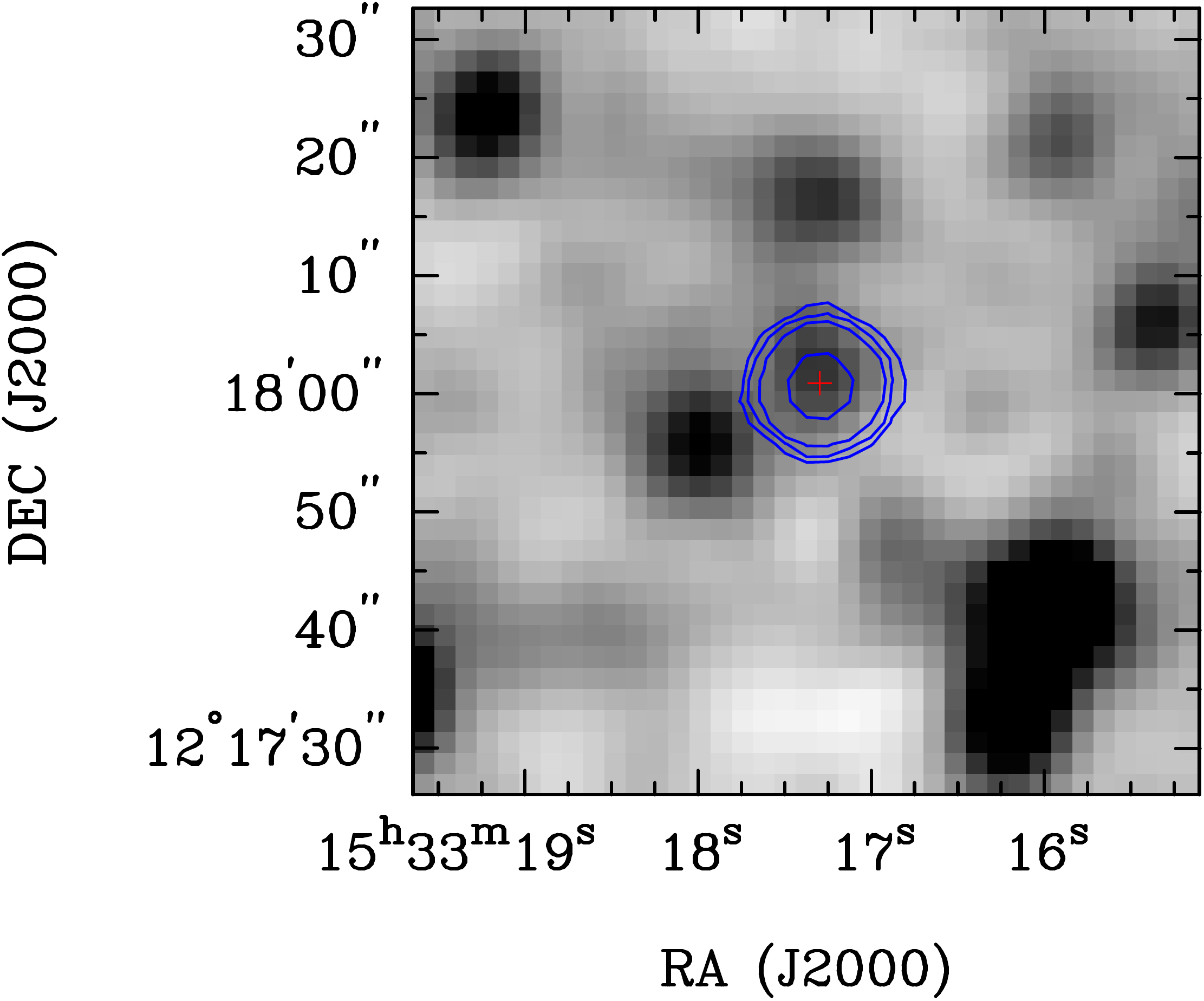}\\
\vspace{4mm}
{{\bf Fig.~\ref{Spectra}.} (continued)}
\end{figure*}

\begin{figure*}
\includegraphics[scale=0.215,trim=0mm 0mm 30mm 10mm]{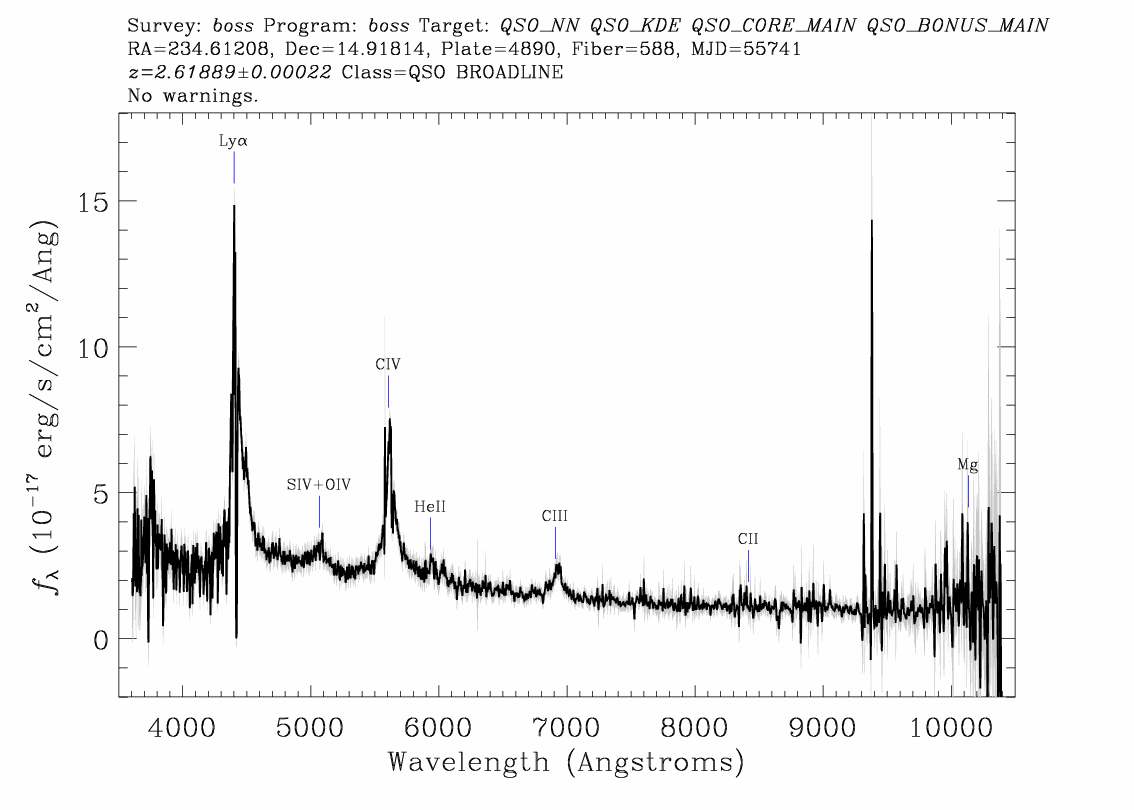}
\includegraphics[scale=0.3]{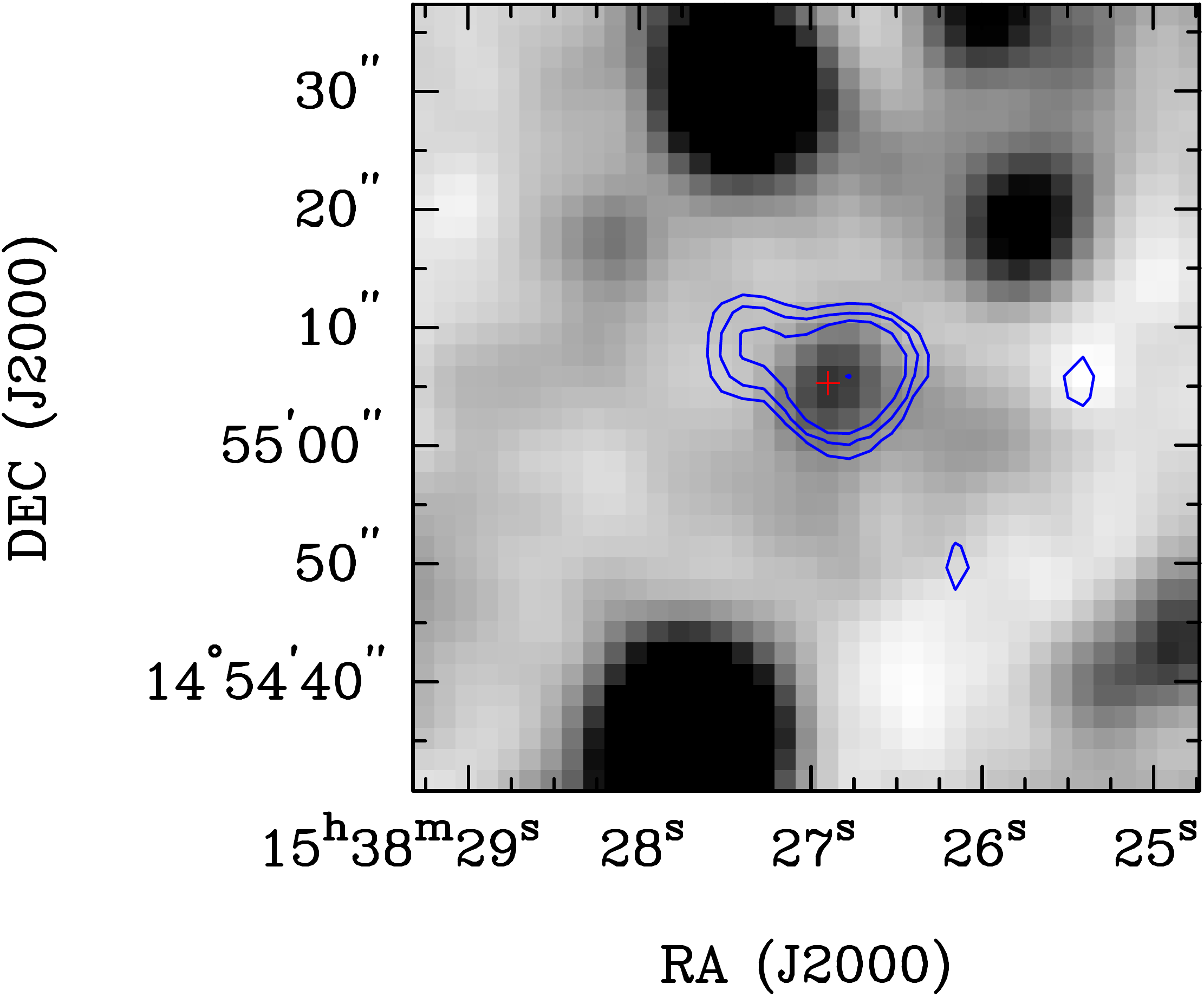}
\includegraphics[scale=0.215,trim=0mm 0mm 30mm 10mm]{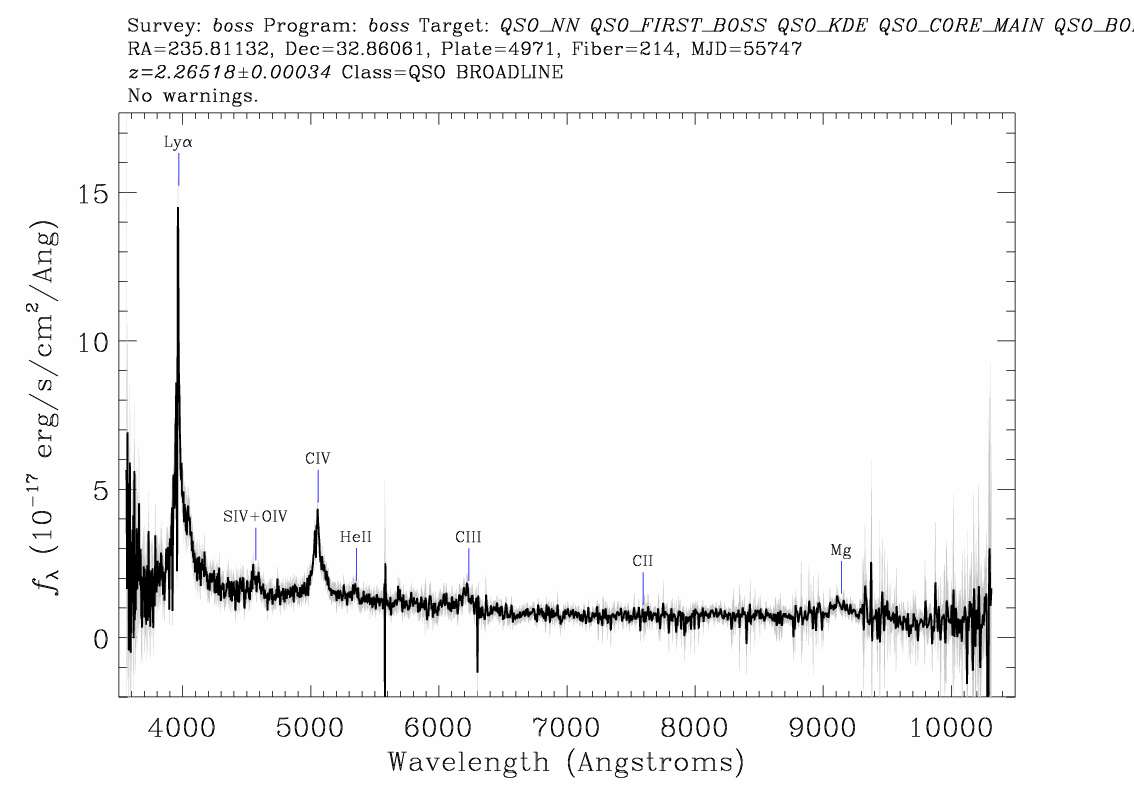}
\includegraphics[scale=0.3]{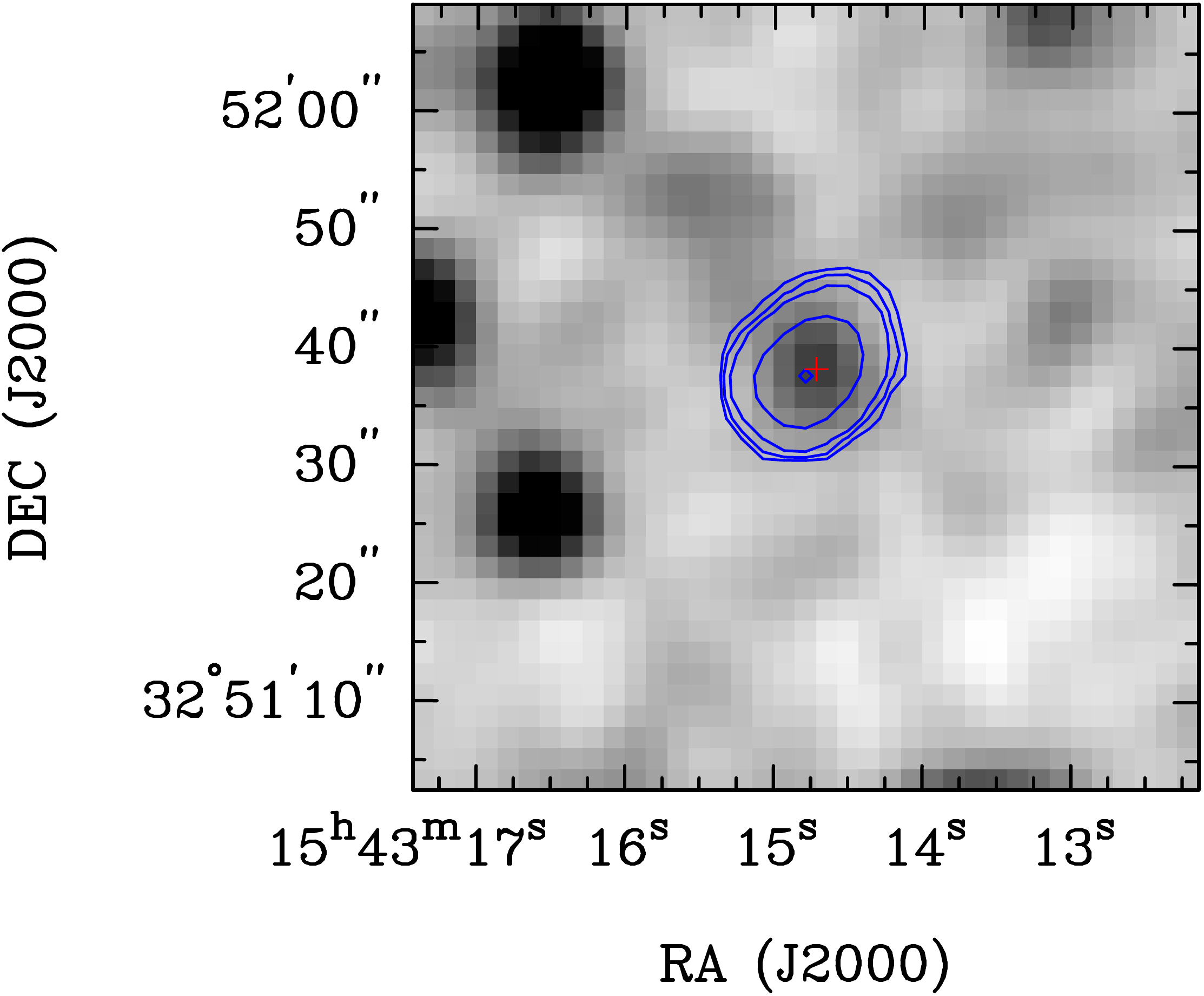}
\includegraphics[scale=0.215,trim=0mm 0mm 30mm 10mm]{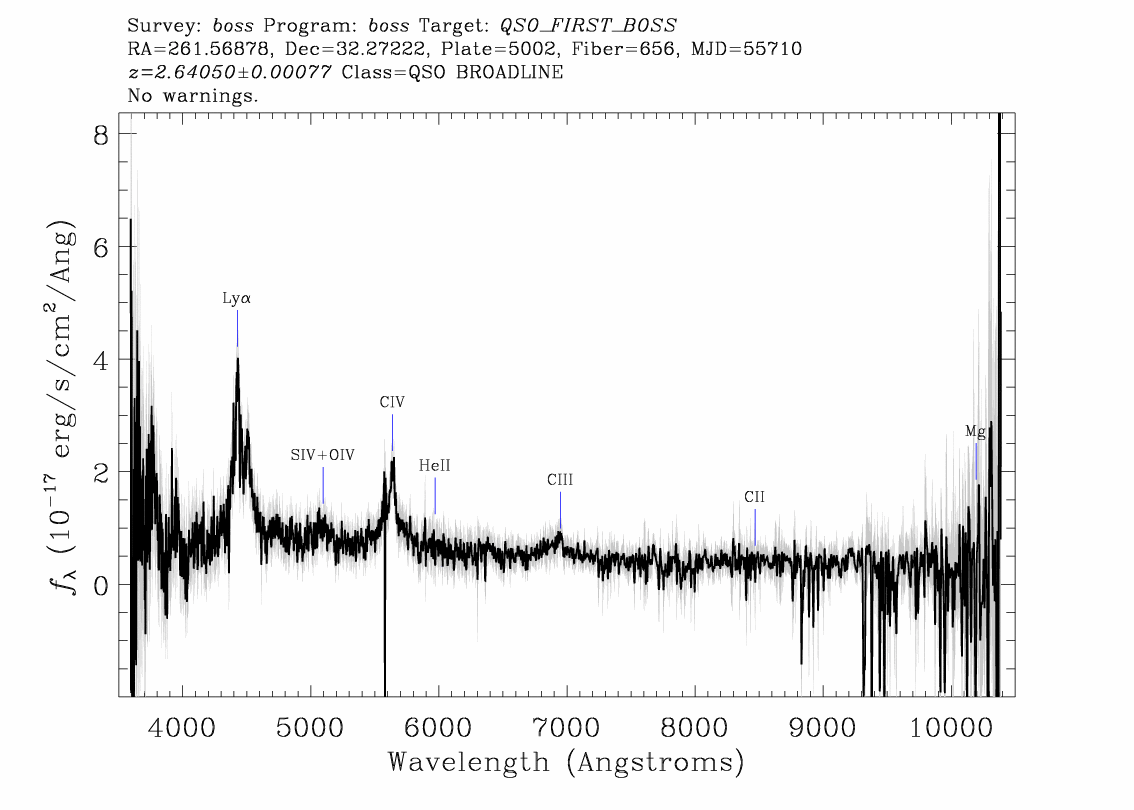}
\includegraphics[scale=0.3]{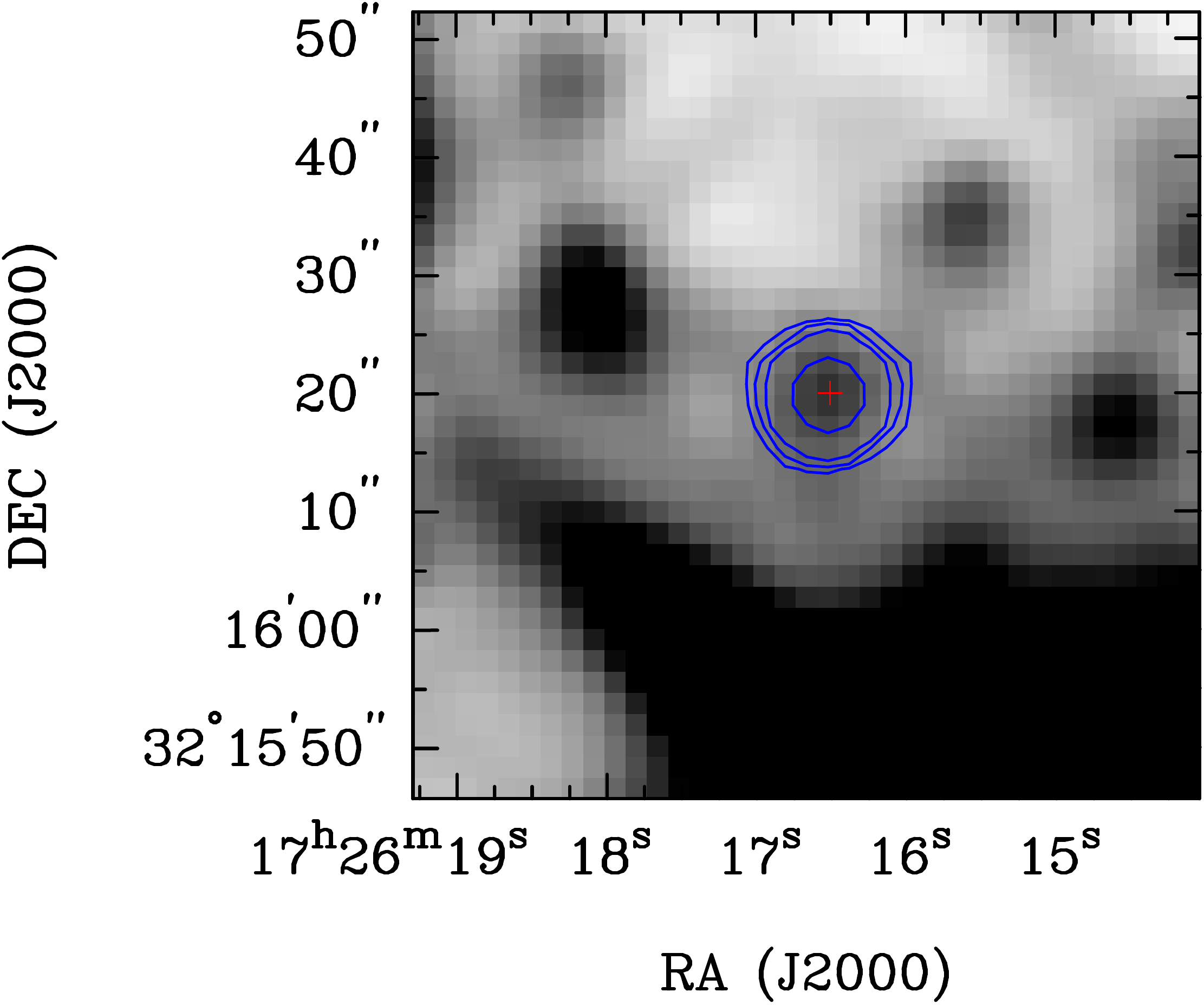}\\
\vspace{4mm}
{{\bf Fig.~\ref{Spectra}.} (continued)}
\end{figure*}

\label{lastpage}
\end{document}